\documentclass{aa}  

\usepackage{graphicx}
\usepackage{txfonts}
\usepackage{hyperref}
\usepackage{xcolor}
\usepackage{orcidlink}

\usepackage{bm}
\usepackage[para]{threeparttable}

\definecolor{Blue}{rgb}{0,0.08,0.65}
\definecolor{Blue2}{rgb}{0,0.4,0.6}

\hypersetup{
    colorlinks,
    linkcolor={red!60!black},
    citecolor={Blue2},
    urlcolor={Blue}
}

\begin{document} 

   \title{Large-scale geometry and topology of gas fields: \\ Effects of AGN and stellar feedback}
   \titlerunning{Large-scale geometry and topology of gas fields}
   
   \authorrunning{C. Schimd, K. Kraljic, R. Dav\'e \& C. Pichon}
  
   \author{
          Carlo~Schimd\inst{1}\orcidlink{0000-0002-0003-1873} \and
          Katarina~Kraljic\inst{2}\orcidlink{0000-0001-6180-0245}\and
          Romeel~Dav\'e\inst{3,4,5} \orcidlink{0000-0003-2842-9434}\and
          Christophe~Pichon\inst{6,7,8}\orcidlink{0000-0003-0695-6735}
          }

   \institute{
            Aix Marseille Univ, CNRS, CNES, LAM, IPhU, Marseille, France \,\,
            \email{carlo.schimd@lam.fr}
         \and
            Observatoire Astronomique de Strasbourg, Universit\'e de Strasbourg, CNRS, UMR 7550, F-67000 Strasbourg, France \,\,
            \email{kraljic@unistra.fr}
         \and
             Institute for Astronomy, University of Edinburgh, Royal Observatory, Blackford Hill, Edinburgh, EH9 3HJ, United Kingdom   
         \and
             University of the Western Cape, Bellville, Cape Town 7535, South Africa
         \and
             South African Astronomical Observatories, Observatory, Cape Town 7925, South Africa
         \and
             Institut d'Astrophysique de Paris, CNRS and Sorbonne Universit\'e, UMR 7095, 98 bis Boulevard Arago, F-75014 Paris, France
        \and
             IPhT, DRF-INP, UMR 3680, CEA, L'Orme des Merisiers, B\^at 774, 91191 Gif-sur-Yvette, France
        \and
             Korea Institute for Advanced Study, 85 Hoegi-ro, Dongdaemun-gu, Seoul 02455, Republic of Korea
             }

    \date{Received January 10, 2024; accepted June 4, 2024}

% Abstract of the paper
\abstract{Feedback from stars and active galactic nuclei (AGNs) primarily affects the formation and evolution of galaxies and the circumgalactic medium, leaving some kind of imprint on larger scales. Based on the {\sc Simba} hydrodynamical simulation suite and using the full set of Minkowski functionals (MFs), this study systematically analyses the time evolution of the global geometry and topology of the gas temperature, pressure, density (total, {\sc Hi}, and {\sc H$_2$}), and the metallicity fields between redshifts $z=5$ and $z=0$. The MFs show that small-scale astrophysical processes are persistent and manifest on larger, up to tens of Mpc scales, highlighting the specific morphological signatures of the relevant feedback mechanisms on these scales in the last $\sim12$~Gyr. In qualitative terms, we were able establish a ranking that varies according to the field considered: stellar feedback mostly determines the morphology of the pressure and density fields and AGN jets are the primary origin of the morphology of the temperature and metallicity fields, while X-ray heating and AGN winds play the second most important role in shaping the geometry and topology of all the gaseous fields, except metallicity. 
Hence, the cosmic evolution of the geometry and topology of fields characterising the thermodynamical and chemical properties of the cosmic web offers complementary, larger scale constraints to galaxy formation models.
}

\keywords{Galaxies: evolution, formation, intergalactic medium -- Cosmology: large-scale structure of Universe}

\maketitle

% ========================================
\section{Introduction}

Astrophysical thermal and non-thermal processes due to stellar and active galactic nucleus (AGN) activity, jointly with hydrodynamical and tidal gravitational interactions acting on top of gravitational collapse, determine the physical evolution of galaxies and their local environments \citep{Hummels2013,CGM2017,Appleby2023M}. On such small scales, even though the very intricate physics yields highly non-Gaussian random fields, low-order statistics such as number counts or univariate statistics and two-point correlation functions are the measures routinely adopted to characterise the spatial distribution of matter fields \citep{Scannapieco2006}. Still, the exquisite quality and spatial coverage of current and forthcoming data sets are pushing towards the investigation of the relationship between the local and larger scale dynamics, including possible signatures of feedback physics. This improvement 
necessarily opens to the use of more sophisticated statistical analyses.

Since the advent of cosmological surveys, many higher order statistics have been used extensively to study the geometry and topology of the matter density field on large scales, from a few megaparsecs up to gigaparsec. Among them, Minkowski functionals (MFs) are of particular interest as they offer a complete and global description of the morphology of spatial patterns, largely employed in several domains of natural sciences and for image analysis \citep[e.g.][]{MeckeStoyan2002,MantzJacobsMecke2008}. Using integral rather than differential expressions, they are therefore robust against small-scale spatial fluctuations and have a very simple interpretation in two and three dimensions \citep{Adler1981}. Introduced in cosmology by \citet{MeckeBuchertWagner1994}, MFs supplemented the seminal studies by \citet{GottMellotDickinson1986} and \citet{Coles+1993} on the topology of the large-scale structure (LSS) based on the genus curve and Euler-Poincar\'e characteristic \citep[see also][]{Melott+1989,Gott+1990,ParkGott1991,MeckeWagner1991,Park+1992,Vogeley1994b,ColesDaviesPearson1996,SchmalzingBuchert1997,Park+2001,BerianJames+2009,ParkKim2010,Zunckel+2011,Choi+2013}. Moreover, MFs include the percolation analysis \citep{Shandarin1983,ShandarinZeldovich1983,Yess+1997,ShandarinYess1998,Colombi2000} and the void probability function \citep[e.g.][]{White1979,Vogeley+1994a}. They are also intimately related to alpha-shapes and Betti numbers \citep[][]{vandeWeygaert+2011,Park+2013,Pranav+2019,Feldbrugge+2019}.

Encompassing the two-point correlation functions \citep[e.g.][]{KerscherSchmalzingBuchert1996,Kerscher+1998}, MFs have been used in cosmology to investigate the primordial non-Gaussianity of the matter field producing CMB radiation \citep[e.g.][]{SchmalzingGorski1998,SchmalzingTakadaFutamase2000,HikageMatsubara2012,Ducout+2013,Modest+2013,Munshi+2013,BuchertFranceSteiner2017} and affecting the galaxy distribution on large scales \citep[e.g.][]{Hikage+2006}. They have also been used to study the gravitational non-Gaussianity of the late-time galaxy density field for cosmography applications \citep[e.g.][]{BlakeBerianJamesPoole2014,WiegandBuchertOstermann2014,FangLiZhao2017,SullivanEisensteinWiegand2019,Liu+2020,Appleby+2022} and the projected dark-matter field from weak-lensing convergence maps \citep[e.g.][]{Kratochvil+2012,Petri+2013}, as well as in extragalactic physics to assess the galaxy assembly history \citep{Hikage+2003,Einasto+2011} and the morphology of the {\sc Hi} density field during and after the epoch of reionisation \citep[e.g.][]{Gleser+2006,Yoshiura+2017,Chen+2019,SpinaPorcianiSchimd2021}.

All the aforementioned applications concerned the morphology of total or baryonic (biased) matter density fields. The question is whether the geometric and topological content of other physical fields that characterise the thermodynamical and chemical properties of the gaseous component offer relevant and complementary information on the process of galaxy formation in general and on the impact of feedback on a larger scale in particular. These fields and their associated observables, such as the X-ray brightness \citep[e.g.][]{Gilli2007,2020MNRAS.496.4221J}, Sunyaev?Zel'dovich Compton-$y$ maps \citep[e.g.][]{Bleem2022,Yang2022}, Lyman-$\alpha$ absorptions \citep[e.g.][]{LyMAS2014,Lee2018A,Japelj2019A&A,Kraljic2022}, or the CO and $K$-band luminosities \citep[e.g.][]{BolattoWolfireLeroy2013,Garratt+2021} are usually considered on galactic or galaxy clusters scales, since they are associated with local hydrodynamical and feedback processes. The properties of MFs make them privileged statistics to evaluate whether their small-scale effect manifests itself on larger scales and with what delay; namely, they are used to assess how and when the physics of galaxies and circumgalactic medium affect the physics of the intergalactic medium. Indeed, MFs are robust to small-scale fluctuations, so their signal is plausibly measurable even when integrated on scales larger than a few megaparsecs. When compared with suitable observables, this can in turn put additional valuable constraints on 
often poorly understood baryonic processes, such as stellar and AGN feedback. The latter, namely the energy release from black holes, is expected to have a significant impact on the properties of galaxies, galaxy groups, and galaxy clusters, but also on their outskirts in the circumgalactic medium and beyond, as hot gas bubbles percolate \citep[e.g.][]{Fabian2012,Sorini2022,Ayromlou+2023}.
Indeed, the sources are mainly distributed along the cosmic web, so hot baryons are expected to percolate faster along filaments; the percolation process should then be boosted when the ionisation front enters an already ionised region.

Large-scale cosmological hydrodynamic simulations such as {\sc Horizon-AGN}~\citep{Dubois2014}, {\sc Illustris}~\citep{vogelsbergeretal14}, {\sc Eagle}~\citep{schayeetal15}, {\sc IllustrisTNG}~\citep{pillepichetal18}, {\sc Simba}~\citep{Dave2019}, {\sc Extreme-Horizon}~\citep{chabanieretal20}, and {\sc Horizon Run 5}~\citep{leeetal21} are well suited to assess the impact of feedback processes on the large-scale morphology of the matter fields. Among them, the cosmological hydrodynamic simulation suite {\sc Simba} stands out, as it implements several recipes of stellar and AGN feedbacks on top of hydrodynamical interactions; thus, monitoring the gas temperature, pressure, neutral atomic and molecular hydrogen density, and metallicity.

Here, {\sc Simba} is used to address the question whether the MFs of thermodynamical fields can be used to distinguish between various feedback processes, in particular stellar feedback from various components of AGN feedback such as AGN winds, jets, and X-ray heating. The global MFs of the excursion sets of the 3D gas fields are analysed as a function of the field value (or threshold) and redshift from $z=5$ to $z=0$, separately considering the effect of feedback recipes. Lacking motivated models of the spatial statistics, the analysis is limited to qualitative considerations; some quantitative summary statistics extracted from the Minkowski curves are then considered to describe the morphological time evolution of gas fields, beyond the evident observation that they are not Gaussian. The investigation of observed fields, usually projected on the celestial sphere (therefore requiring MFs in two dimensions) or in the 3D redshift space (and thereby affected by redshift-space distortions) is left for future studies.

The purpose of this study is to set up a physical framework in which to address such questions as how feedback impacts the geometry of the intergalactic medium (IGM); how  the thermodynamic state and chemical content of the gas impact its topology; how the large-scale morphology is correlated with the major epochs of the cosmic evolution of star formation and active galactic nucleus feedback; whether there is any delay between the phenomena on large and small scales; and, finally, whether this delay simply reflects a propagation time or whether there are  non-trivial percolation effects related to the known anisotropy of the cosmic web. 

The remainder of this paper is structured as follows. 
The {\sc Simba} suite and the mathematics of MFs are introduced in Section~\ref{sec:method}. The analysis of MFs as a function of redshift, separately for individual feedback processes is presented in Section~\ref{sec:results}. The overall picture is discussed in Section~\ref{sec:discussion}. Finally, Section~\ref{sec:conclusions} presents our conclusions.

% ========================================
\section{Method}
\label{sec:method}

% ----------------------------------------
\subsection{The {\sc Simba} suite: baryonic physics on large scales}
\label{sec:method:simba}

The {\sc Simba} simulation is fully described in \cite{Dave2019}; therefore, here, we provide  only its summary and focus on its features relevant to this study. 
{\sc Simba} was run with a modified version of the gravity plus hydrodynamics solver {\sc Gizmo} \citep{Hopkins2015}, employing the {\sc Gadget-3} tree-particle-mesh gravity solver \citep{Springel2005} and a meshless finite mass solver for hydrodynamics. 

Radiative cooling and photo-ionisation heating models make use of the {\sc Grackle-3.1} library \citep{Smith2017} accounting for metal cooling and non-equilibrium evolution of primordial elements. A spatially uniform UV ionising background follows the \cite{HaardtMadau2012} model, modified to account for self-shielding based on \cite{Rahmati2013}. The neutral hydrogen ({\sc Hi}) content of gas particles is modelled self-consistently.  

{\sc Simba} implements star formation based on molecular hydrogen ({\sc H$_2$}) in a Jeans mass-resolving pressurised interstellar medium (ISM; with the hydrogen density $n_\mathrm{H} \geq 0.13$~cm$^{-3}$), following \citet{Dave2016}.  The computation of the H$_2$ fraction is based on the \citet{KrumholzGnedin2011} prescription. The chemical enrichment model tracks eleven different elements (H, He, C, N, O, Ne, Mg, Si, S, Ca, and Fe) from type Ia and II supernovae (SNe), and asymptotic giant branch (AGB) stars,  following the yield tables of \cite{Iwamoto1999}, \cite{Nomoto2006}, and \cite{OppenheimerDave2006}, respectively.
{\sc Simba} also tracks dust growth and destruction for each individual element on the fly (see \citealt{Li2019} for a detailed investigation of the dust model, and \citealt{Donevski2020} for comparison with observations). The stellar feedback is modelled using decoupled galactic winds, namely, with the hydrodynamics turned off in the winds until they leave the ISM, metal-loaded kinetic two-phase galactic winds, and with 30\% of wind particles being ejected with the temperature set by the kinetic energy of the wind subtracted from the supernova energy.

{\sc Simba} includes black hole (BH) particles, employing a specific two-mode accretion model. Hot gas ($T > 10^5$~K) is accreted in a spherically symmetric fashion following the \citet{Bondi1952} formula, while cold gas accretion is modelled via a torque-limited sub-grid prescription describing the response of gas inflows near the BH to angular momentum loss due to dynamical instabilities \citep{HopkinsQuataert2011,Angles-Alcazar2017}. 
As Bondi accretion models gravitational capture from a dispersion-dominated medium, it is more appropriate for hot gas. The torque-limited accretion mode is instead well suited for the growth of BHs in rotationally supported disks.  
This unique combination of BH accretion modes determines the implementation of feedback from active galactic nuclei (AGNs) in the form of two-mode kinetic feedback, the so-called radiative mode and jet mode feedback:
\begin{itemize}
    \item \textit{Radiative mode:} BHs with high accretion rates (above 0.2 times Eddington rate) and mass above 10$^{7.5}~\rm{M}_\odot$ eject material in $\sim 1000$-km/s winds without changing its temperature. This is consistent with observations of ionised multi-phase gas outflows \citep{Perna2017}.   
    \item \textit{Jet mode:} As the BH accretion rate drops below 0.2 of the Eddington rate, jet feedback mode turns on and is fully achieved below 0.02. Gas is ejected at much higher velocities compared to the radiative mode, with a velocity increment proportional to the logarithm of the inverse of the accretion rate and capped at 7000 km/s. Another difference related to the radiative mode is the increase of temperature of the ejected particles in the jet mode, consistently with observations \citep[][]{Fabian2012}. 
\end{itemize}
Ten percent of the material accreted into the central region is assumed to fall onto the BH, and the gas elements are immediately ejected in the purely kinetic and bipolar way (namely with zero opening angle w.r.t. the angular momentum of the inner disk) according to these two modes. In addition, {\sc Simba} includes a third AGN-driven feedback channel:
\begin{itemize}
    \item \textit{X-ray heating:} X-ray radiation pressure feedback is activated only in galaxies with low cold gas content and when the jet mode is active. The effect of this X-ray radiation pressure feedback is to push outwards the gas surrounding the accretion disk based on the high-energy photon momentum flux generated in the black hole accretion disk. X-ray heating is proportional to the inverse square of the distance of the gas element with respect to the BH. Its implementation broadly follows the model of \cite{Choi2012} and works as follows. Non-ISM gas ($n_{\mathrm{H}} < 0.13$ cm$^{-3}$) is heated by directly increasing its temperature; whereas for ISM gas, one-half of the X-ray energy is applied kinetically to give the gas particles a radial outwards kick and the second half is added as heat.
\end{itemize}

The combination of these three forms of feedback produces a population of quenched and star-forming galaxies and their black holes in good agreement with local studies of black hole properties \citep{Thomas2019}, with galactic size-mass relation and radial profiles \citep{Appleby2020}, and with 1.4~GHz radio luminosities \citep{Thomas+2021}.
Overall, while radiative AGN feedback only mildly affects the galaxy properties, the jet mode is mainly responsible for the quenching of galaxies. The X-ray feedback plays globally a small, but important role in suppressing residual star formation, and seems to be crucial for reproducing the observed centrally suppressed specific star formation rate and gas profiles of star-forming and green valley galaxies. These ingredients were included to improve the model of galactic-scale processes but potentially leave larger-scale signatures, which this paper will quantify using MFs.

% ----------------------------------------
\subsection{{\sc Simba} runs}
\label{sec:method:runs}

% ----------------
\begin{figure*}
\centering
\includegraphics[width=\textwidth]{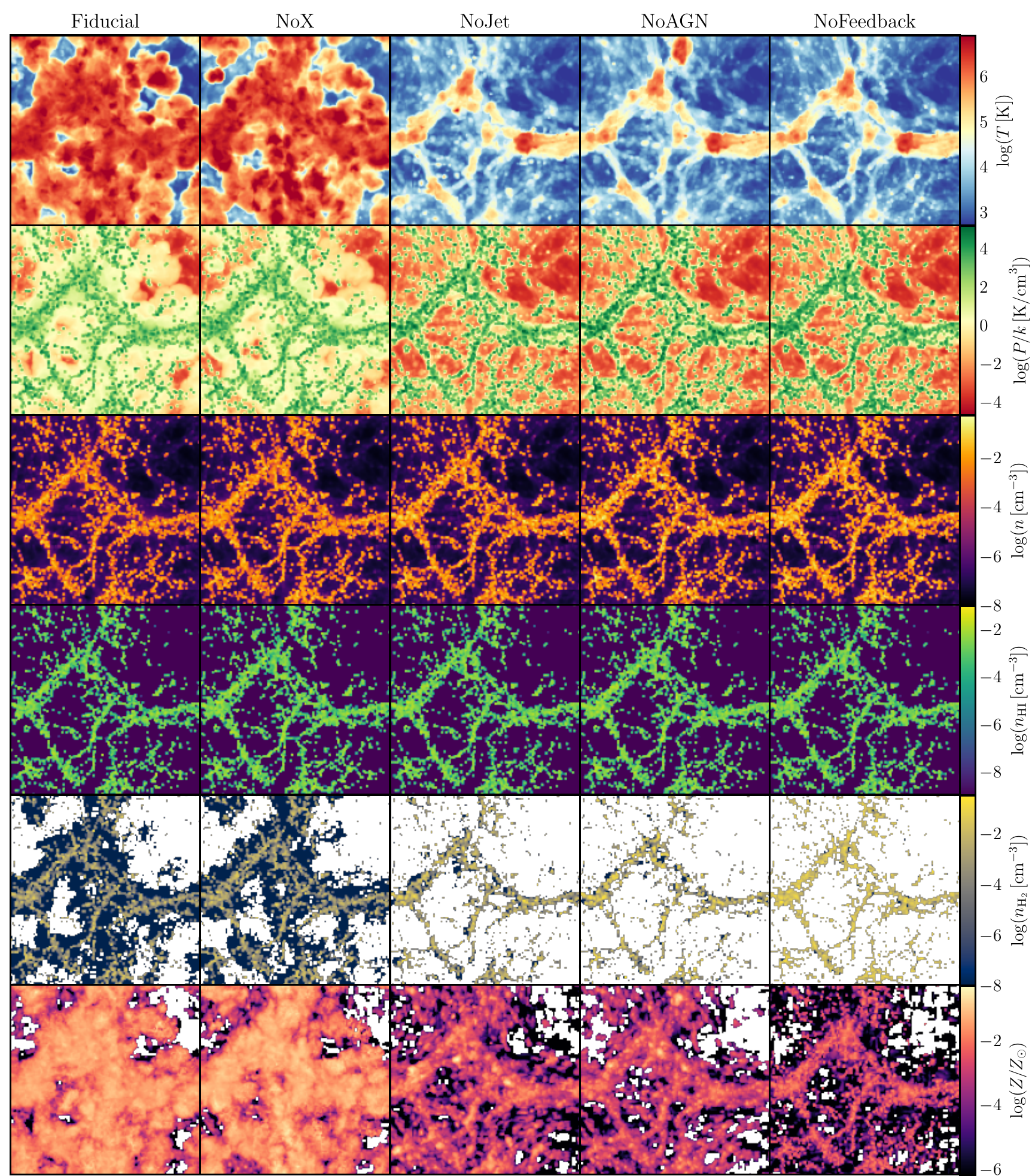}
\caption{Visualisation of full box-size sections ($8h^{-1}$Mpc thick and $50h^{-1}$Mpc wide in both directions) of the unsmoothed 3D temperature, pressure, total density, {\sc Hi} density, {\sc H$_2$} density, and metallicity fields (from top to bottom) at $z=0$, for different {\sc Simba} models progressively excluding feedback mechanisms (from left to right; see Table~\ref{tab:models}). It clearly shows a high level of morphological diversity. It also illustrates how individual feedback processes operating on galactic scales leave different morphological imprints on the gas fields at large scales; note for instance the influence of jets (second column) on the morphology of the $T$, $P$, $n_{\mathrm{H}_2}$, and $Z$ fields, but not on $n$ and $n_\mathrm{HI}$ fields. The corresponding 3D smoothed fields at $z=0$ and $z=5$ as used for the computation of the MFs are shown in Appendix~\ref{app:MF2dsections} as well as Figures~\ref{fig:2dmaps_smooth_z0} and \ref{fig:2dmaps_smooth_z5}. The purpose of this paper is to compress the information content of all maps into a few numbers extracted from the MFs of their excursion and quantify how feedback impacts them across cosmic time.
}
\label{fig:2dmaps_nosmooth_z0}
\end{figure*}

%% ------------------------------
%\iffalse
%\begin{table}
%\centering
%\caption{Summary of main ingredients of different {\sc Simba} runs used in this study. `Fiducial' model includes all feedback processes, `NoFeedback' model includes only hydrodynamics and no feedback processes. See \S\ref{sec:method:runs}.}
%\begin{tabular}{lcccc}
%\hline
%& \textbf{Fiducial} & \textbf{NoX} & \textbf{NoJet}& \textbf{NoAGN} & \textbf{NoFeedback} \\
%\hline
%Stellar feedback & \checkmark & \checkmark & \checkmark & \checkmark &  \\
%AGN winds & \checkmark & \checkmark & \checkmark &  &  \\
%Jets & \checkmark & \checkmark &  &  &  \\
%X-ray heating & \checkmark &  &  &  &  \\
%\hline
%\end{tabular}
%\label{tab:models}
%\end{table}
%\fi

% ------------------
\begin{table}
\centering
\caption{Summary of main ingredients of %different 
{\sc Simba} runs. % used in this study. %`Fiducial' model includes all feedback processes, `NoFeedback' model includes all the hydrodynamical but feedback processes. See \S\ref{sec:method:runs}.
}
\begin{threeparttable}
\begin{tabular*}{\columnwidth}{@{\extracolsep{\fill}}lccccc}
\hline
 & \textbf{\!\!\!\!Fiducial} & \textbf{\!\!\!\!NoX} & \textbf{\!\!\!\!NoJet}& \textbf{\!\!\!\!NoAGN} & \textbf{\!\!\!\!NoFeedback} \\
\hline
Stellar feedback & \checkmark & \checkmark & \checkmark & \checkmark &  \\
AGN winds & \checkmark & \checkmark & \checkmark &  &  \\
Jets & \checkmark & \checkmark &  &  &  \\
X-ray heating & \checkmark &  &  &  &  \\
\hline
\end{tabular*}
\end{threeparttable}
\tablefoot{`Fiducial' model includes all feedback processes, `NoFeedback' model includes all the hydrodynamical but feedback processes. See \S\ref{sec:method:runs}.}
\label{tab:models}
\end{table}

To examine the effect of stellar feedback and different types of AGN feedback on the global geometry and topology of various gas fields, we use five runs of the {\sc Simba} suite in 50~$h^{-1}$~Mpc comoving volumes with 512$^3$ gas elements and 512$^3$ dark matter particles with the mass resolution of 1.82$\times$10$^7~\rm{M}_\odot$ for gas and 9.6$\times$10$^7~\rm{M}_\odot$ for dark matter particles.
Cosmology adopted in {\sc Simba} is a standard $\Lambda$CDM compatible with results from \cite{PlanckCollaboration2016}, namely $\Omega_m=0.3$, $\Omega_{\Lambda}=0.7$, $\Omega_b=0.048$, $H_{0}=68$ km s$^{-1}$ Mpc$^{-1}$, $\sigma_8=0.82$ and $n_s=0.97$.

Five different runs of the {\sc Simba} suite correspond to different variants of AGN feedback adopted following a strategy where one input physics element is turned off at a time, to which a model without AGN feedback, and without stellar feedback are added, as summarised below (see also Table~\ref{tab:models}):
\begin{itemize}
    \item \textbf{Fiducial} - denotes a run with all forms of feedbacks,
    \item \textbf{NoX} - denotes a run with X-ray AGN feedback turned off, but including radiative and jet mode AGN feedback and stellar feedback,
    \item \textbf{NoJet} - denotes a run with both X-ray and jet AGN feedback turned off, but including AGN winds, namely radiative AGN feedback, and stellar feedback,
    \item \textbf{NoAGN} - denotes a run without any form of AGN feedback, but including stellar feedback,
    \item \textbf{NoFeedback} - denotes a run without any feedback.
\end{itemize}
Figure~\ref{fig:2dmaps_nosmooth_z0} offers an initial insight into the impact of the different feedback mechanisms on the six gas fields, showing two-dimensional sections $8h^{-1}$Mpc thick at $z=0$.

% ----------------
\begin{figure*}
\centering\includegraphics[width=0.33\textwidth]{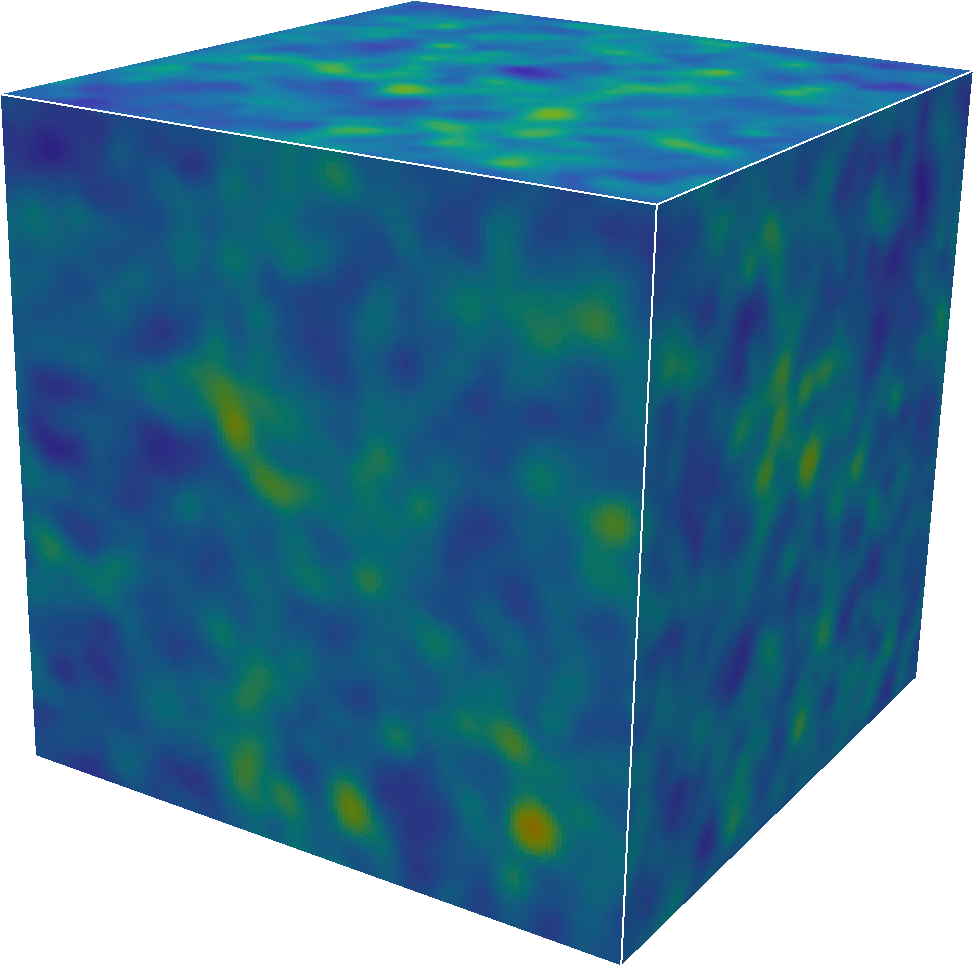}
\centering\includegraphics[width=0.33\textwidth]{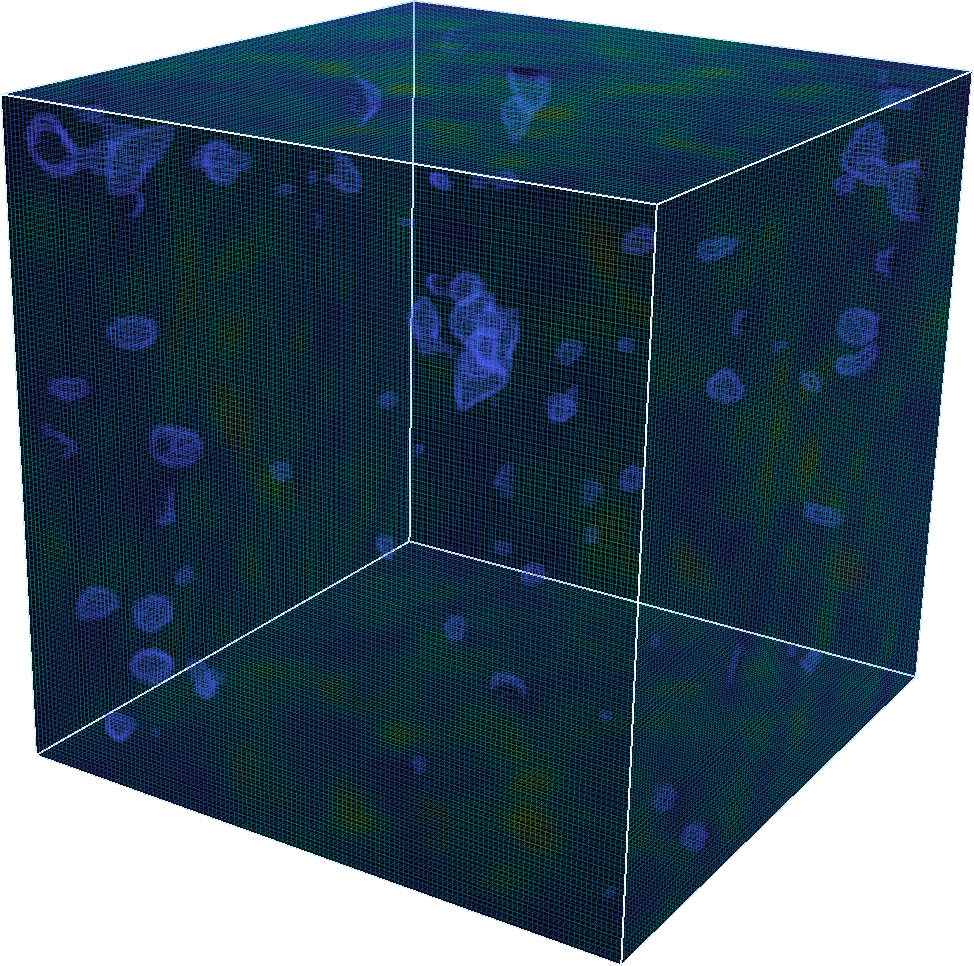}
\centering\includegraphics[width=0.33\textwidth]{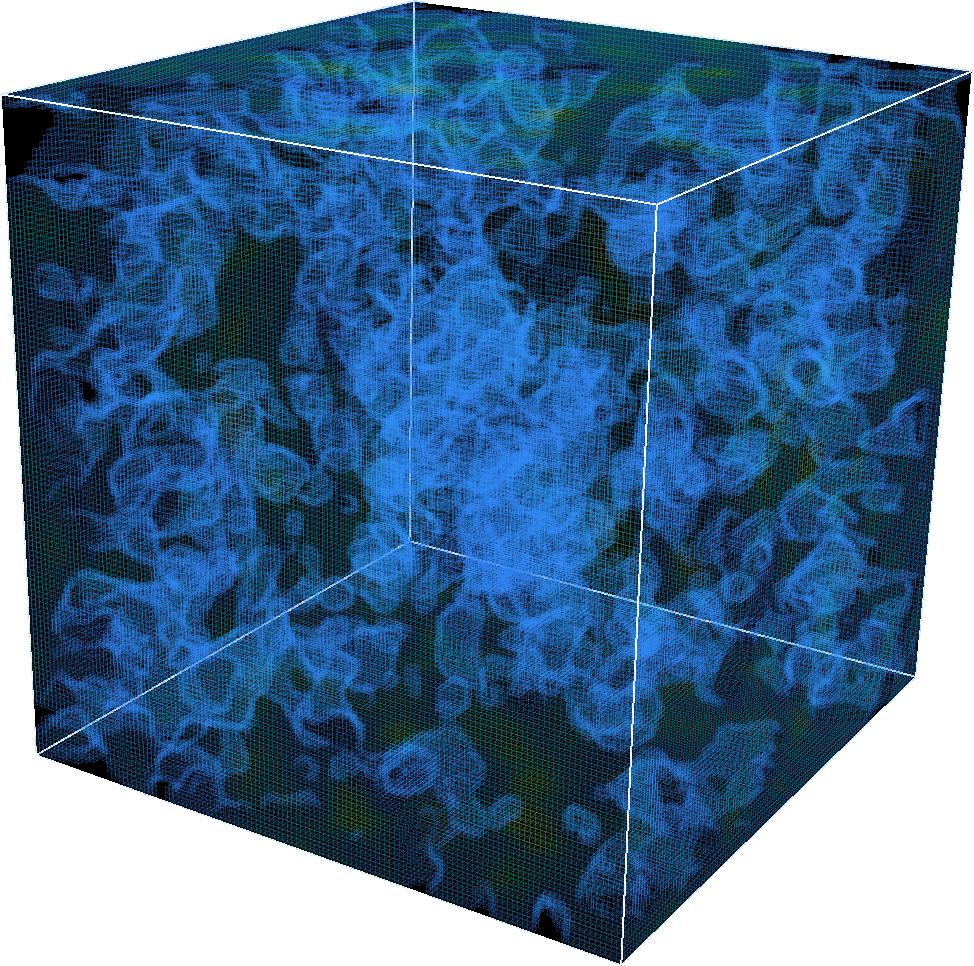}\\
\centering\includegraphics[width=0.33\textwidth]{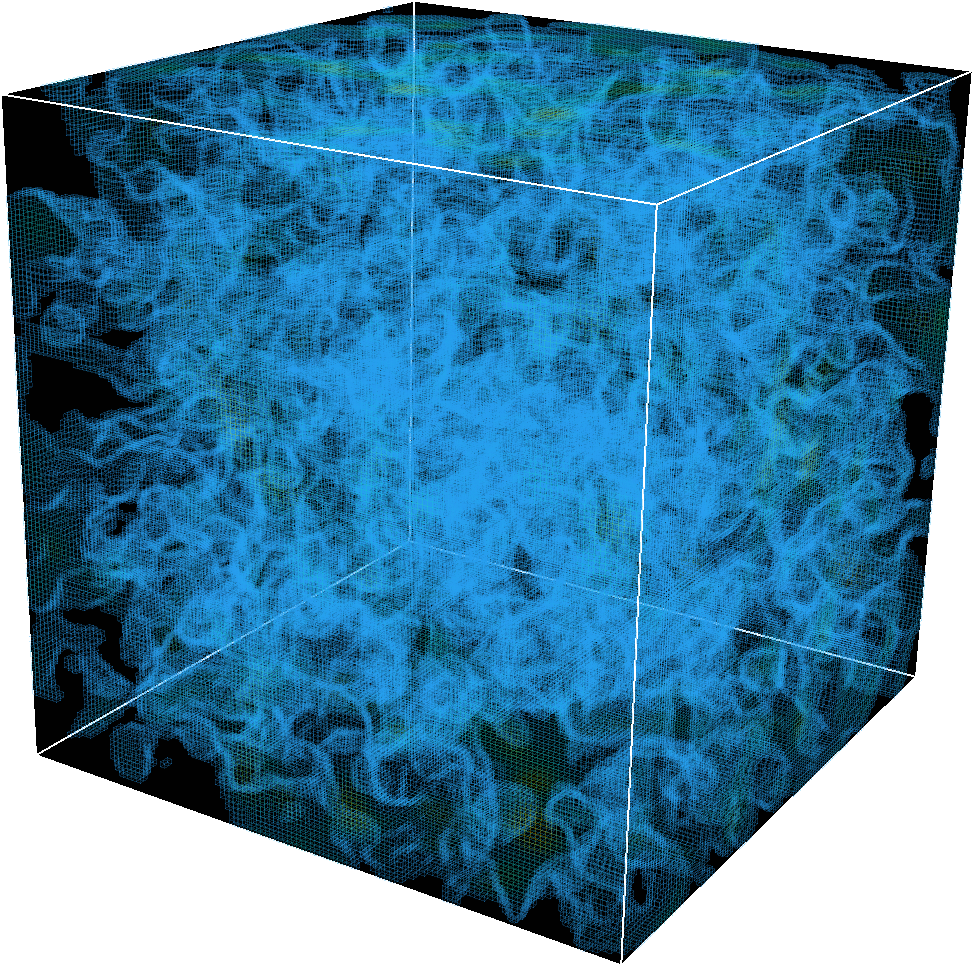}
\centering\includegraphics[width=0.33\textwidth]{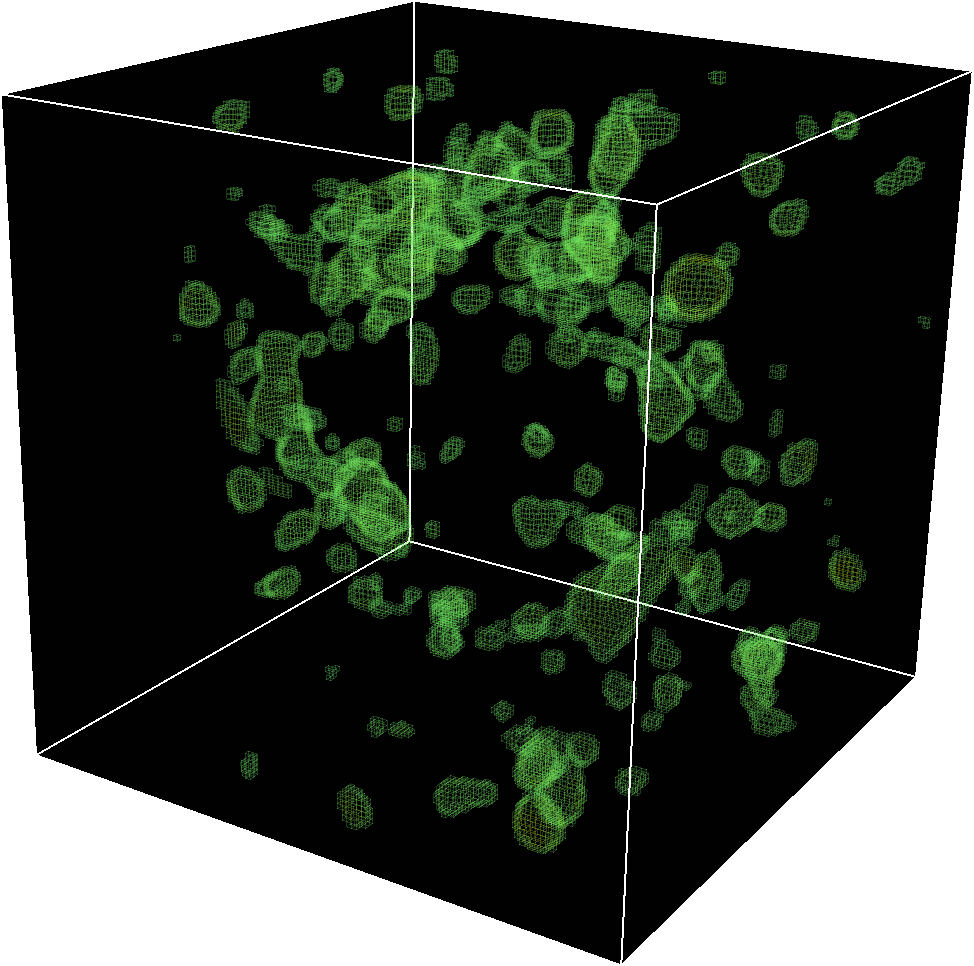}
\centering\includegraphics[width=0.33\textwidth]{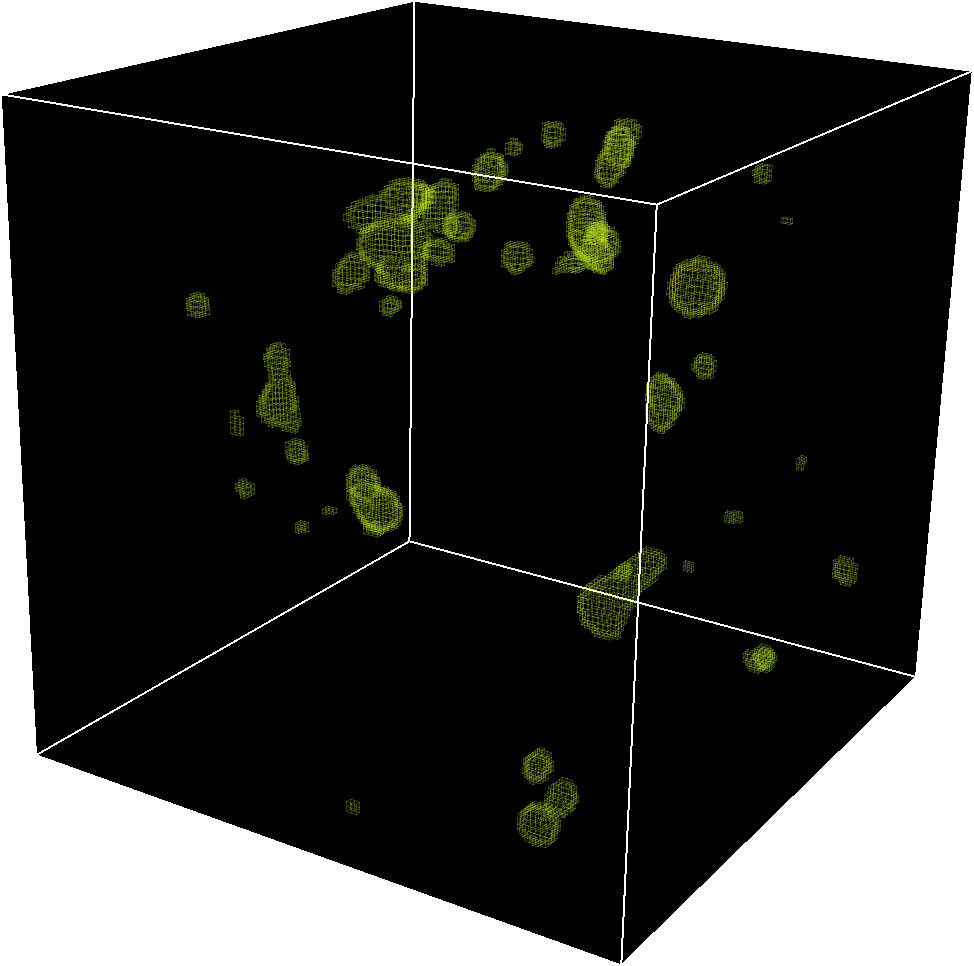}
\caption{Excursion-sets of the gas temperature for the fiducial model at $z=5$ for different values of the threshold: whole field (top left), $T\geq 6\times10^3$~K (top middle), $T\geq 8\times10^3$~K (top right), $T\geq 1\times10^4$~K (bottom left), $T\geq 3\times10^4$~K (bottom middle) and $T\geq 5\times10^4$~K (bottom right). For illustrative purposes, 3D maps correspond to fields before applying additional Gaussian smoothing of 1.56 $h^{-1}$Mpc. Low values of the threshold reveal cold isolated cavities, at intermediate thresholds an interconnected filamentary structure emerges, while higher thresholds delimit hot regions forming isolated clumps. The Minkowski functionals computed for these thresholded 3D maps quantify the complex morphology of the underlying temperature field. See Figure~\ref{fig:excursion_set_T_fiducial_z0} for analogous excursion-sets at $z=0$.
}
\label{fig:excursion_set_T_fiducial_z5}
\end{figure*} 

% ----------------
\begin{figure*}
\centering\includegraphics[width=0.33\textwidth]{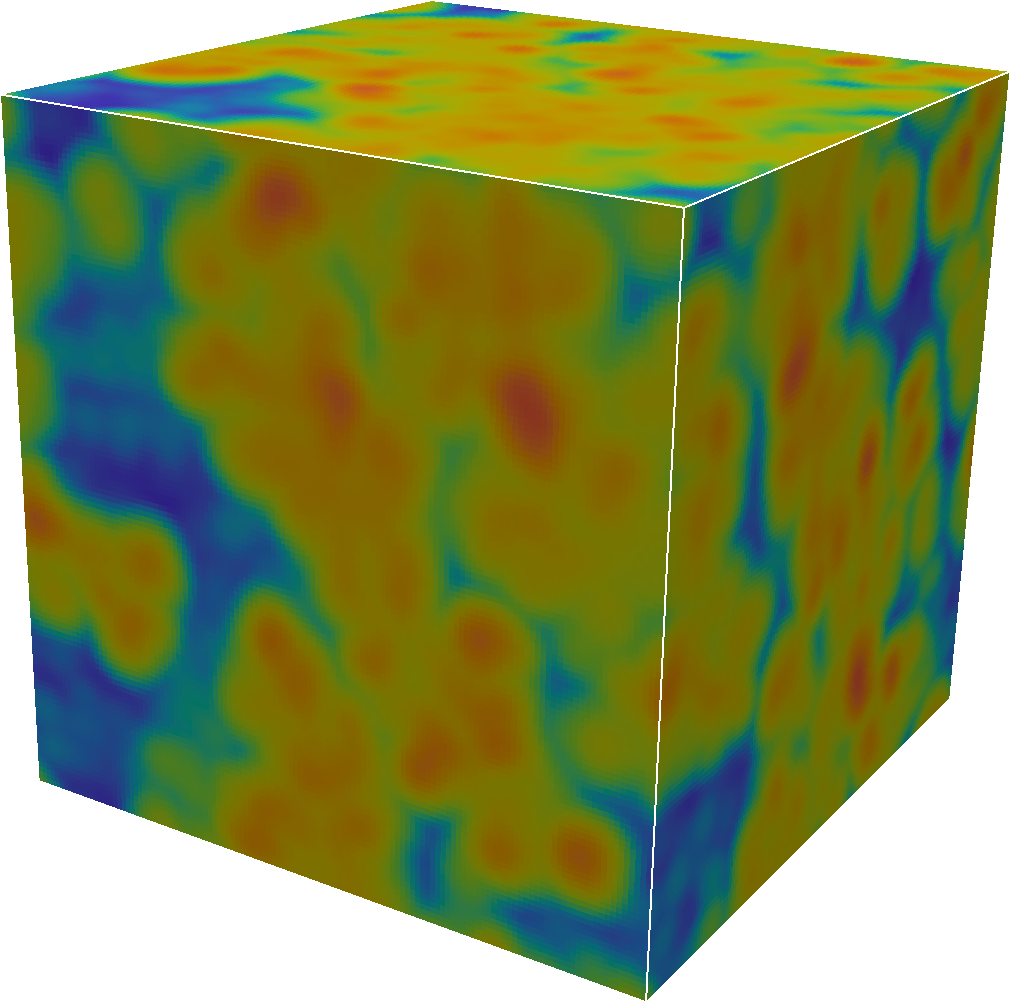}
\centering\includegraphics[width=0.33\textwidth]{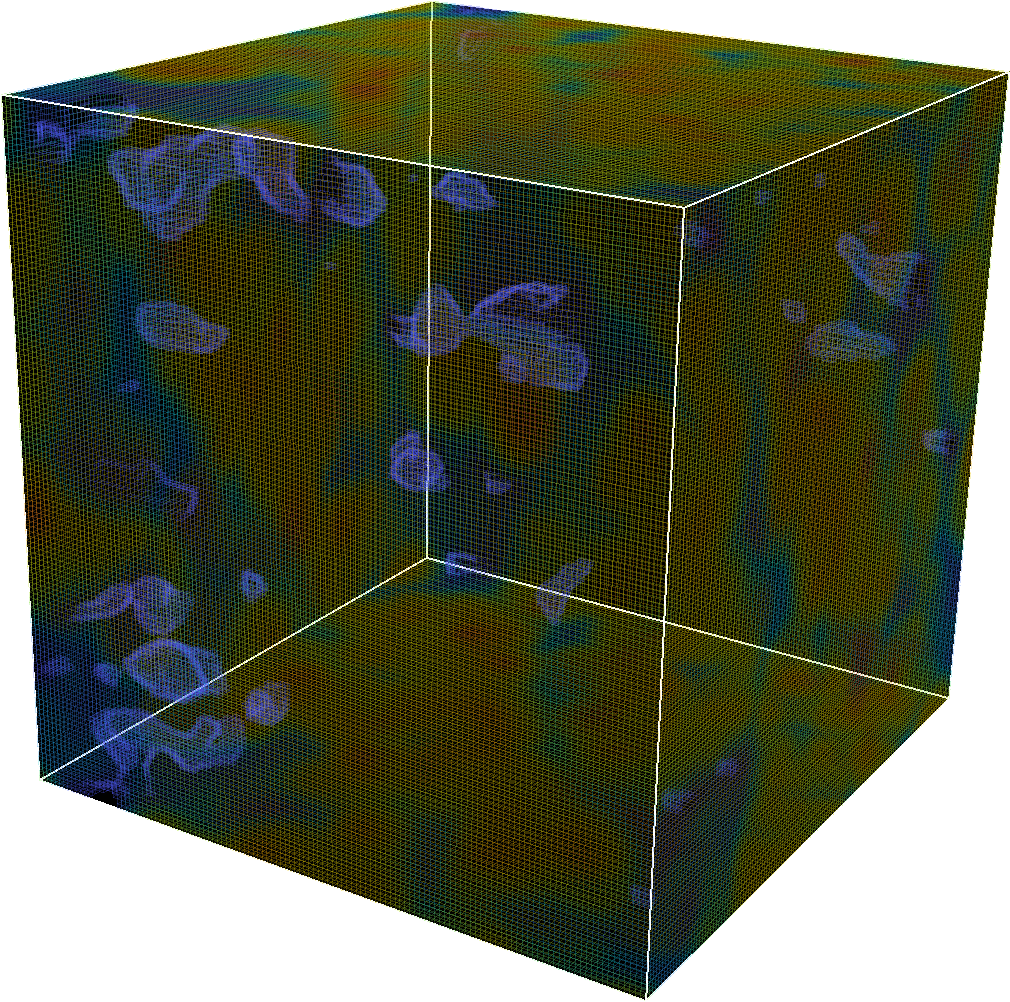}
\centering\includegraphics[width=0.33\textwidth]{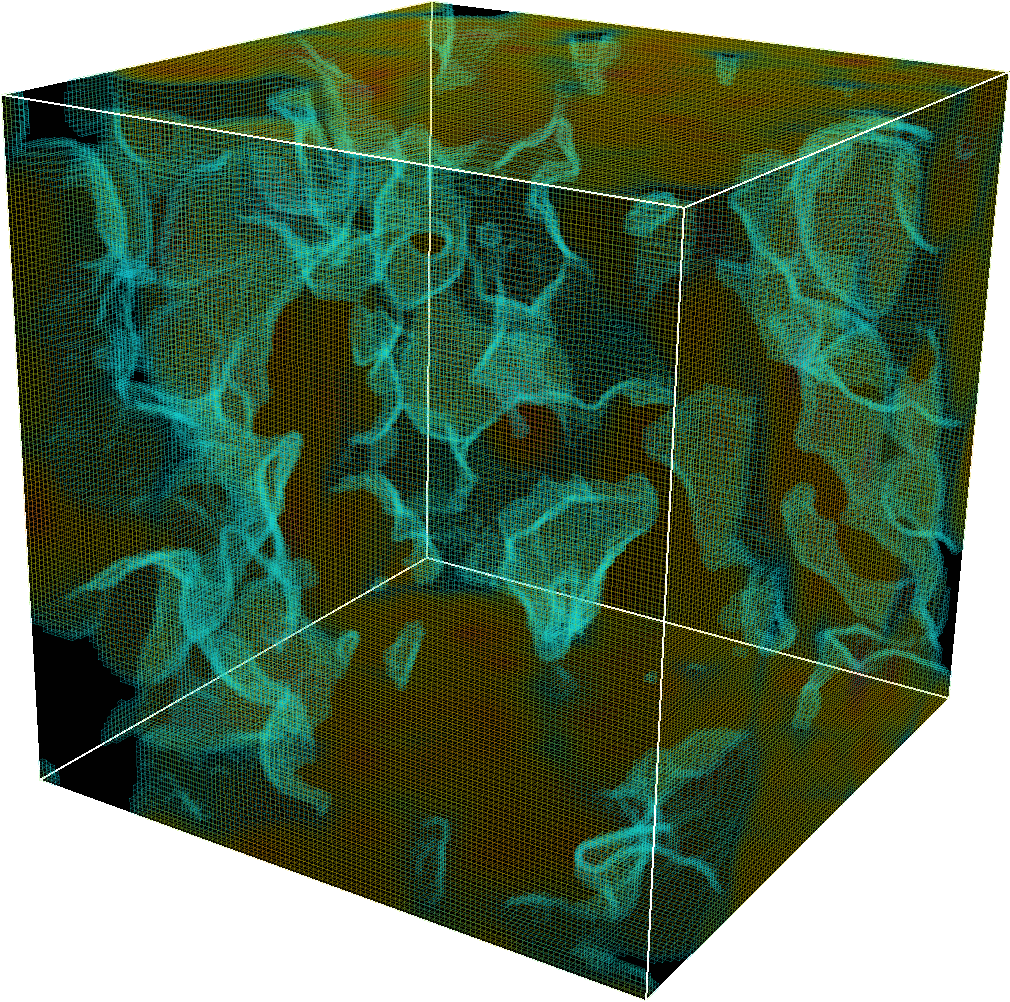}\\
\centering\includegraphics[width=0.33\textwidth]{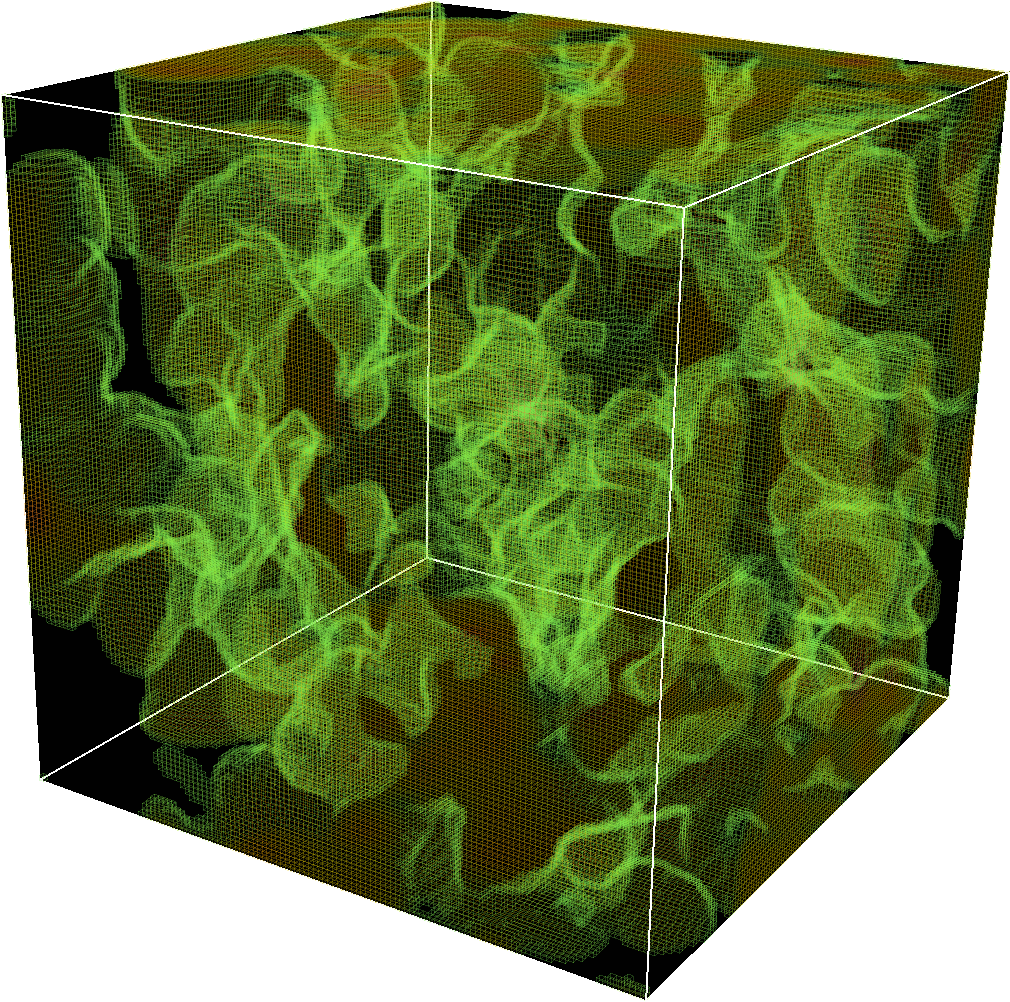}
\centering\includegraphics[width=0.33\textwidth]{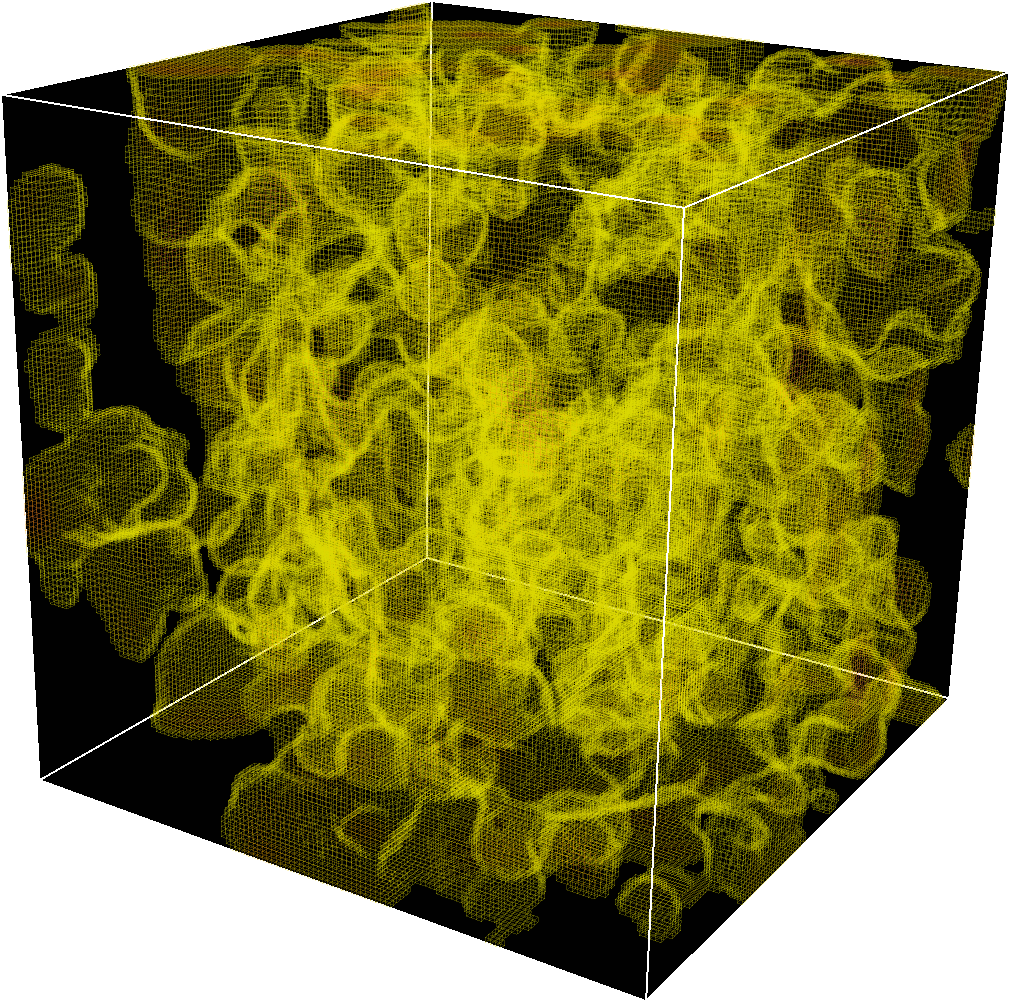}
\centering\includegraphics[width=0.33\textwidth]{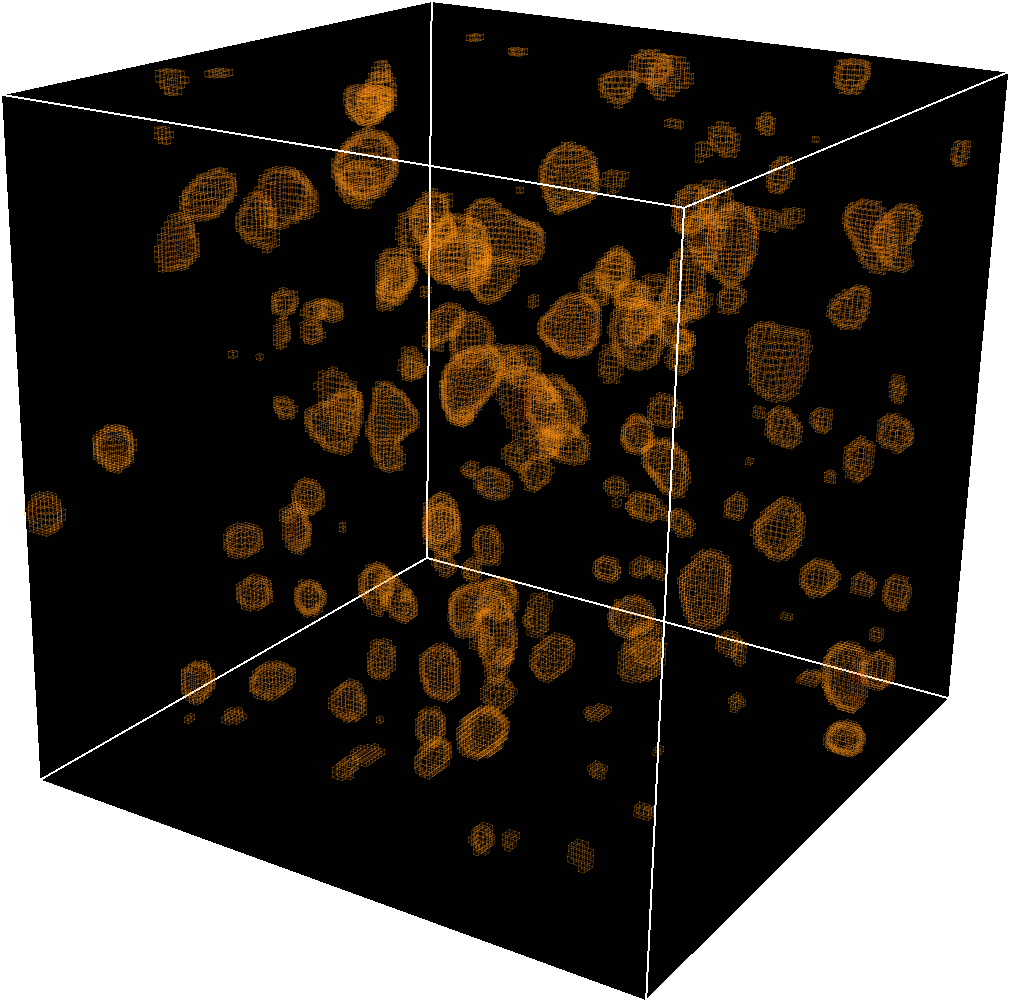}
\caption{Excursion set of temperature map for the fiducial model at $z=0$ for different values of the threshold: whole field (top left), $T\geq 7\times10^2$~K (top middle), $T\geq 10^4$~K (top right), $T\geq 10^5$~K (bottom left), $T\geq 10^6$~K (bottom middle) and $T\geq 10^7$~K (bottom right). As for Figure~\ref{fig:excursion_set_T_fiducial_z5}, these maps do not include the additional Gaussian smoothing of 1.56 $h^{-1}$Mpc and adopt the same colour code.
}
\label{fig:excursion_set_T_fiducial_z0}
\end{figure*}

% ----------------------------------------
\subsection{Minkowski functionals}
\label{sec:method:MF}

Minkowski functionals (MFs) are integral measures that generalise the notion of volume accounting for the size, shape (geometry) and connectivity (topology) of bodies or spatial patterns fully characterising their morphology, namely being subadditive, invariant under Galilean transformations, and continuous \citep{Hadwiger1957}. Dealing with random fields like the density, temperature, pressure, and metallicity of the gaseous component of the LSS, it is natural to consider their excursion set or isocontours, $\mathcal{C}_\theta = \{ \bm{x}\in D\, |\, f(\bm{x})\geqslant\theta\}$, where $f$ denotes a generic field defined over a spatial domain $D$ and $\theta$ the threshold (see Figures~\ref{fig:excursion_set_T_fiducial_z5} and \ref{fig:excursion_set_T_fiducial_z0} for illustration). In three dimensions, the MFs of $\mathcal{C}_\theta$ correspond to its volume ($V$), surface area ($A$), integral mean curvature ($H$), and integral Gaussian curvature ($G$), the two latter measuring the extrinsic and intrinsic curvatures, respectively, of the body integrated over its surface. Owing to the Gauss-Bonnet theorem, the Gaussian curvature is proportional to the Euler characteristic $\chi=G/4\pi$ and linearly related to the genus $g=1-\chi/2$ of $\mathcal{C}_\theta$, both usually adopted as more directly related to the counts of connected regions, tunnels, and cavities of the excursion-set, or maxima, minima, and saddle points of the thresholded field $f(\bm{x})$, accounting for the topology of $\mathcal{C}_\theta$. Analytical expressions exist for simple configurations such as triaxial ellipsoids, cylinders, or trivial unions thereof describing the idealised isocontours of isolated or merging clusters with filaments or bridges \citep{SchimdTereno2021} and for the expectation value of MFs of Gaussian and Gaussian-related random fields (see Appendix~\ref{app:analyticMFs}).

Here, we adopt the normal based on the so-called intrinsic volumes $(V_0,V_1,V_2,V_3)=(V,A/6,H/3\pi,\chi)$ and compute the MFs as the spatial average of Koenderink invariants using the code presented in \citet{SchmalzingBuchert1997}, which yields the MFs' densities as a function of the threshold, or Minkowski curves, as:
\begin{subequations}
    \begin{align}
    \varv_0(\theta) &= \frac{1}{\mathcal{V}}\int_{\mathcal{C}_\theta}\mathrm{d}V(\bm{x}),\\
    \varv_1(\theta) &= \frac{1}{6} \frac{1}{\mathcal{V}}\int_{\partial\mathcal{C}_\theta}\mathrm{d}A(\bm{x}),\\
    \varv_2(\theta) &= \frac{1}{6\pi} \frac{1}{\mathcal{V}}\int_{\partial\mathcal{C}_\theta}\mathrm{d}A(\bm{x})\left(\frac{1}{R_1(\bm{x})}+\frac{1}{R_2(\bm{x})}\right),\\
    \varv_3(\theta) &= \frac{1}{4\pi} \frac{1}{\mathcal{V}}\int_{\partial\mathcal{C}_\theta}\mathrm{d}A(\bm{x})\frac{1}{R_1(\bm{x})R_2(\bm{x})},
\end{align}
\end{subequations}
where $\mathrm{d}V(\bm{x})$ and $\mathrm{d}A(\bm{x})$ are the elementary volume and area of the excursion sets of the six random fields $f=T, P, n_\mathrm{gas}, n_\mathrm{HI}, n_{\mathrm{H}_2}, Z$ (hereafter referred as $f$-fields), which have boundary surface $\partial\mathcal{C}_\theta$ with principal local curvature radii $R_{1,2}(\bm{x})$. Also, $\mathcal{V}$ is the total sample volume, namely the simulation box. The fields are sampled at $128^3$ Cartesian regularly spaced lattice points using a cloud-in-cell kernel, to limit the computational load, and folded with a Gaussian kernel with standard deviation corresponding to a resolution of 4 lattice points, equivalent to about $1.56 h^{-1}$Mpc, largely serving the Nyquist limit. Dealing with simple (periodic) boundaries, the results are consistent with those obtained using the Crofton formula (see the discussions in \citealt{Schmalzing+1999} and \citealt{Hikage+2003}).

% ========================================
\section{Morphological analysis} 
\label{sec:results}

% ----------------------------------------
\subsection{Guidelines}
\label{sec:results:guidelines}

Figures~\ref{fig:MFs_T_nonorm}-\ref{fig:MFs_metal_nonorm} which are discussed in the following subsections show the four MFs per unit volume, $\varv_\mu$ ($\mu=0,1,2,3$), of the excursion-set of the three-dimensional temperature, pressure, density (total, {\sc Hi}, and {\sc H$_2$}), and metallicity fields for all the models, as a function of the threshold and at different redshift. Each row focuses on the effect of one component at a time, namely X-ray heating, jets, AGN winds, and stellar feedback (from top to bottom), by comparing each time the two models (solid and dotted lines on the top panels) differing by this same component. That is, the impact of individual feedback mechanisms is assessed by the differences between MFs densities (quoted as $\Delta$ in the bottom sub-panels) as follows:
\begin{itemize} 
    \item impact of X-ray heating: $\varv_\mu[\mathrm{Fiducial}]-\varv_\mu[\mathrm{NoX}]$,
    \item impact of jets: $\varv_\mu[\mathrm{NoX}]-\varv_\mu[\mathrm{NoJets}]$,
    \item impact of AGN winds: $\varv_\mu[\mathrm{NoJets}]-\varv_\mu[\mathrm{NoAGN}]$,
    \item impact of stellar feedback: $\varv_\mu[\mathrm{NoAGN}]-\varv_\mu[\mathrm{NoFeedback}]$.
\end{itemize}
The full model accounting for all the feedback mechanisms and the simplest model with no feedback (only hydrodynamics) correspond to the solid line in the top panels and to the dashed line in the bottom panels, respectively. In each panel, symbols pinpoint the MFs of the excursion set corresponding to the mean value of the field, $\bar{f}(\bm{x})$; we will refer to these domains as $\bar{f}$-regions. 

The Minkowski curves of the excursion sets $C_\theta$ have similar qualitative trends regardless of the gas property one does consider. The following outline guides their interpretation:
\begin{itemize}
\item The volume filling fraction $\varv_0$ is a monotonically decreasing function of the threshold $\theta$, tending to zero if the highest values of the field $f$ are concentrated in point-like regions. 
Likewise, $1-\varv_0$ measures the volume fraction of the complementary domain $D\setminus\mathcal{C}_\theta = \{ \bm{x}\in D\, |\, f(\bm{x})<\theta\}$. We note that for the fixed value of the threshold, $\varv_0$ increases (decreases) with time when the corresponding domains expand (contract).

When approximated by a step function $H(\theta_*-\theta)$, like the temperature field at high redshift, $\varv_0(\theta)$ accounts for a homogeneous (single-phase) field with characteristic value $f(\bm{x})=\theta_*$. Correspondingly, the density area $\varv_1$ is close to a Dirac-delta, indeed $\varv_1(\theta) \approx \mathrm{d}\varv_0(\theta)/\mathrm{d}\theta$. Consistently, a multi-phase field with sharp domains is accounted for by a piecewise constant function, namely, $\varv_0(\theta)=\sum_i\varv_{0,i}H(\theta_i-\theta)$ with $\sum_i\varv_{0,i}=1$, the $i$-th phase occupying a fractional volume $\varv_{0,i}$.
\item According to the isoperimetric inequality \citep{Schmalzing+1999}, the surface area of fixed volume is minimum for a sphere, namely, a wrinkly surface has more area than a smooth one. All the excursions-sets with threshold different from the maximum point of $\varv_1(\theta)$ have therefore on average a more regular surface than domains with maximum $\varv_1$. Note also that the apparent skewness of $\varv_1(\theta)$ is due to the logarithmic scale of the threshold and typically grows with time.
\item The integrated-mean-curvature density $\varv_2(\theta)$ is positive (negative) for domains $C_\theta$ that are on average convex (concave), which are dominated by lumps (cavities). The opposite is true for the complementary domains $D\setminus\mathcal{C}_\theta$.
\item The Euler characteristic is the alternate sum of Betti numbers, counting the number of topologically invariant domains, namely: connected regions, tunnels, and cavities. Consistently, the $\varv_3(\theta)$ curve has a characteristic M-shape: it is positive for small values of the threshold $\theta$, which are typically smaller than the mean value $\bar{\theta}$ (marked by symbols), and attains a local maximum that measures the largest possible number of cavities per unit volume in the excursion set; it becomes negative for intermediate values of $\theta$, with a minimum counting (in absolute value) the largest number density of tunnels in the excursion set, corresponding to a `sponge-like' topology; and it is again positive for larger values of the field, attaining a second (often global) maximum that indicates the largest number density of isolated lumps in the system. When the negative peak occurs for $\theta>\bar{\theta}$ or $\theta<\bar{\theta}$ (for instance, as in the $T$-field at low redshift for the NoAGN model with stellar feedback only; see Figure~\ref{fig:MFs_T_nonorm}, bottom-right panel) the topology of the field is respectively interpreted as `bubble-like' (or `Swiss-cheese-like'), namely dominated by cavities, or `meatball-like', namely composed of isolated regions \citep[see][]{ColesDaviesPearson1996,Choi+2013}. The rightmost zero of $\varv_3(\theta)$ defines an estimation of the threshold for continuous percolation \citep{MeckeWagner1991}.
\item Gaussian random fields, the usual reference for the LSS, are described by analytical MFs' mean values depending on the variances of the field and its gradient \citep{Tomita1986}, with even $\varv_1(\theta)$ curve peaked at $\theta=\bar{\theta}$, odd $\varv_2(\theta)$ with equal absolute amplitude of extrema and vanishing for $\theta=\bar{\theta}$, and even $\varv_3(\theta)$ with a negative minimum at $\theta=\bar{\theta}$ and two local maxima, the amplitude of $\varv_{1,2,3}$ being determined by the first spectral moment of the field (equivalent to the rms-variance of its gradient $\nabla f$). Weakly non-Gaussian and log-normal random fields also admit analytical MFs' expectation values, with coefficients depending respectively on the generalised skewness parameters and on the variance of the field \citep{Matsubara2003,Hikage+2003,Hikage+2006,GayPichonPogosyan2012,Matsubara+2022}. More details are given in Appendix~\ref{app:analyticMFs}. Not surprisingly, the MFs of the gas fields considered in this study are strongly non-Gaussian, as indicated by the strong deviation from the M-shape around the minimum of $\varv_3(\theta)$, especially at low redshift as a result of non-linear gravitational evolution and shot noise (in observed fields, additional deviations are due to redshift space distortions and, if any, primordial non-Gaussianities). Apart from some specific deviations and especially at high redshift, $z>4$, their profile resembles that of log-normal random fields.
\end{itemize}

Qualitatively, we can recognise the following spatial patterns from the values of MFs densities:

\textbullet~ \textit{Isolated regions} with field values larger than in the surrounding environment, $f>f_\mathrm{env}$ (e.g. hot bubbles in cold or warm IGM), have small $\varv_0$, large (small) $\varv_1$ if they have smooth (wrinkly) boundary surface, $\varv_2>0$, and large $\varv_3$;

\textbullet~ \textit{Isolated bubbles} with $f<f_\mathrm{env}$ (e.g. cold bubbles in warm IGM) have large values for $\varv_0$, surface density $\varv_1$ like for isolated regions, while $\varv_2<0$, and large values for $\varv_3$ like for isolated regions;

\textbullet~ \textit{Network of filaments} with $f>f_\mathrm{env}$ (e.g. hot, dense, or metal-rich filaments alimented by local stellar activity) have large (small) $\varv_0$ if on average filaments are thick (thin),  $\varv_1$ as above, $\varv_2>0$ and smaller for cylindrical shape (the largest radius of curvature diverges), and $\varv_3\approx0$ if the network is percolating;

\textbullet~ \textit{Network of filaments} with $f<f_\mathrm{env}$, namely, an excursion set $\mathcal{C}_\theta$ dominated by tunnels (e.g. cold filaments in warm IGM, or metal-poor filaments in which metals are expelled by supernovae or stellar winds) have opposite volume filling fraction than in the previous case, namely, large (small) $\varv_0$ if filaments are on average thin (thick),  $\varv_1$ as before, $\varv_2<0$ and larger for cylindrical shape, and $\varv_3<0$ and large in absolute value, and vanishing if the network is percolating as before. The directions of the filamentary network and the anisotropy of the excursion set cannot be captured by MFs, but by higher-rank Minkowski valuations \citep[e.g.][]{Beisbart+2002}.

Finally, we note that one probes the morphology of fields that are effectively smoothed twice, first by grid sampling and then by the Gaussian filter needed for the calculation of the covariant derivatives in Koenderink invariants. The resulting spatial resolution of the excursion sets is about $1.56 \, h^{-1}$Mpc, largely coarser than the original resolution of {\sc Simba}. The numerical accuracy of MFs' densities is also limited for excursion sets with extreme values of the field at odds with their environment, as expected in very sparse or non-resolved regions. The global morphological analysis of fields in domains of a size typical of galaxy groups or smaller and in very diffuse domains is computationally demanding and, therefore not addressed in this study. We note that in the following subsections, all the quoted values of fields must therefore be considered carefully, namely, as spatially-averaged values. A more detailed analysis is presented in Appendix~\ref{app:bias}.

% ----------------------------------------
\subsection{Temperature field}
\label{sec:results:T}

% ----------------
\begin{figure*}
\centering\includegraphics[width=17.cm]{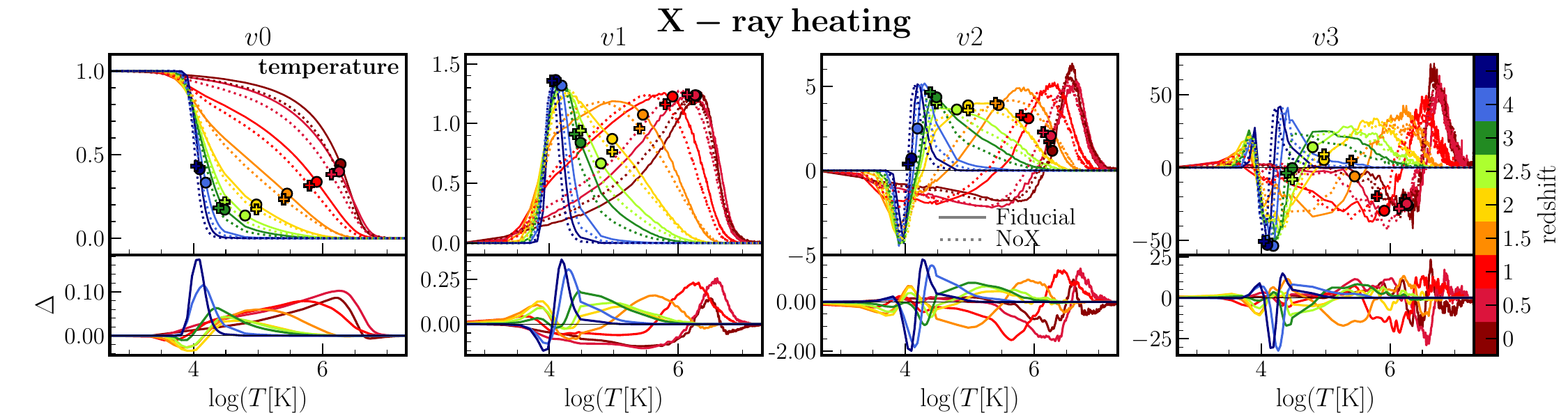}
\centering\includegraphics[width=17.cm]{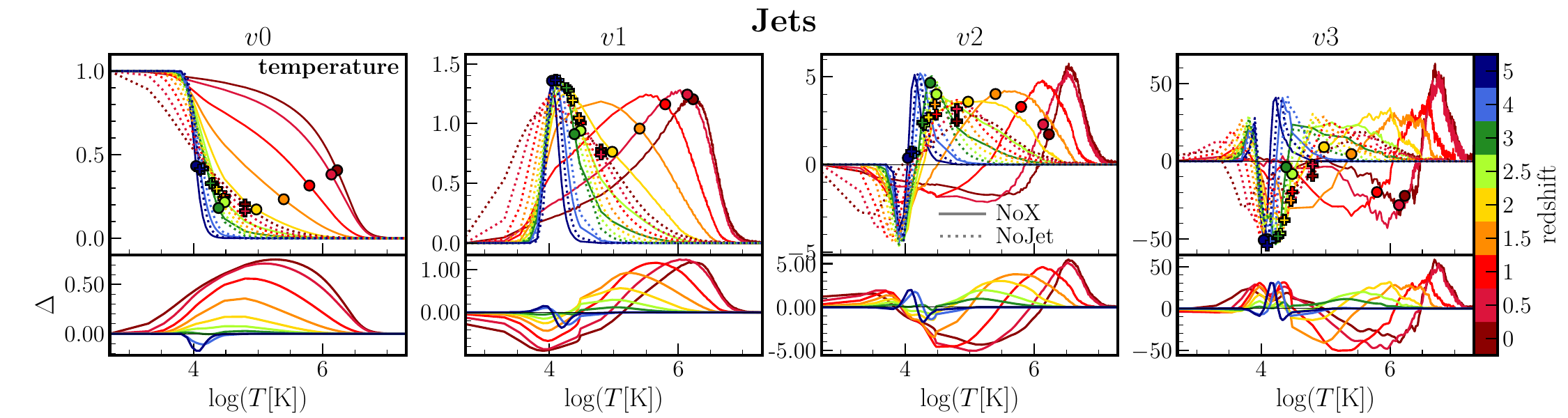}
\centering\includegraphics[width=17.cm]{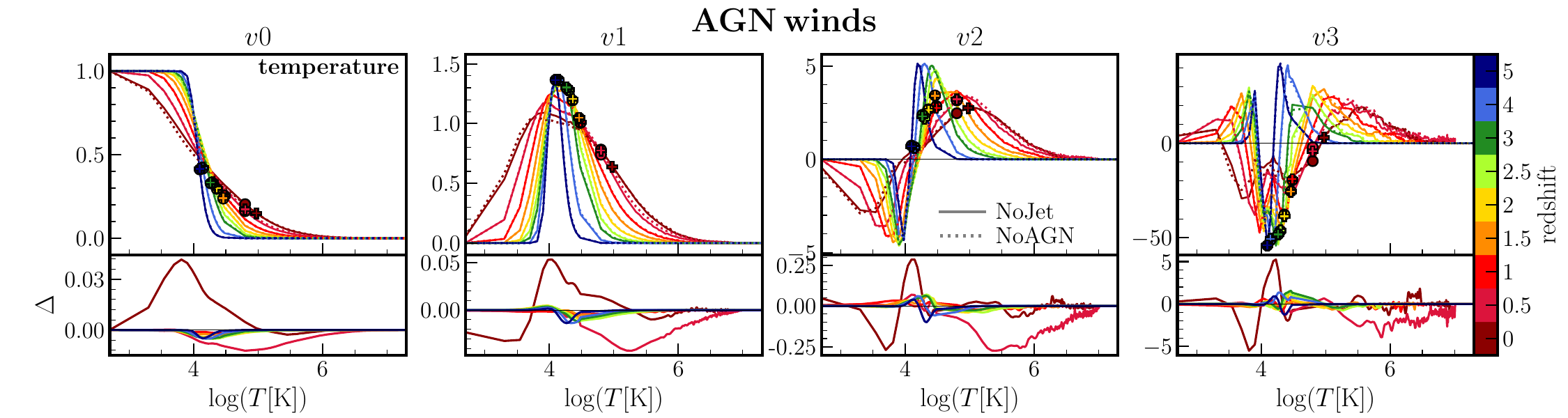}
\centering\includegraphics[width=17.cm]{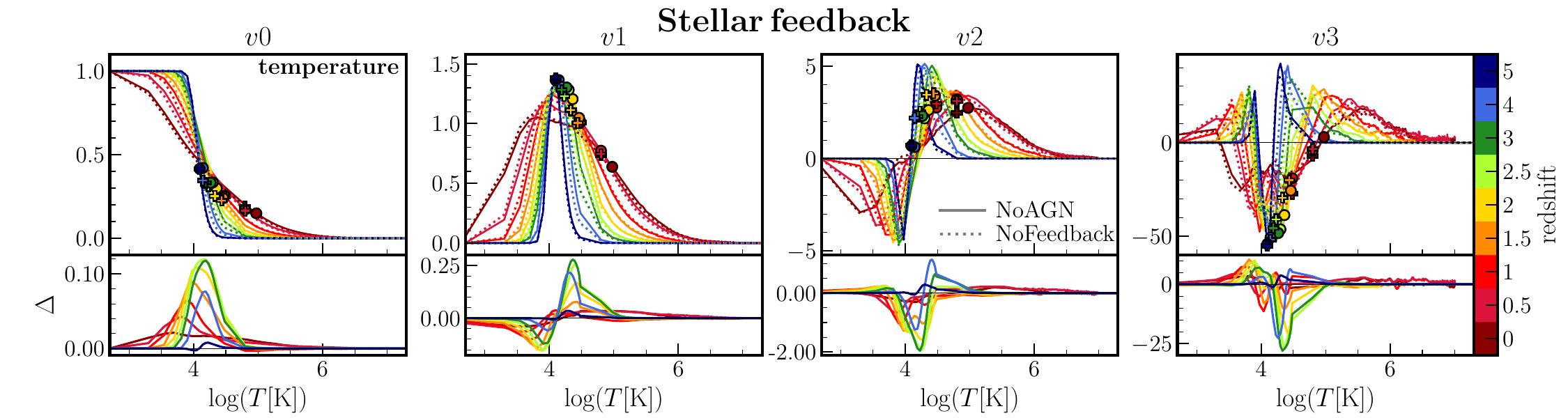}
\caption{\textbf{Temperature.} Minkowski curves (columns) of gas temperature for different models (top to bottom) and redshift (colours). Each row shows a comparison of two models differing by one ingredient at a time (indicated in the plot titles), probing its impact on Minkowski functionals. In each row, the upper panel shows the two models (solid and dashed lines) and the bottom panel shows their differences, $\Delta$. Filled coloured circles and plus symbols indicate the mean value of the field at each redshift for two models shown with solid and dashed lines, respectively.
\textit{First row:} Fiducial model is compared to the NoX, showing the effect of X-ray heating.  \textit{Second row:} NoX model is compared to NoJet, showing the effect of jets.
\textit{Third row:} NoJet model is compared to NoAGN, highlighting the effect of AGN winds. \textit{Fourth row:} NoAGN model is compared to NoFeedback, showing the effect of stellar feedback. 
The morphology of the $T$-field is most strongly impacted by AGN jets at low redshift ($z \lesssim 2$), but also by the X-ray heating at both high ($z\sim4-5$) and low redshift ($z\lesssim 2-3$), though to a somewhat lesser extent. Interestingly, the strongest impact of the stellar feedback (with a strength comparable to X-ray heating) is seen near the epoch of cosmic noon $z\sim2-3$.}
\label{fig:MFs_T_nonorm}
\end{figure*} 

Figure~\ref{fig:MFs_T_nonorm} shows the MFs of the excursion sets of the smoothed $T$-field, whose three-dimensional representations are illustrated in Figures~\ref{fig:excursion_set_T_fiducial_z5} and \ref{fig:excursion_set_T_fiducial_z0}. 

When all the feedback mechanisms are included (solid lines in first-row panels), at redshift $z>4$ the volume filling fraction $\varv_0(T)$ is a step-like function with the transition occurring around warm regions with $T\simeq10^4$~K, indicating that at that epoch the universe was approximately in a single phase, with hot regions occupying a tiny fraction of the total volume. The main effect of setting this temperature is photo-ionisation heating within the IGM. At low redshifts, due to the dropping meta-galactic photo-ionisation rate, gas can reach temperatures down to $\sim 10^3$~K.  However, in the case of full {\sc Simba} physics, the IGM is heated strongly by AGN feedback~\citep{Christiansen2020}, resulting in a much smaller fraction of photo-ionised IGM. Another effect is that {\sc Simba}'s pressure floor does not allow cooling in the interstellar medium below $10^4$~K, but the fraction of overall baryons in the ISM is small in any case, particularly with full {\sc Simba} physics.

Correspondingly, the surface density $\varv_1(T)$ is symmetric and very peaked around $10^4$~K, resembling a Dirac delta, indicating a single-phase $T$-field at high redshifts. At later times, hotter regions occupy larger and larger volume as time passes due to the combined effect of shock heating on the one hand, and star formation and AGN feedback on the other hand, where the latter typically dominates \citep[e.g.][]{SomervilleDave2015}. The surface density curve $\varv_1(T)$ becomes and remains right-tailed until $z\sim2$ and left-tailed afterwards, in favour of regions with temperature smaller than the mean value at late times, namely, for $T<\bar{T}\simeq10^6$~K at $z=0$ (marked by a circle). These regions are concave ($\varv_2<0$) and dominated by tunnels ($\varv_3<0$), especially at redshift $z=1$ for $T\gtrsim10^{4}$~K and at $z=0.5$ for $T\gtrsim10^{5}$~K, {suggesting a network of cold filaments in a warm percolating IGM.} Interestingly, at $z=0$ concave regions, namely tunnels and bubbles, with $T\lesssim10^{5.5}$~K stop to negatively contribute to the Euler characteristic, which is almost vanishing; this is compatible with the late-time appearance of cold bubbles, as illustrated in Figure~\ref{fig:excursion_set_T_fiducial_z0} (top middle panel).

Without AGN and stellar feedbacks, namely with only pure hydrodynamics (dashed line in bottom line panels), the morphological evolution of the temperature field is very different; namely, it is qualitatively close to a log-normal random field at all redshifts, except $z=0$ (not shown; see Appendix~\ref{app:analyticMFs}). MFs indicate regions with temperature slightly larger than $T=10^4$~K as very special: besides marginal deviations at $z<0.5$, at all times they fill about one-third of volume ($\varv_0\simeq 0.35$), have the maximum surface density, vanishing mean curvature, and minimum (negative) Euler characteristic, suggesting a network of cold filaments set by pure gravity and hydrodynamics. 

Overall, AGN jets have the strongest impact on the morphology of the temperature field (second-row panels), as can be seen in the amplitude of $\Delta$. Starting at redshift $z~=~2$, namely $\sim$ 3.3~Gyr after the Big Bang, they substantially change the profile of MFs at all the values of temperature explored in {\sc Simba} simulations: they increase by more than 50 percent the volume filling fraction, especially of regions with $T>10^4$~K, skew the $\varv_1(T)$ curve by decreasing the surface area of cold ($T\simeq10^4$~K) regions by $\sim$ 7 percent and increasing that of hot regions ($T~\gtrsim~10^6$~K) by almost a factor of 10, and almost reverse the convexity of $T~>~10^4$~K regions. In the same epoch, AGN jets also strongly alter the relative abundance of clumps, tunnels, and bubbles. In particular, as the universe evolves with time, bubbles are restricted to regions with temperature $T\lesssim10^4$~K, without jets they would occupy also cooler regions (the first zero of $\varv_3(T)$ decreases with time, especially for $z>1.5$, as shown by the dashed line in the second-row or solid line in third-row panels). Because of jets, at the same time, clumps form in regions with larger and larger temperatures (the second zero of $\varv_3(T)$ increases almost linearly with redshift excluding the redshift range $1<z<1.5$, when it is nearly constant). Accordingly, the temperature range concerned by negative values $\varv_3$, or tunnels, increases with time, reaching the minimum value at $z=0$ for $T=10^6$~K regions. Without jets, the topology of the $T$-field would be dominated by tunnels over a very narrow range of temperatures at $z=5$ ($10^4~\mathrm{K}\lesssim T\lesssim 10^{4.2}~\mathrm{K}$) and a much broader range at $z=0$ ($10^{3.3}~\mathrm{K}\lesssim T\lesssim 10^5~\mathrm{K}$); after $z=2$, because of jets, the percolation threshold increases dramatically, up to two orders of magnitude at $z=0$ from $T\simeq10^{4.5}~\mathrm{K}$ to $T\simeq10^{6.3}~\mathrm{K}$, consistently with a warm-hot (ionised) IGM permeating the largest fraction of the total volume, with less than 10 percent occupied by hot regions, and very few cold cavities.

X-ray heating and stellar feedback have a secondary role with respect to AGN jets in shaping the morphology, by about the same relative amount with respect to their reference model. 
However, while X-ray heating operates in two distinct epochs at high and low redshift ($z>4$ and $z<1.5$), respectively on cold ($T\simeq10^4$~K) and hot ($T\gtrsim10^5$~K), stellar feedback acts at intermediate redshift ($1<z<2.5$) and on warm regions ($10^{3.5}\mathrm{K}\lesssim T\lesssim10^5\mathrm{K}$).  This is because X-ray feedback acts to remove cool gas from the central regions of galaxies with massive black holes, which can alter the morphology of $\sim 10^4$~K gas via heating or removal at high redshift, while at lower redshift it enables the jet feedback to be more effective at redistributing baryons from within galaxies into the hot IGM. Meanwhile, stellar winds have more moderate velocities than jets, so they remove cool gas from galaxies but are only able to heat it to warm-hot temperatures in the intergalactic medium.

AGN winds have the smallest effect on the morphology of the $T$-field, of the order of a few percent, limited to very low redshift ($z<0.5$) for regions with temperatures around $10^{4.5}-10^{5.5}$~K. They non-trivially change the topology: AGN winds increase the value of $\varv_3$ and therefore the number of hot clumps ($T>10^5$~K) at $z=0.5$, and enhance the peculiar topology of regions with $10^{3.7}~\mathrm{K}\lesssim T\lesssim 10^{4.4}~\mathrm{K}$ already set by jets, by increasing (decreasing) the number of tunnels in regions with temperature smaller (larger) than $T=10^4$~K. In general, radiative AGN winds are seen to have very minor effects on galaxy properties~\citep{Dave2019}, and this appears to be reflected in the global gas morphologies as well.

A complementary way to interpret the Minkowski curves is looking at Figure~\ref{fig:MFs_T_nonorm} from top to bottom and considering the excursion set defined by the mean temperature, hereafter dubbed $\bar{T}$-regions.\footnote{We propose this kind of analysis for the temperature field only, for illustrative purposes.} Focusing on the volume filling fraction, the $\bar{T}$-regions occupy less than half the volume when all the AGN and stellar feedback mechanisms are included, with $\varv_0(\bar{T})$ decreasing from about 50 percent at $z=5$ to 15-20 percent at $z=2.5$ and rising back to about 50 percent at a later time. X-ray heating qualitatively does not change this trend (filled circles and crosses in the first row panels are almost superposed), modified by 10-13 percent at $z>4$ for $T=10^4$~K regions and by less than 10 percent at $z<1.5$ in hotter regions. Jets strongly heat the gas, especially at $z<2$, pushing the $\bar{T}$-regions to occupy more than 20 percent of the full volume. Stellar feedback and AGN winds do not change qualitatively the trend of $\varv_0(\bar{T})$ established by pure hydrodynamics, the former mainly affecting the volume filling fraction of $T=10^{4.2}$~K regions from $z=3$ down to $T=10^{3.8}$~K regions at $z=0.5$. AGN winds only increase $\varv_0$ of $T\simeq10^4$~K regions by about 3 percent and only at $z=0$. 

% ------
\begin{figure*}
\centering\includegraphics[width=17cm]{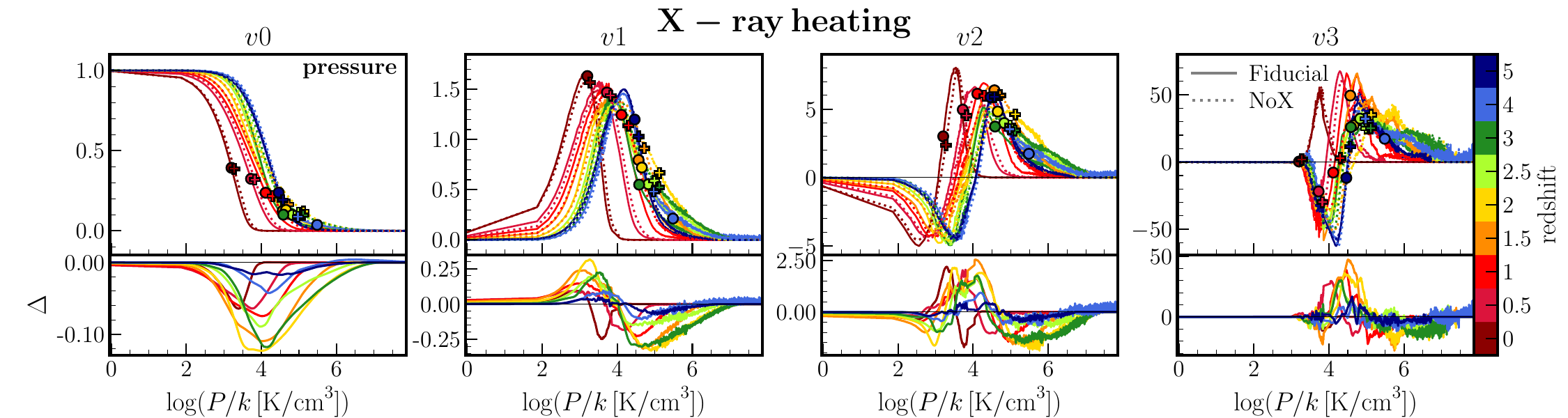}
\centering\includegraphics[width=17cm]{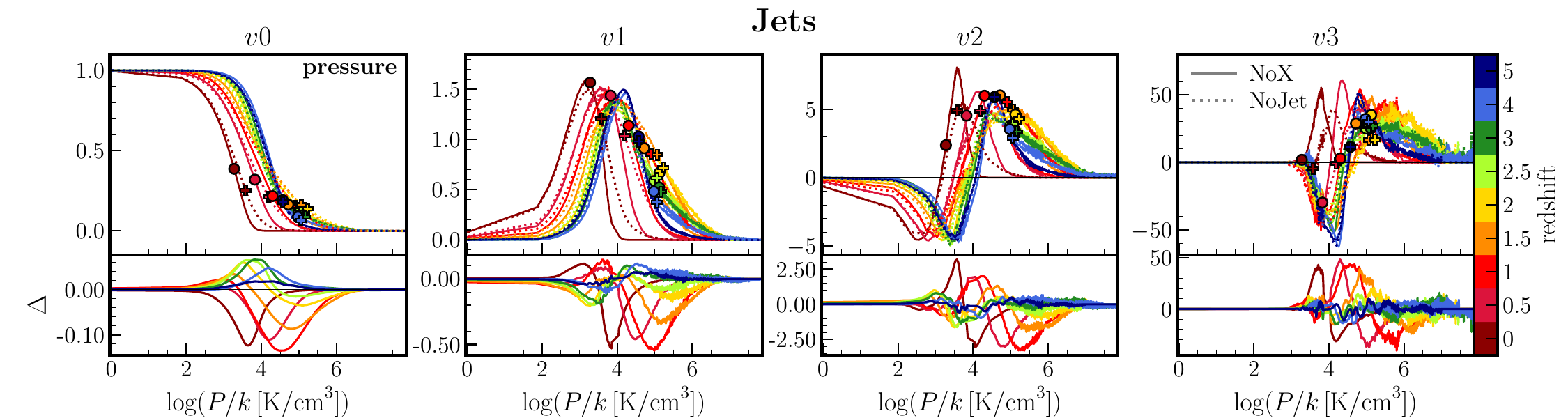}
\centering\includegraphics[width=17cm]{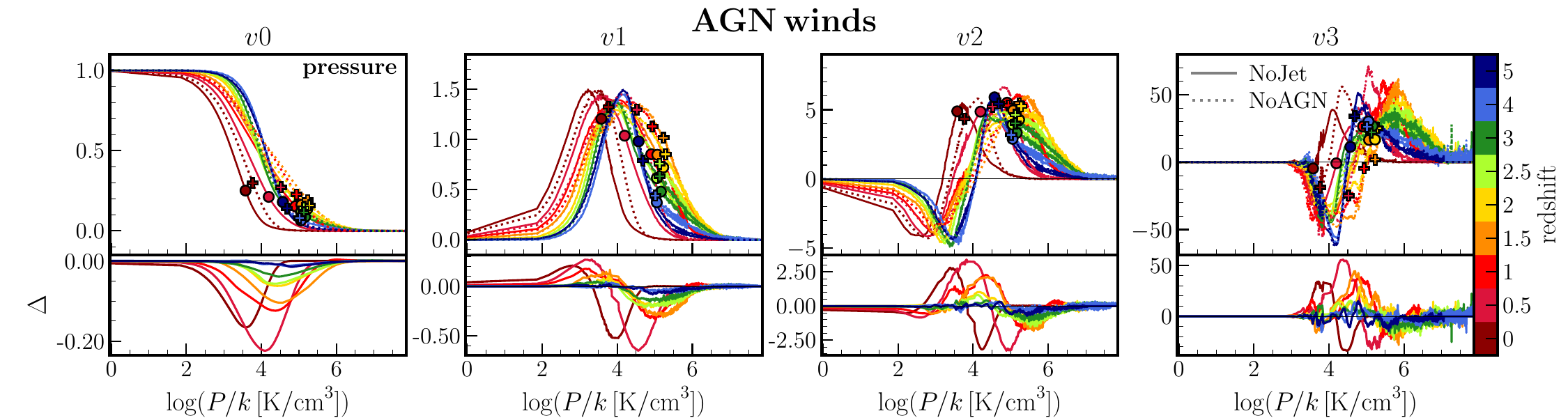}
\centering\includegraphics[width=17cm]{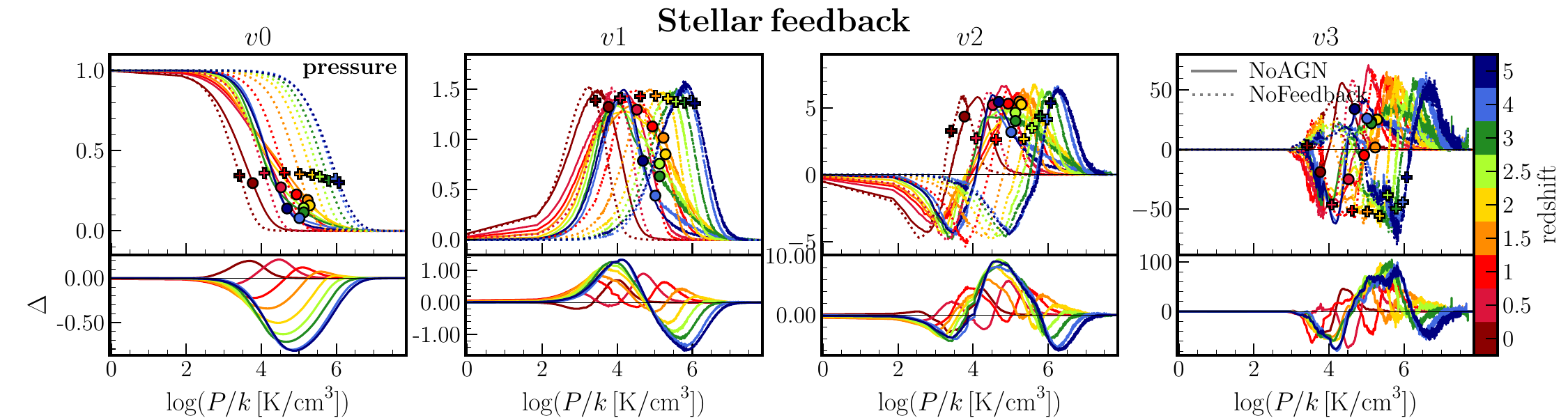}
\caption{\textbf{Pressure.} 
Minkowski curves for the gas pressure or $P$-field (in units of Boltzmann constant, noted $k$), analogous to Figure~\ref{fig:MFs_T_nonorm}. 
Stellar feedback is the first cause of geometrical and topological change of the $P$-field set by gravitational and hydrodynamical interactions, in particular at high redshift ($z\gtrsim2$). At low redshift ($z\lesssim 1.5$), it is AGN winds and AGN jets that have a dominant impact on the morphology of the $P$-field; see \S\ref{sec:results:P}.
}
\label{fig:MFs_p_nonorm}
\end{figure*}

In the full model, the density of surface area of $\bar{T}$-regions, $\varv_1(\bar{T})$, is maximum at very high redshift ($z>4$), consistent with a single-phase (uniform) fluid at mean temperature, and at very low redshift ($z<0.5$), indicating that the corresponding excursion set is jagged, namely, non-spherical, disturbed by late-time or integrated energy injections. Indeed, as reminded in the Guidelines (see \S~\ref{sec:results:guidelines}), spheres are the bounded regions with minimum surface for a fixed volume. Consistently, in these two epochs $\varv_2(\bar{T})$ and $\varv_3(\bar{T})$ indicate non-maximal convexity and minimum Euler characteristic, namely, spongy-like topologies dominated by tunnels. The morphological transition occurs between $z=2.5$ and $z=1.5$, the $\bar{T}$-regions with $\bar{T}=10^5$~K at $z=2$ having a smoother shape with the number of isolated regions being almost equal to the number of tunnels ($\varv_3(\bar{T})\simeq0$); namely, the $\bar{T}$-regions percolate at the peak of the star formation rate. At that epoch, the temperature field is strongly non-Gaussian, with regions with $T<\bar{T}$ dominated by tunnels. This corresponds to a network of cold filaments surrounded by hotter gas resulting from the energy injection by X-ray heating and AGN jets as expected by the percolation of shock waves. As time goes by, the effect of jets, stellar feedback, and AGN winds (in this order) is to make cool regions with $T<\bar{T}$ less and less round as time goes on (larger surface area, negative integrated mean curvature, negative Euler characteristic), consistent with a spongy-like topology.  Finally, by $z=0$, these cooler regions become very sparse as most of the IGM becomes heated via AGN feedback and gravitational shock heating.

% ----------------
\begin{figure*}
\centering\includegraphics[width=17cm]{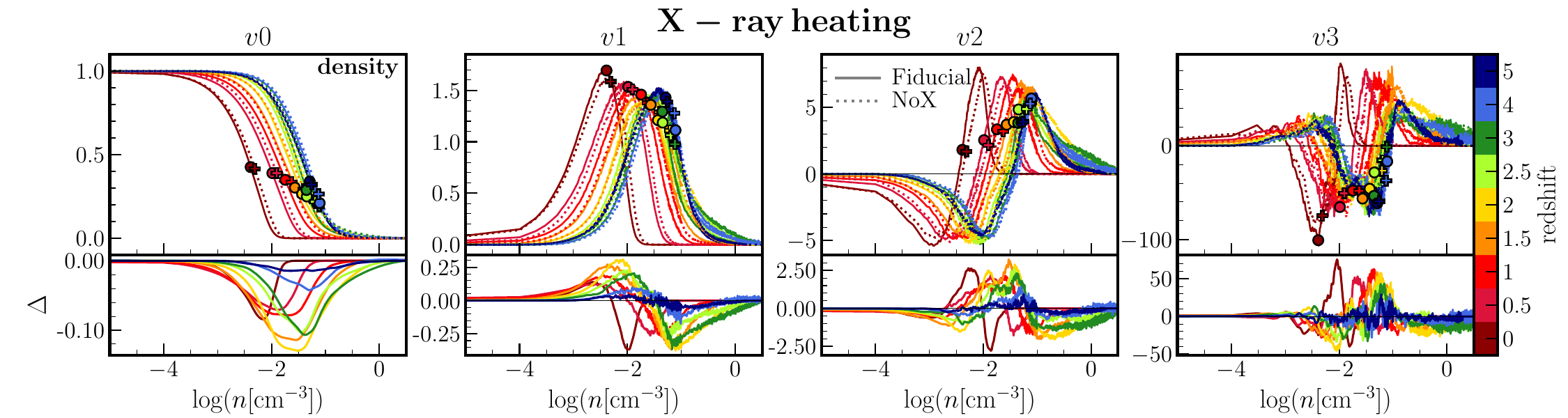}
\centering\includegraphics[width=17cm]{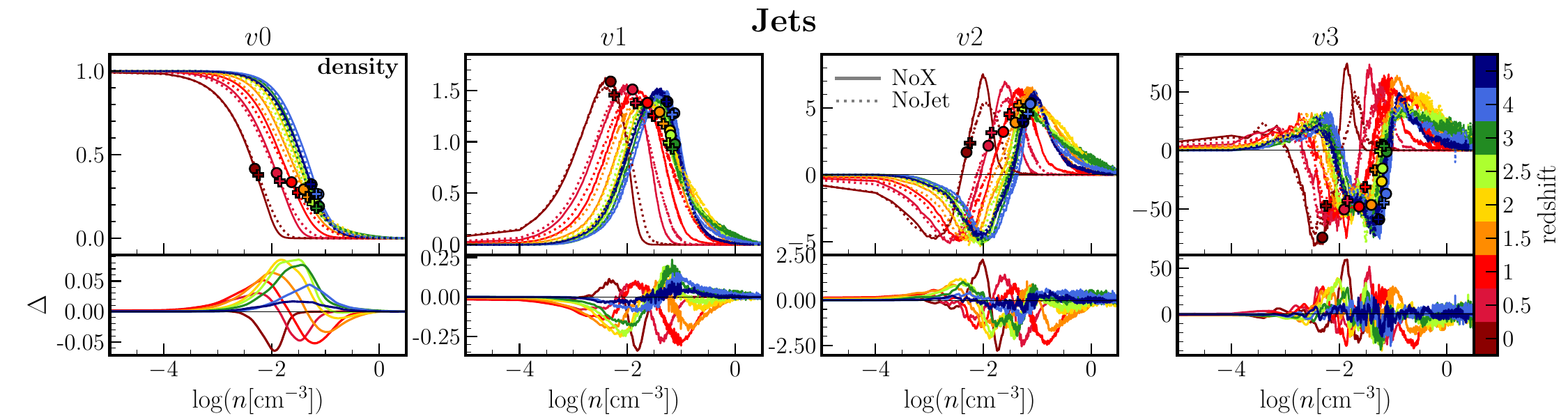}
\centering\includegraphics[width=17cm]{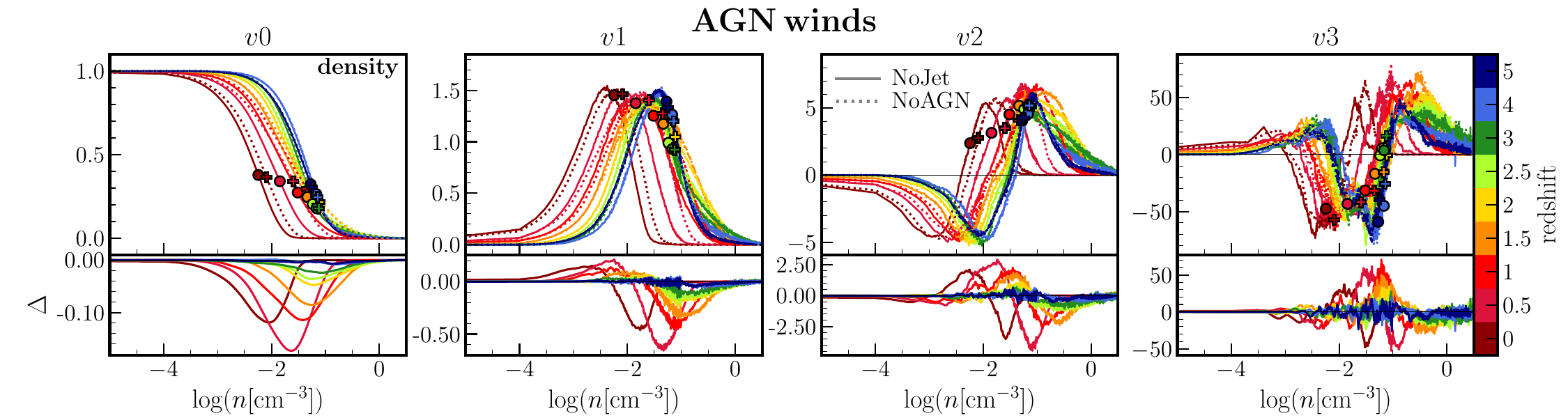}
\centering\includegraphics[width=17cm]{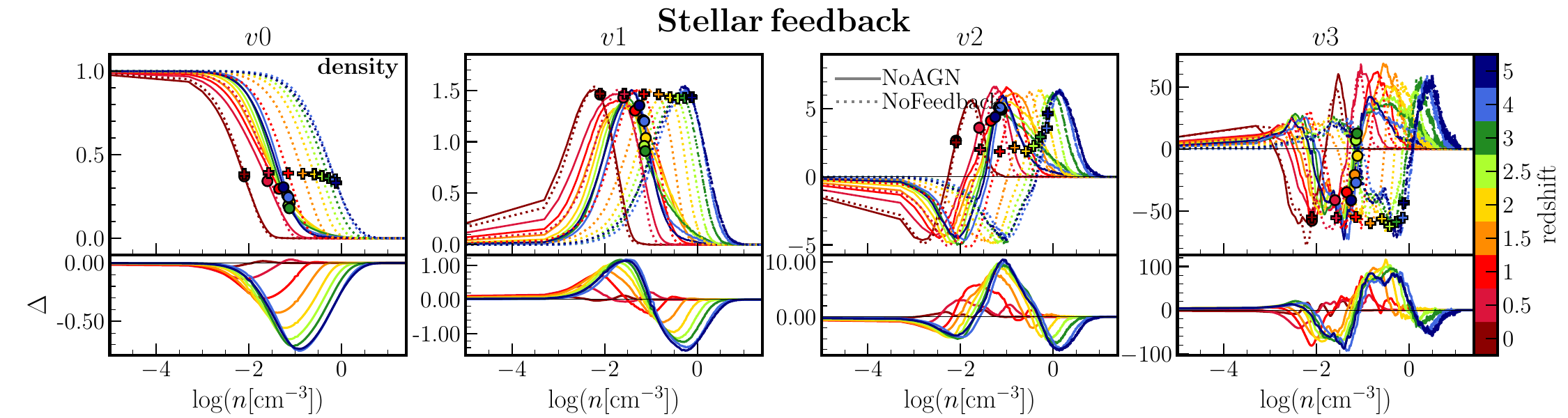}
\caption{\textbf{Total density.} Minkowski curves for the total gas number density or $n$-field, analogous to Figure~\ref{fig:MFs_T_nonorm}.
Until $z\simeq1$ stellar feedback modifies morphology by the largest amount, progressively weakening until $z\simeq0.5$; at later time AGN feedback mechanisms become the main driver shaping the $n$-field.
}
\label{fig:MFs_rho_all_nonorm}
\end{figure*} 

% ----------------------------------------
\subsection{Pressure}
\label{sec:results:P}

Figure~\ref{fig:MFs_p_nonorm} shows the MFs of the $P$-field, showing a redshift evolution qualitatively opposite to that of the $T$-field, consistent with a universe that evolves towards a thermodynamical vacuum, namely, with lowering pressure. We note that the multi-phase nature of the gas components invalidates the assumption of a single equation-of-state relating $P$, $T$, and $n$ (see Appendix~\ref{app:phasediagrams}). Therefore, the morphology of the $P$-field cannot be deduced from that of the $T$ and $n$ fields.

The Fiducial model (solid lines, first-row panels) manifests a smooth and weak evolution of the morphology until approximately $z=1.5$, as indicated by a small translation of the Minkowski curves that essentially maintain their shape, followed by a sudden evolution occurring between $z=0.5$ and $z=0$ (the abrupt change of $\varv_0$ and $\varv_1$ curves at $P>10^2k_\mathrm{B}\mathrm{cm}^{-3}$K, especially visible at $z<0.5$, is a numerical artefact). At very early times, between $z=5$ and $z=4$, the four Minkowski curves do not change, apart from a slight increase of $\varv_3$ for the highest pressure regions ($P>10^5k_\mathrm{B}\mathrm{cm}^{-3}$K). However, during this epoch the $\bar{P}$-regions tend to occupy only a small fraction of the total volume ($\varv_0(\bar{P})\simeq0.2$ at $z=5$ and $<0.05$ at $z=4$), suggesting that apart from sparse, small, and convex (spherical) high-pressure clumps most of the volume is low pressure with $P<\bar{P}$. For $z<4$ and until $z\sim1.5$ the fraction of the total volume occupied by high-pressure domains progressively decreases, the $\bar{P}$-regions and furthermore the highest pressure domains forming a network with meatball topology. Between $z=1$ and $z=0.5$ the volume occupied by the $\bar{P}$-field and its surface density continue to increase while its convexity decreases, which is compatible with the progressive appearance of tunnels as indicated by the negative Euler characteristic. In the last time step, for $z<0.5$, all regions with $P>10^{3.9}k_\mathrm{B}\mathrm{cm}^{-3}$K disappear; regions with $P<\bar{P}=10^{3.4}k_\mathrm{B}\mathrm{cm}^{-3}$K occupy more than 60 percent of the volume and are essentially concave with $\varv_3\simeq0$, compatible with either a network with an equal number of bubbles and tunnels or a homogeneous distribution; and regions with $10^{3.4}<P/k_\mathrm{B}\mathrm{cm}^{-3}\mathrm{K}<10^{3.9}$ form convex, isolated clumps.

Stellar feedback is the main cause of deviation from the morphology of the $P$-field set by the no feedback case (dotted lines in bottom-row panels). In the absence of any feedback, the geometry and topology of the $P$-field smoothly evolve over the full redshift range $0<z<5$ studied here, the MFs curves shifting from high to low values of $P$. Stellar feedback limits high-pressure domains with $P\gtrsim10^{4.9}k_\mathrm{B}\mathrm{cm}^{-3}$K to occupy less than 15-20 percent of the total volume, quenching the time evolution of their shape and topology until $z\sim2$. This is visible by comparing the MFs of the $\bar{P}$-field, which move towards smaller values only for $z<2$ when stellar feedback is active.

X-ray heating and jets have a quantitatively similar impact on the MFs, the former in the redshift range $1.5<z<3$ and mostly concerning the domains with $P\gtrsim10^{2.9}k_\mathrm{B}\mathrm{cm}^{-3}$K, the latter at $z<1.5$ (not considering the highest pressure regions with $P/k_\mathrm{B}\gtrsim10^{5.9}\mathrm{cm}^{-3}$K, very likely not resolved because of smoothing; see Appendix~\ref{app:phasediagrams}).
Among the AGN feedback mechanisms, winds have the largest impact on the global morphology but are limited to the lowest redshift, $z<1$, changing $\varv_0$ by 20 percent, $\varv_1$ by more than 40 percent, $\varv_2$ by 50 percent, and $\varv_3$ by up to 100 percent. At higher redshifts ($z\gtrsim1.5$) AGN winds modify the geometry of the $P$-field approximately at the same level as the other AGN feedback mechanisms.

% ----------------
\begin{figure*}
\centering\includegraphics[width=17cm]{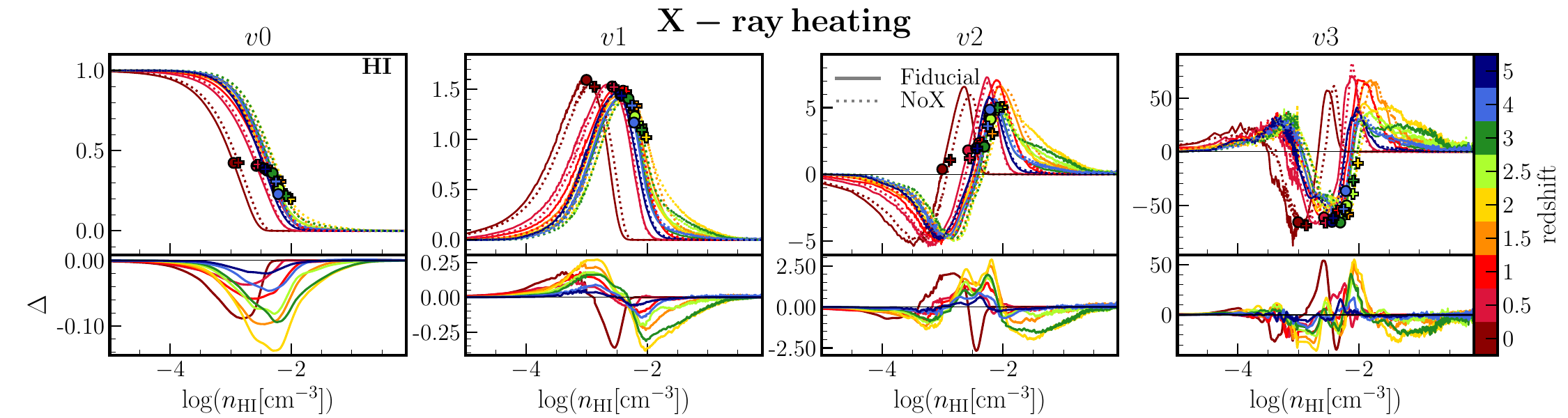}
\centering\includegraphics[width=17cm]{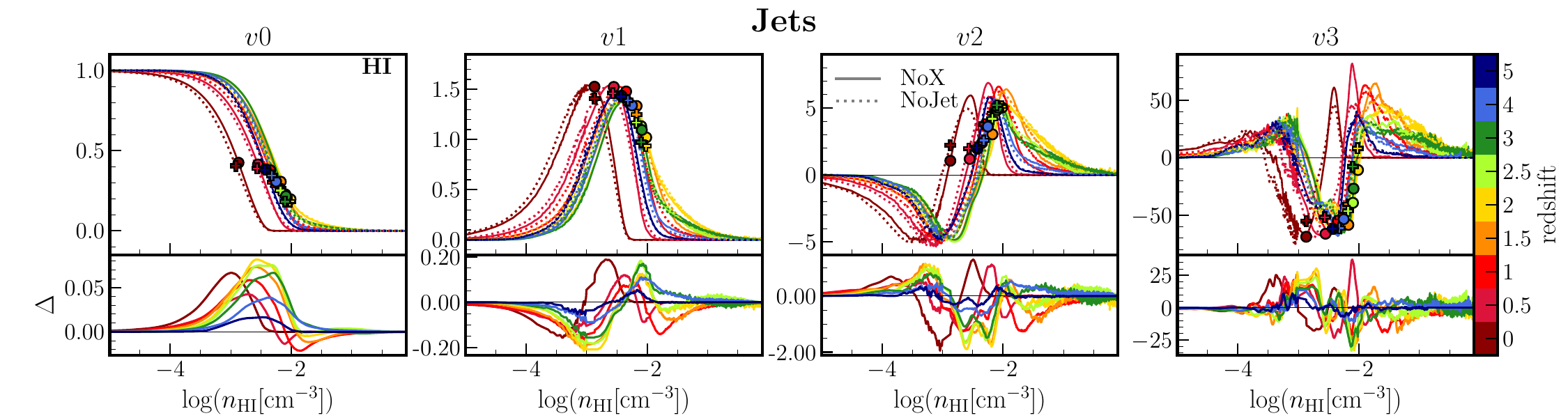}
\centering\includegraphics[width=17cm]{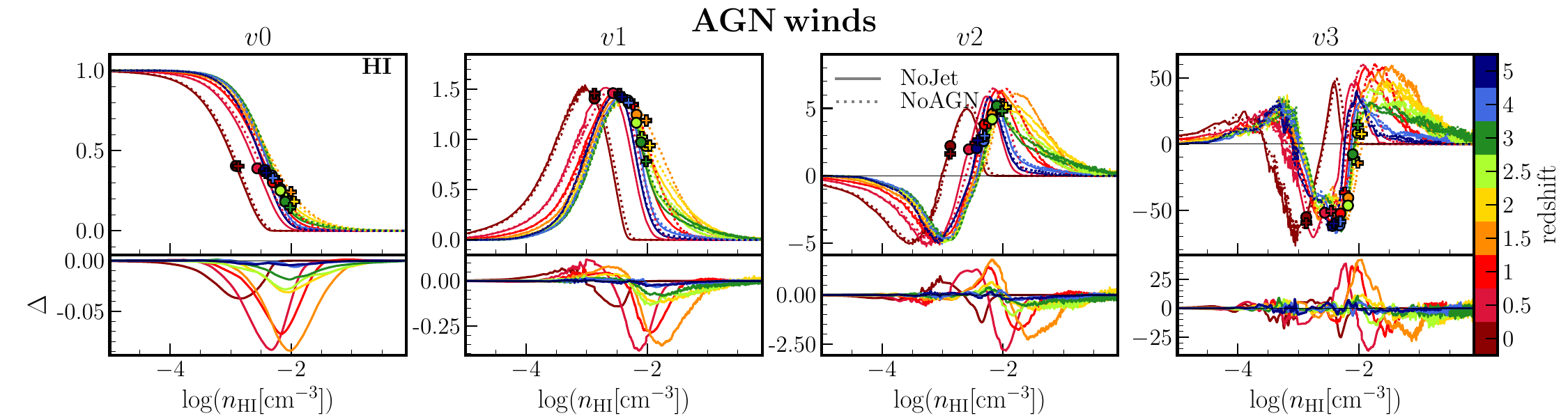}
\centering\includegraphics[width=17cm]{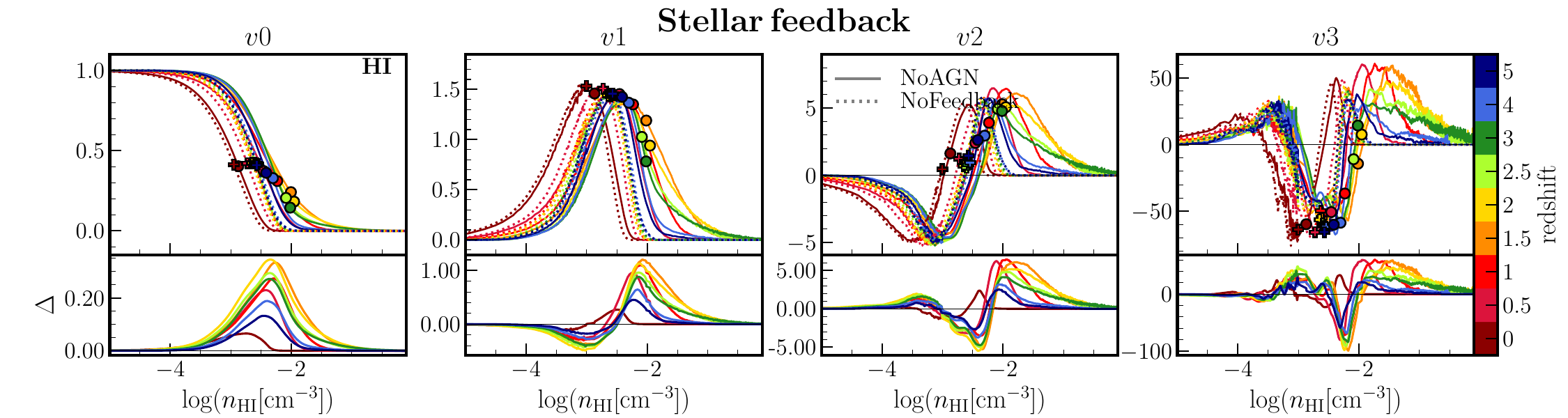}
\caption{\textbf{HI density.} Minkowski curves for the neutral atomic hydrogen number density, or $n_{\mathrm HI}$-field, analogous to Figure~\ref{fig:MFs_T_nonorm}. They are similar to the Minkowski curves of the $n$-field at $z>2.5$, and likewise modified at a later time by stellar feedback but in the opposite way; see \S\ref{sec:results:n}.
}
\label{fig:MFs_rho_HI_nonorm}
\end{figure*}

% ----------------
\begin{figure*}
\centering\includegraphics[width=17cm]{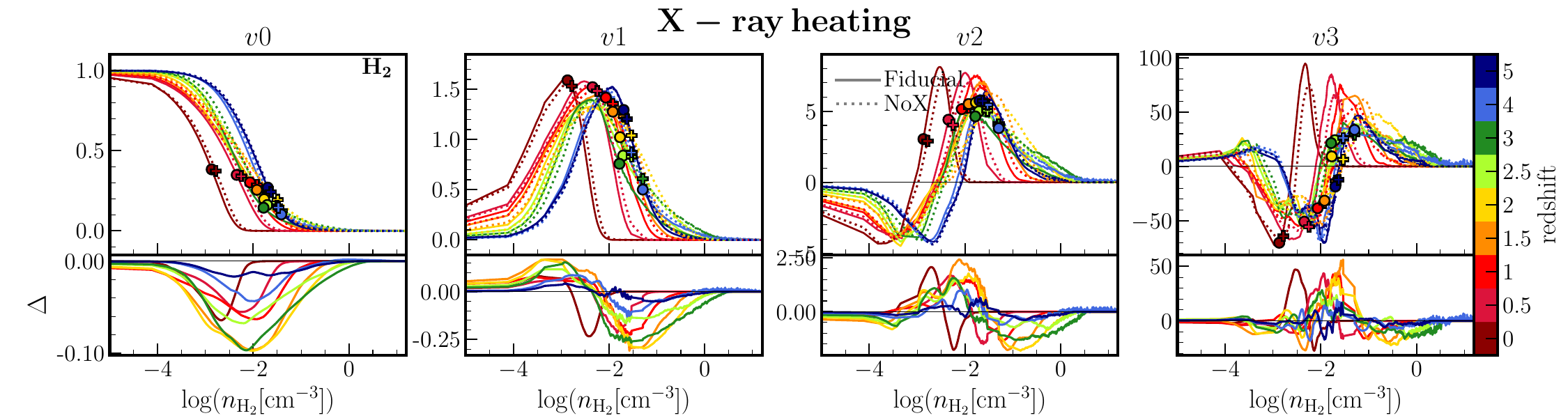}
\centering\includegraphics[width=17cm]{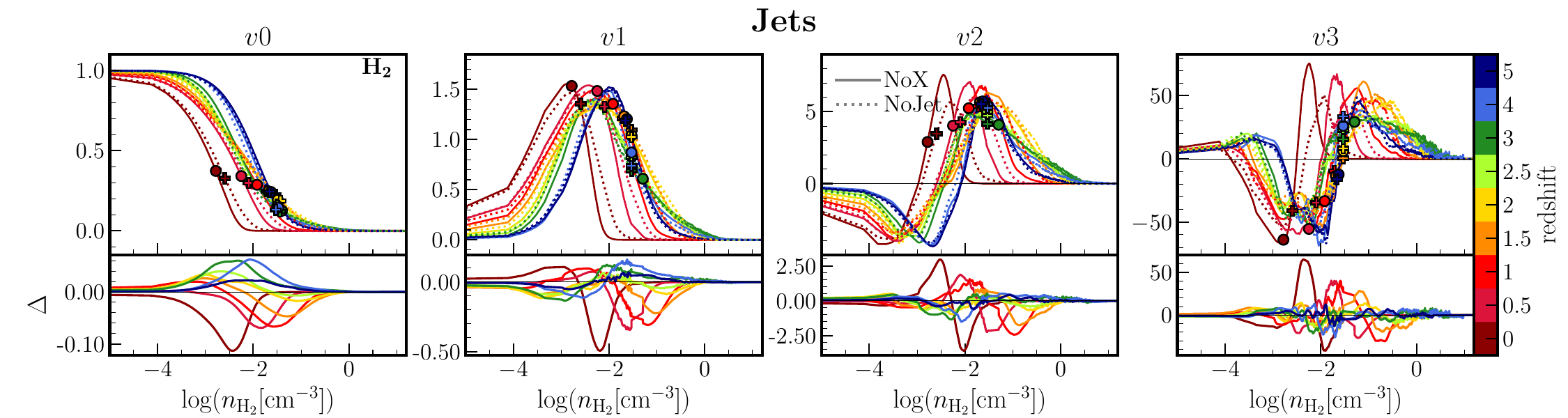}
\centering\includegraphics[width=17cm]{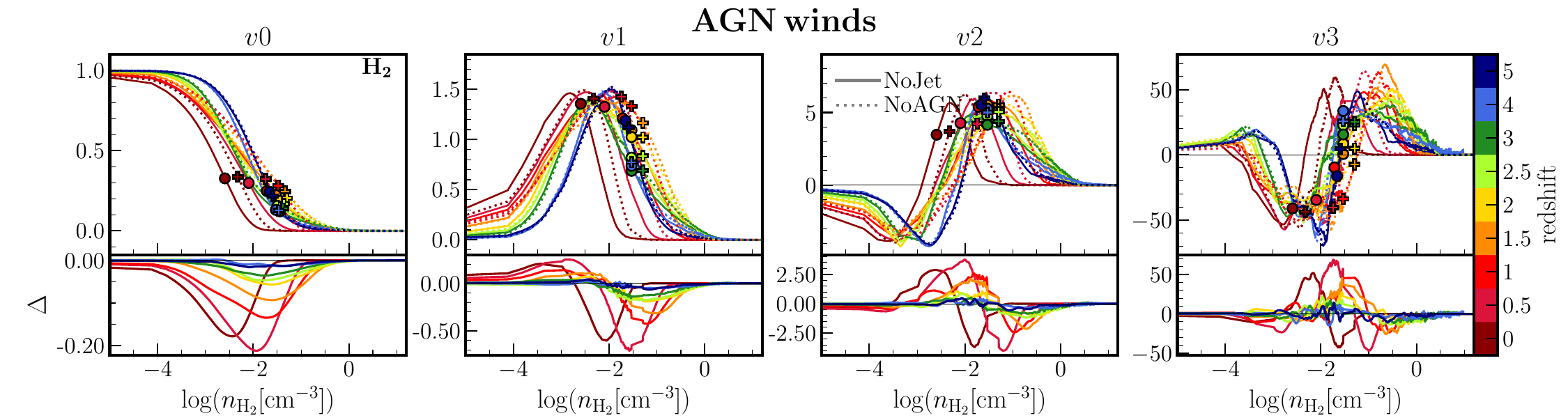}
\centering\includegraphics[width=17cm]{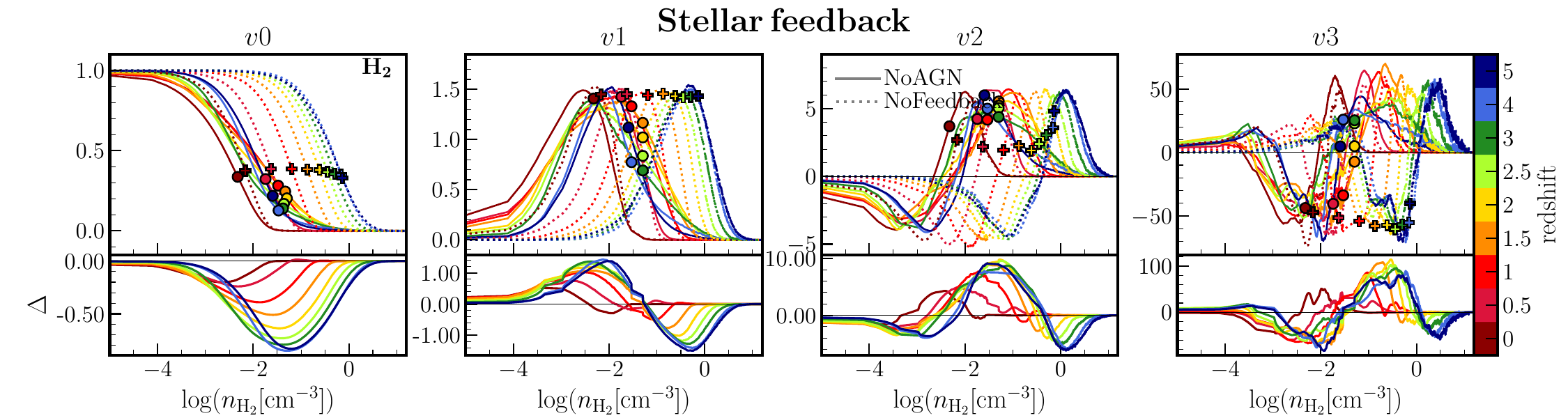}
\caption{\textbf{{\sc H$_2$} density.} Minkowski curves for the molecular hydrogen number density or $n_{\rm{H}_2}$-field, analogous to Figure~\ref{fig:MFs_T_nonorm}. 
Its evolution is very similar to that of the total gas density; see \S\ref{sec:results:n}.
}
\label{fig:MFs_rho_H2_nonorm}
\end{figure*}

% ----------------------------------------
\subsection{Density fields}
\label{sec:results:n}

% ------
\subsubsection{Total density field}

Figure~\ref{fig:MFs_rho_all_nonorm} shows the MFs for the total gas number density field $n$. Overall, the redshift evolution of the total and individual density fields, namely neutral ({\sc Hi}) and molecular ({\sc H$_2$}) gas, is qualitatively similar to that of the $P$-field; as suggested by the volume filling fraction, high-density regions progressively disappear in agreement with the global expansion of the universe and the dilution of matter.

When all the feedback mechanisms are included (solid lines in top row panels), the global morphology at high redshift vaguely reminds that of a log-normal random field. 
Between $z=5$ and $z=0$, the domains with mean density or $\bar{n}$-regions (circles) occupy a fraction of volume that increases from 20 to 40 percent; starting from $z=1.5$, they have the largest surface density, and approximately at the same epoch their convexity begins to decrease at a lower rate than before, though always being positive. A network of tunnels defined by $\bar{n}$-regions, or equivalently a network of filaments with $n<\bar{n}$, is already in place at $z=5$; it grows until $z=3$ while the abundance of voids is almost constant, then it stops growing between $z=3$ and $z=1$ (the minimum of $\varv_3$, marked by circles is almost constant in this period); after $z=1$ it grows again, at the expenses of higher density clumps ($n\simeq10^{-2}\mathrm{cm}^{-3}$) that occupy a very tiny fraction of the total volume, and at the expenses of voids that instead occupy the largest fraction of the volume.

The largest modification to the global morphology of the $n$-field is caused by stellar feedback (bottom-row panels). This is expected because, while AGNs (particularly jets) dominate the global feedback energy input, stellar winds displace a far larger amount of mass. Without stellar feedback (dotted lines) its morphological evolution would smoothly evolve from $z=5$ to $z=0$; the values of all the MFs but $\varv_2$ of $\bar{n}$-regions (marked by crosses) are constant over time, and their global topology is dominated by tunnels. With stellar feedback (solid lines), the time evolution of $\bar{n}$ and, correspondingly, the morphological evolution of $\bar{n}$-domains are instead quenched until $z\simeq1-1.5$, namely until the late-end of the star-formation-rate peak \citep{MadauDickinson2014}. MFs suggest that at $z>2$ stellar feedback limits the abundance and size of gas regions to lower values and reduces the number of tunnels, but does not modify their convexity ($\varv_2$). Below $z=1$, stellar feedback is not morphologically efficient anymore; the geometry and topology of the total density field evolve controlled by AGN feedback mechanisms, especially AGN winds, as also indicated by the MFs of $\bar{n}$-regions (circles in bottom-row panels).

Overall, AGN feedback mechanisms mildly modulate the geometry and topology of the gas density field set by pure hydrodynamics and stellar feedback until $z=0.5-1$. Afterwards, they start tempering or reinforcing the existing morphology,  competing with stellar feedback.

AGN winds and X-ray heating have a qualitatively similar influence on the MFs of regions with density $10^{-3} \lesssim n/\mathrm{cm}^{-3} \lesssim 10^{-0.5}$.
Both decrease the volume filling fraction by more than 10 percent and alter the convexity by $\sim40$ percent in the same way. Nonetheless, AGN winds mainly operate at $z<1$ and produce rounder overdensity regions with a smaller surface, while X-ray heating has the same effect also in the redshift range $1<z<2$.

Jets modify the volume filling fraction of density regions by only 5 percent but in a more subtle way: they increase $\varv_0$ of both overdense and underdense regions at $z>2$, increase (decrease) $\varv_0$ of overdense (underdense) regions in the range $0.5<z<2$, and decrease $\varv_0$ of both overdense and underdense regions at $z<0.5$. The effect on $\varv_1$ and $\varv_2$ is qualitatively and quantitatively similar to AGN winds.

As for the topology of the $n$-field, all the AGN feedback mechanisms have qualitatively and quantitatively similar impact on the topology of the density field, limited to $z<3$. X-ray heating and jets decrease the number of tunnels with a density slightly larger than the mean value and more intensively increase the number of high-density clumps. AGN winds instead decrease the number of high-density clumps. Both jets and winds are moderately more efficient in shaping the topology of the density field at redshift $z<0.5$.

The morphology of low-density excursion sets with $n\leq10^{-5}\mathrm{cm}^{-3}$ (not shown in figures) is worth commenting on. Focusing on the Fiducial model at $z=0$, these regions form a pattern made of very small and isolated domains with overall almost vanishing volume-filling-fraction (so that the complementary excursion set has $\varv_0\approx1$), very small surface density, and positive integrated mean curvature (correspondingly, the complementary excursion sets with $n>10^{-5}\mathrm{cm}^{-3}$ have $\varv_1\gtrsim0$ and $\varv_2\lesssim0$), and very small and positive Euler characteristic density, $\varv_3\gtrsim0$, namely, they form a diffuse ensemble of disconnected spheres. This morphology is qualitatively unchanged by feedback mechanisms and qualitatively similar for domains with larger density at an earlier time. Although restrained by the spatial smoothing of the field, similar conclusions concern very dense excursion sets with $n>10^{-1.5}\mathrm{cm}^{-3}$ at $z=0$ ($n\gtrsim 2\mathrm{cm}^{-3}$ at $z=5$).

% ------
\subsubsection{HI density field}

The MFs for the number density field of the atomic neutral hydrogen gaseous component are shown in Figure~\ref{fig:MFs_rho_HI_nonorm}. Similarly to the total gas density field, when all the feedback mechanisms are included (solid lines in top row panels) the global morphology of the $n_\mathrm{HI}$-field is qualitatively similar to that of a log-normal random field at very high redshift, $z=5$, and evolves in a qualitatively similar way afterwards, though less linearly with redshift. Three epochs can be identified: \textit{(i)} between redshift $z=4$ and $z=2.5$ the surface density, curvature, and number density of very dense clumps ($10^{-2}\mathrm{cm}^{-3}<n_\mathrm{HI}<10^{-0.5}\mathrm{cm}^{-3}$) increase, indicating an evolution compatible with a local gravitational collapse, like for the total gas $n$-field; \textit{(ii)} from $z=2.5$ to $z=0.5$, unlike the $n$-field, the regions with similar and smaller density ($10^{-2.5}\mathrm{cm}^{-3}<n_\mathrm{HI}<10^{-1.5}\mathrm{cm}^{-3}$) become more spherical (smaller $\varv_1$, larger $\varv_2$) and more abundant (larger $\varv_3$); \textit{(iii)} an abrupt change of the global morphology occurs at $z=0.5$, the volume filling fraction $\varv_0$ being almost unchanged until this epoch like for the total gas $n$-field. However, unlike the latter, at $z=0$ the mean {\sc Hi} field occupies half of the volume ($\varv_0(\bar{n}_\mathrm{HI})\simeq0.5$) and is on average flat ($\varv_2(\bar{n}_\mathrm{HI}) \simeq 0.5$), namely, it is much more Gaussian although the maxima of $\varv_3$ have again unequal amplitude.

The stellar feedback is the main source of disturbance, like for the $n$-field. However, it works in the opposite way and especially at $1<z<3$, not progressively at all redshifts: it does increase $\varv_0$ by up to 30 percent, especially for intermediate density regions; it increases (decreases) the surface density $\varv_1$ of regions with higher (lower) density than the critical value $n_\mathrm{HI}~\simeq~10^{-2.5}~\mathrm{cm}^{-3}$ by up to 70 percent; it does change the density of integrated mean curvature $\varv_2$, again with respect to the critical value $n_\mathrm{HI} \simeq 10^{-2.5}$cm$^{-3}$; and increases the number of tunnels of excursions sets for a narrow range of thresholds, $10^{-2.5} < n_\mathrm{HI}/\mathrm{cm}^{-3} \simeq 10^{-2}$.

The three AGN feedback mechanisms affect the morphology of the $n_\mathrm{HI}$-field to a smaller extent and with similar amplitude. X-ray heating modifies the same densities as stellar feedback but with the opposite effect and is limited to $1.5<z<3$ and, suddenly, at $z=0$. Among the AGN feedback mechanisms, it has the largest impact on the topology. AGN winds have an analogous effect but at a later time, in the redshift range $0.5<z<1.5$. Jets behave like stellar feedback, apart from a less trivial impact on the topology of the field and a very specific and strong impact on the geometry and topology of domains with $n_\mathrm{HI} < 10^{-2.5}$cm$^{-3}$ at $z=0$, like X-ray heating.

% ------
\subsubsection{H$_2$ density field}

Figure~\ref{fig:MFs_rho_H2_nonorm} shows the MFs for the molecular hydrogen number density or $n_{\mathrm{H}_2}$-field. Very likely because of its lower fraction, they are very similar to that of the total gas density (apart from the obvious rescaling, i.e. the shift along the horizontal axis because of the lower density) and the feedback mechanisms operate in a very similar way. Therefore, one can come to the same conclusions discussed above for the $n$-field.

% ----------------------------------------
\subsection{Metallicity}
\label{sec:results:Z}

% ----------------
\begin{figure*}
\centering\includegraphics[width=17cm]{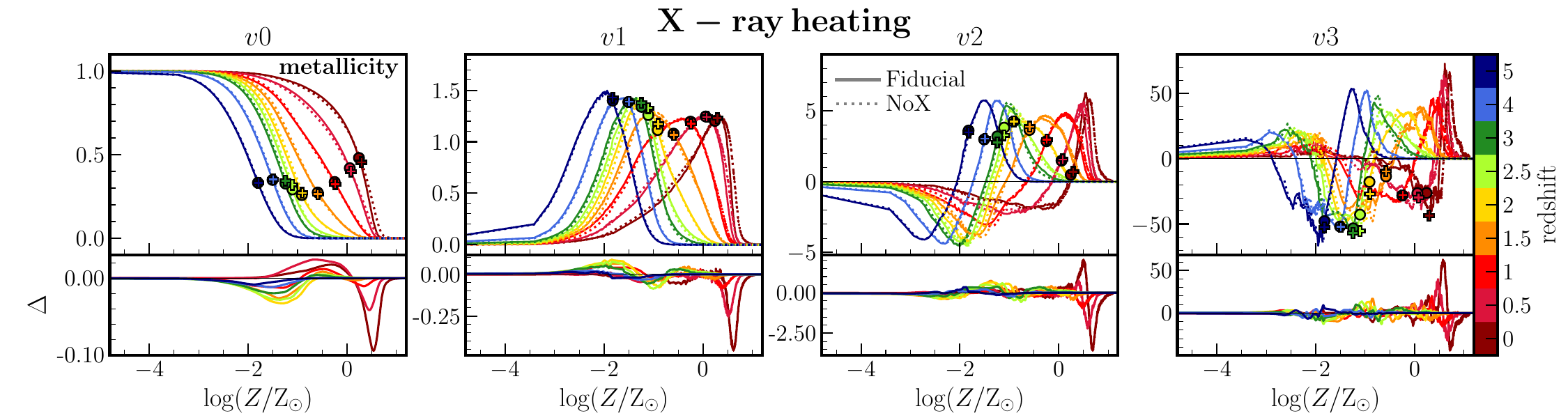}
\centering\includegraphics[width=17cm]{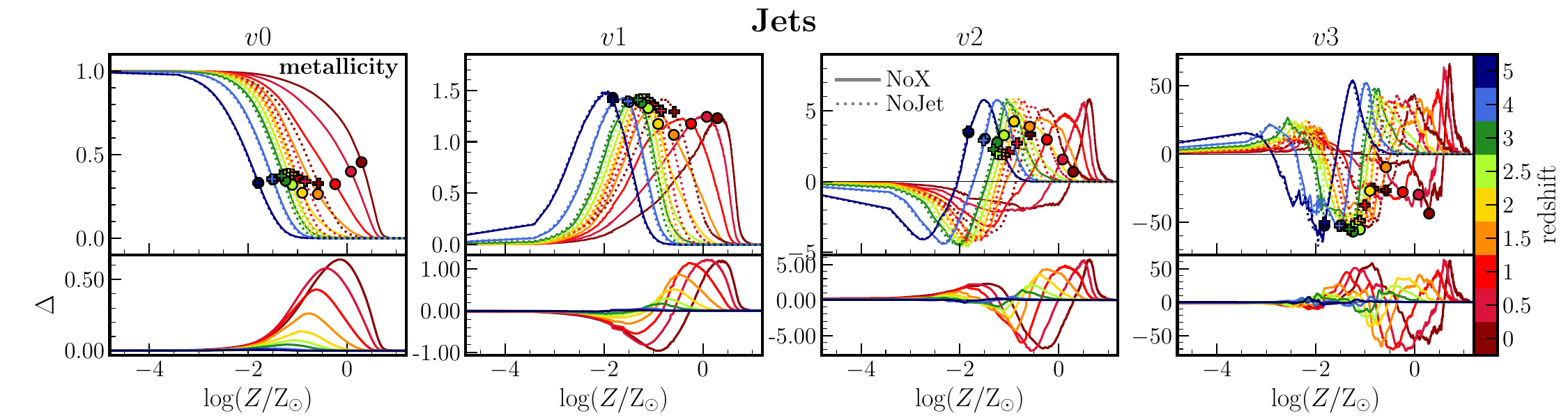}
\centering\includegraphics[width=17cm]{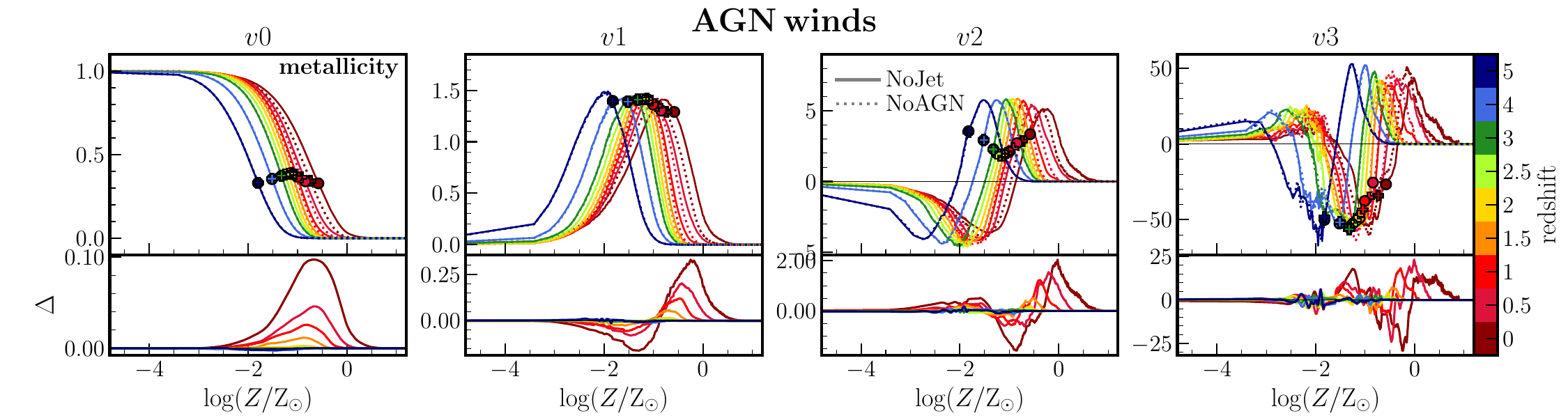}
\centering\includegraphics[width=17cm]{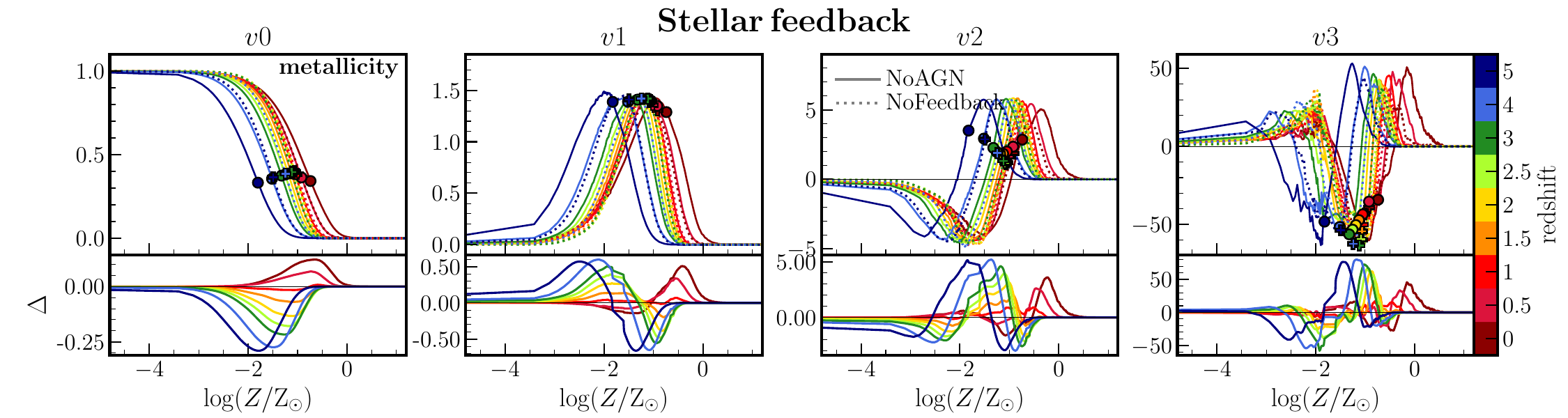}
\caption{\textbf{Metallicity.} Minkowski curves for the gas metallicity or $Z$-field, analogous to Figure~\ref{fig:MFs_T_nonorm}.
AGN feedback mechanisms, especially jets, modify the morphology of regions with $Z\gtrsim0.01Z_\odot$ for $z<1.5$, while at earlier time the morphology is maximally influenced by
stellar feedback, especially in regions with $Z\lesssim0.1Z_\odot$; see \S\ref{sec:results:Z}.
}
\label{fig:MFs_metal_nonorm}
\end{figure*} 

Figure~\ref{fig:MFs_metal_nonorm} shows the MFs for the gas metallicity or $Z$-field. The redshift evolution of the MFs, qualitatively similar to that of the $T$-field shown in Figure~\ref{fig:MFs_T_nonorm} and at odds with that of the $n$-field shown in Figure~\ref{fig:MFs_rho_all_nonorm}, supports a global metal enrichment over time, with a morphological transition occurring around $z=1.5$. {Overall, the $Z$-field is composed by \textit{i)} small, spheroidal, metal-poor bubbles at early time that progressively disappear; \textit{ii)} a network of thin filaments with $Z\simeq Z_\odot$; and \textit{iii)} a collection of small bubbles with metallicity increasing with time from $10^{-2.5} Z_\odot$ at $z=5$ to $10^{0.5} Z_\odot$ at $z=0$.}

When all the feedback mechanisms are considered, since the mean metallicity grows with time, the domains with fixed $Z$ continue to expand, occupying larger and larger volumes, with $\bar{Z}$-regions (symbols) filling about 30 percent of the volume until $z=2.5$, then slightly shrink until $z=1.5$, and strongly expanding afterwards up to 50 percent at $z=0$. The surface density of metal-poor regions ($Z/Z_\odot \lesssim 0.018$) decreases with time, while for regions with intermediate metallicity ($0.01 < Z/Z_\odot \lesssim 1$) it attains a maximum and then decreases; at late time ($z=0$) both attain negative $\varv_2$ and form a network dominated by tunnels. Metal-rich regions with $Z/Z_\odot \gtrsim 1.8$ are instead characterised by $\varv_1$ increasing with time, and positive $\varv_2$ and $\varv_3$, namely, they form a network of sparse and small spherical metal-rich clumps.

For $z>1.5$, the morphology of the $Z$-field is not set by feedback (dotted lines in bottom-row panels), and is maximally influenced by stellar feedback, especially regions with $Z/Z_\odot \lesssim 0.1$. Their volume filling fraction of is reduced by 5 to 25 percent, and their surface density and mean curvature are substantially altered. This is expected since stellar winds are the main distributor of metals into the IGM at high redshifts.

For $z<1.5$, all the feedback mechanisms coherently establish the morphology of regions with $Z/Z_\odot > 0.01$, and especially those with $Z/Z_\odot \simeq0.1$, inducing modifications that increase with time, qualitatively opposed to stellar feedback whose effect is largest at high redshift. During this epoch, AGN jets play the major role: they increase the volume filling fraction by up to 60 percent at $z=0$, decrease (increase) the surface density of regions with $Z/Z_\odot<0.89$ ($Z/Z_\odot>0.89$) by 80 percent at $z=0$, changing their mean curvature and topology, enhancing that of metal-rich regions with $Z/Z_\odot > 3.2$. Since the mass ejected in jets is small, much of this effect is expected arise due to the interaction of the jets with surrounding gas and pushing it outwards, which is consistent with findings that {\sc Simba}'s jets strongly evacuate halos~\citep[e.g.][]{Appleby2020,Sorini2022}. AGN winds have the same qualitative effects as jets, by a factor of approximately two times smaller, and very similar to stellar feedback. X-ray heating only affects the very late-time morphology of metal-rich regions, in a narrow range of metallicities peaked at $Z/Z_\odot \simeq 3.2$ and at redshift $0<z<0.5$. This is because the direct effect of {\sc Simba}'s X-ray feedback is limited to the central regions of halos.

% ========================================
\section{Discussion}
\label{sec:discussion}

% ----------------------------------------
\subsection{Ranking of feedback mechanisms}
\label{sec:discussion:ranking}

Qualitative inspection of MFs amplitudes shown in Figures~\ref{fig:MFs_T_nonorm}-\ref{fig:MFs_metal_nonorm} (differences in the lower part of each panel) suggests the ranking of feedback effects reported in Table~\ref{tab:comparativemorpho}, which takes into account all the four MFs altogether (not only $\varv_0$, as Figure~\ref{fig:2dmaps_nosmooth_z0} would wrongly suggest) at $z\lesssim1.5$ and $z\gtrsim1.5$; note that it reflects global, integrated effect over all possible threshold values. The following discussion distinguishes the four feedback mechanisms in the same order shown in the aforementioned figures:
\begin{table}
\centering
\caption{Ranking of feedback mechanisms w.r.t. their impact on the global morphology of the gas fields at $z\lesssim1.5$ and $z\gtrsim1.5$ (first and second number).
% excursions sets of temperature, pressure, densities, and metallicity fields at $z\lesssim1.5$ and $z\gtrsim1.5$ (first and second number). For instance, the morphology of the $T$-field at $z\lesssim1.5$ is mostly affected by jets, then by X-ray heating, stellar feedback, and AGN winds in this order.
}
\begin{threeparttable}
\begin{tabular*}{\columnwidth}{@{\extracolsep{\fill}}lcccccc}
\hline
 & $T$ & $P$ & $n_\mathrm{gas}$ & $n_\mathrm{HI}$ & $n_{\mathrm{H}_2}$ & $Z$ \\
\hline
X-ray heating & 2-2 & 3-2 & 4-2 & 2-2 & 4-2 & 3-4 \\
Jets & 1-1 & 2-3 & 3-3 & 4-3 & 3-3 & 1-2 \\
AGN winds & 4-4 & 1-4 & 2-4 & 3-4 & 2-4 & 4-3 \\
Stellar feedback & 3-3 & 1-1 & 1-1 & 1-1 & 1-1 & 2-1 \\
\hline
\end{tabular*}
\end{threeparttable}
\tablefoot{For instance, the morphology of the $T$-field at $z\lesssim1.5$ is mostly affected by jets, then by X-ray heating, stellar feedback, and AGN winds in this order.}
\label{tab:comparativemorpho}
\end{table}

%\textbullet~
\textit{X-ray heating:} None of the explored gas fields is impacted by X-ray heating in a dominant way over other feedback mechanisms. It is the second most important process in shaping the global morphology of $T$ and $n_\mathrm{HI}$ fields at all studied redshift. The $n_\mathrm{gas}$ and $n_{\mathrm{H}_2}$ fields show quantitatively similar differences as for $n_\mathrm{HI}$-field, but given the relatively stronger impact of other feedback mechanisms, X-ray heating is ranked only as third and fourth. The morphology of $Z$-field is impacted the least by X-ray heating and this is limited to low redshift ($z\lesssim1$).

%\textbullet~
\textit{Jets:} The impact of AGN jets is the largest, and dominant over other feedback mechanisms, on the $T$-field, in particular at $z\lesssim1.5-2$. This is somewhat expected as jets, when active, deposit large amounts of energy at substantial distances owing to their high velocities and hence large kinetic energies~\citep{Christiansen2020}. Jets also impact the $Z$-field very significantly, again dominantly compared to other explored feedback mechanisms, by pushing hot metal-enriched outwards from massive halos~\citep{Borrow2020}. As for the $T$-field, the effect of jets on the $Z$-field is strongest at $z\lesssim1.5-2$. AGN jets impact the morphology of the $n_\mathrm{gas}$, $n_\mathrm{HI}$, and $n_{\mathrm{H}_2}$ fields to a lesser extent compared to other feedback mechanisms, while they are competitive with AGN winds in shaping the $P$-field. 

%\textbullet~
\textit{AGN winds:} According to the implementation of the radiative mode of AGN feedback in {\sc Simba} simulations, AGN winds have a negligible impact on the $T$-field. Their strongest effect is at $z\lesssim1.5$ on the $P$-field morphology (comparable in strength to the effect of jets, for all the MFs but the volume filling fraction), and subdominant in the other fields, especially at high redshift. The reason is the typical speed of AGN winds, which in 14~Gyr in vacuum, pushes particles up to about 1~Mpc, namely, below the resolution limit set by effective smoothing.

%\textbullet~
\textit{Stellar feedback:} Star formation-driven winds are found to play a crucial role in modifying the global morphology of the gas even at relatively large scales such as those probed here. While such winds have even lower velocities than the AGN winds, the cumulative amount of mass displaced is very large, substantially exceeding the mass in stars, and weighted towards high redshift since galaxies are small and thus mass loading factors are high. The stellar feedback is the dominant mechanism for all but the $T$ and $Z$ fields. Its impact on the $P$, $n_\mathrm{gas}$, $n_{\mathrm{HI}}$, and $n_{\mathrm{H}_2}$ fields is highest at high redshift, progressively decreases, but stays important down to $z=0$. {\sc Hi} field shows the strongest signature of the stellar feedback between redshift 1.5 and 0.5. As the star formation-driven winds are metal-loaded, it is not surprising to find that they impact the global morphology of the $Z$-field with the strength ranked right after jets although with reversed redshift dependence, namely, the signature of the stellar feedback is strongest at the highest redshift, while that of jets is almost absent at high redshift and becomes strongest at $z=0$. The $T$-field is impacted the least by the stellar feedback and is limited to a relatively narrow range of temperatures around $10^4$~K.

% ----------------------------------------
\subsection{Time evolution of MFs}
\label{sec:results:t-evolution}

\begin{figure*}
\centering\includegraphics[width=17cm]{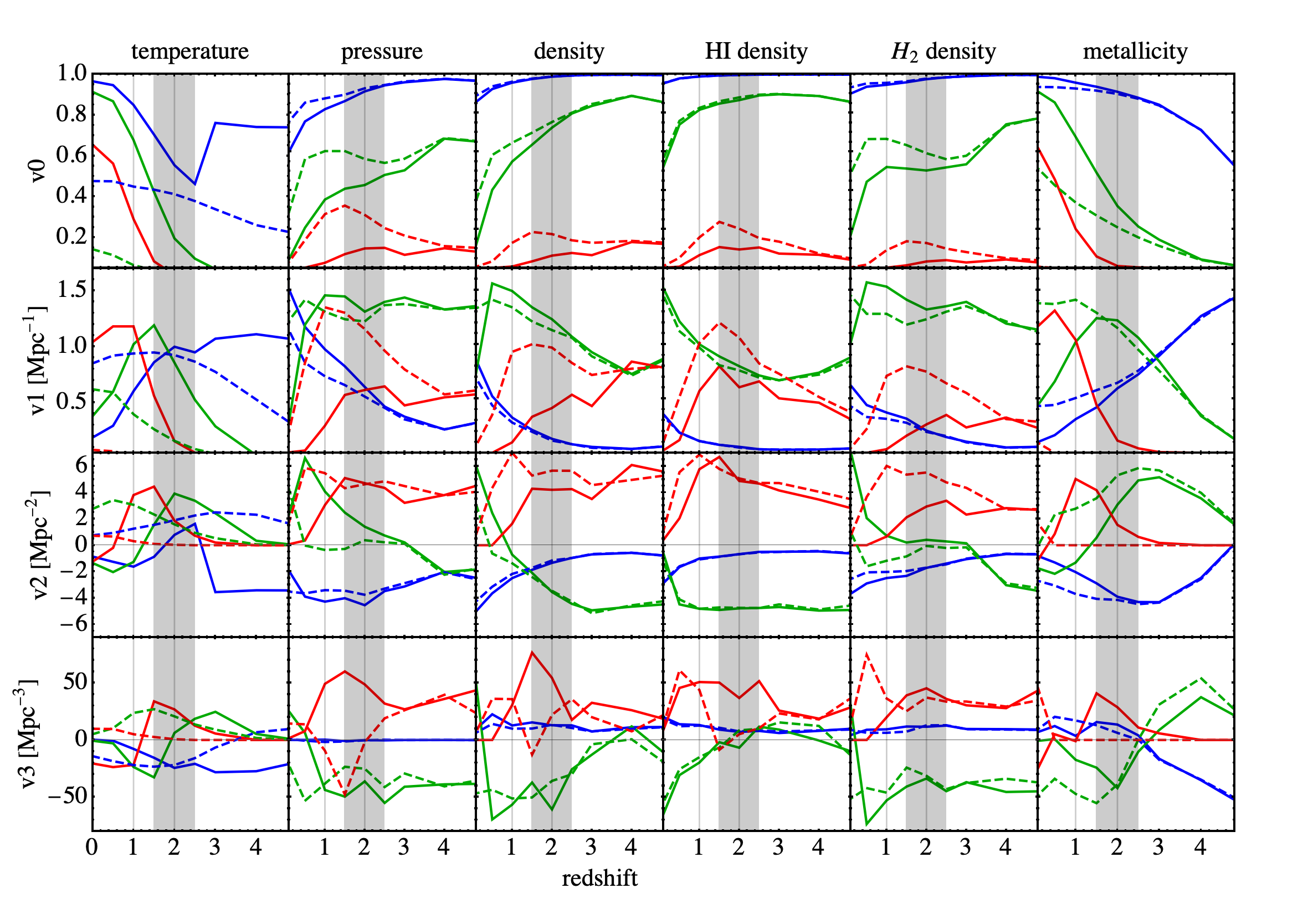}
\caption{Time evolution of MFs (top to bottom) for all the fields at three specific threshold values spanning the full range, for the Fiducial model (solid line) and for the NoAGN model (dashed line). Threshold values (blue, green, red in increasing order): $\log(T[\mathrm{K}])=\{4, 5, 6\}$; $\log(P/k_\mathrm{B}[\mathrm{cm}^{-3}\mathrm{K}])=\{3, 3.9, 4.8\}$; $\log(n[\mathrm{cm}^{-3}])=\{-3,-2,-1\}$; $\log(n_\mathrm{HI}[\mathrm{cm}^{-3}])=\{-4,-3,-2\}$; $\log(n_{\mathrm{H}_2}[\mathrm{cm}^{-3}])=\{-4,-2.5,-1\}$; $\log(Z[Z_\odot])=\{-2,-1,0\}$. The shaded band indicates the redshift range $1.5<z<2.5$ around the peak of star formation rate density, which approximately coincides with the maximum of AGN luminosity function for AGN with bolometric luminosity $10^{45}<L_\mathrm{bol}/(\mathrm{erg~s}^{-1})<10^{46}$; vertical lines mark the maximum of the AGN luminosity function for AGN with $10^{44}<L_\mathrm{bol}/(\mathrm{erg~s}^{-1})<10^{45}$ ($z\simeq1$) and  $10^{46}<L_\mathrm{bol}/(\mathrm{erg~s}^{-1})<10^{48}$ ($z\simeq2$). As discussed in \S\ref{sec:results:t-evolution}, these curves display many distinct features depending on the captured feedback processes, supporting the informative value of the cosmic evolution of morphology  (each feature likely encodes a specific signature of the underlying process).
}
\label{fig:MFtime}
\end{figure*}

Minkowski curves provide a static picture of AGN and stellar feedback on large scales. The time evolution of the MFs of the gas fields for some specific excursion-set thresholds offers a complementary synthesis of the results. It highlights the possible role of stellar or AGN feedback mechanisms on top of hydrodynamics in driving morphological transitions, namely, abrupt or smooth changes of the MFs following immediately or with some delay the peak of stellar and AGN activity.

This analysis is illustrated in Figure~\ref{fig:MFtime}, which focuses on three values of the six gas fields approximately spanning the full range covered by mean-field values $\bar{f}$ (indicated by symbols in Figures~\ref{fig:MFs_T_nonorm}-\ref{fig:MFs_metal_nonorm} and quoted below) for the Fiducial and NoAGN models. This figure compares the role of all AGN and stellar feedback mechanisms altogether (solid lines) against the role of stellar feedback only (dashed lines) for $z<4$, namely in the last 11.8~Gyr. For reference, in {\sc Simba} \citep{Dave2019} the peak of star formation rate density occurs in the redshift range $1.5<z<2.5$ (dark shaded band), in agreement with observations \citep[e.g.][]{MadauDickinson2014}; the maximum of AGN luminosity function occurs approximately in the same redshift range for AGN with bolometric luminosity $10^{45}<L_\mathrm{bol}/\mathrm{erg~s}^{-1}<10^{46}$, and around $z\simeq1$ ($z\simeq2$; vertical lines) for AGN with $10^{44}< L_\mathrm{bol}/\mathrm{erg~s}^{-1}<10^{45}$ ($10^{46}< L_\mathrm{bol}/\mathrm{erg~s}^{-1}<10^{48}$), as measured by \cite{Habouzit2022a} \citep[see also][]{Fontanot+2020}.
Overall, the following can be deduced:

\textbullet~ \textit{Temperature:} AGN activity strongly modifies the morphology of the $T$-field at every redshift \textit{i)} by boosting the occupied volume $\varv_0$ at increasing times for $T>10^4$~K and with a sudden contraction of colder regions between $z=3$ and $z=2$, at the same epoch when the number density of brightest quasars is largest
(afterwards, AGN feedbacks and in particular jets become more ubiquitous and enriched by the faint population of quasars, starting to heat the gas on progressively larger and larger scales);
\ \textit{ii)} by anticipating the shrinkage of the surface area $\varv_1$ at $z=1.5-2$, especially for cold ($T=10^4$~K) and warm ($T=10^5$~K) isothermal surfaces, which are therefore globally rounder; \ \textit{iii)} by simultaneously increasing the convexity of warm and hot ($T=10^6$~K) domains, which are globally less flat at respectively $z>2$ and $z>1$ (larger integrated mean curvature $\varv_2$) and forming a network dominated by tunnels afterwards (negative $\varv_3$); \ \textit{iv)} by changing the sign of the integrated mean curvature for the regions with $T>10^4$~K, which is positive at all times when only stellar feedback is active whilst always negative (apart from the epoch $1.8<z<2.7$) because of AGN feedback, with the complementary domain, namely regions with $T\leq10^4$~K, forming a network of filaments (filled tunnels) at all probed times.

\textbullet~ \textit{Pressure:} The morphology of the $P$-field evolves over time in a non-trivial manner depending on the pressure value and feedback mechanisms. \textit{i)} The geometry of the isobaric domains with low pressure ($P = 10^3 k_\mathrm{B}\mathrm{cm}^{-3}\mathrm{K}$) is only marginally affected after $z\simeq2$ by X-ray heating, AGN jets, and winds (solid lines), occupying slightly smaller volume and becoming moderately more irregular (larger surface density) than in presence of stellar feedback only, and their topology is almost unaltered as indicated by the Euler characteristic that remains vanishing at all epochs. \textit{ii)} When AGN feedback is included, intermediate-pressure domains ($P = 10^{3.9} k_\mathrm{B}\mathrm{cm}^{-3}\mathrm{K}$) instead shrink over time to a much larger extent, already since $z\simeq3$, while keeping the same surface until $z\simeq1$ as in presence of stellar feedback only.
After $z\simeq0.5$, they start forming a network progressively dominated by isolated regions with positive integrated mean curvature and positive Euler characteristic, namely, with meatball topology at the present time. If AGN feedbacks were not active, intermediate-pressure domains would occupy 30 percent of the total volume, have a much more irregular shape, and form today a network still dominated by tunnels.
\textit{iii)} The geometry and topology of high-pressure domains ($P > 10^{4.8} k_\mathrm{B}\mathrm{cm}^{-3} \mathrm{K}$) are instead more strongly affected by AGNs, which firstly operate on small scales by cancelling out the expanding effect of stellar feedback; because of AGN, the high-pressure domains tend indeed to occupy a progressively smaller volume fraction with smaller surface density, namely, they are more clumpy, and form a network with meatball topology (positive $\varv_3$), especially when most quasars of intermediate brightness is in place, which coincides with the epoch around the peak of the star-formation activity.

\textbullet~ \textit{Total, {\sc Hi}, and {\sc H$_2$} densities:} For the lowest-to-intermediate density thresholds of all the three density fields, the global morphology set by hydrodynamics and stellar feedback is unchanged by AGN feedbacks; the reason is that these densities are typical for very large scales, where gravitational interactions dominate. AGN feedback modifies instead the morphology of high isodensity domains, where $n=0.1\ \mathrm{cm}^{-3}$, $n_\mathrm{HI}=0.01\ \mathrm{cm}^{-3}$, and $n_{H_2}=0.1\ \mathrm{cm}^{-3}$, with maximum efficiency at the epoch near the peak of star-formation activity and later, especially for the total and {\sc Hi} gas fields. Interestingly, while AGN feedback prompts the formation of a network with meatball topology at $1<z<2.5$, it does not modify the topology of the {\sc H$_2$} field in the same epoch, already established by hydrodynamics and stellar feedback while driving a spongy-like topology at $z<1$ as deduced by the smaller value of $\varv_3$ indication of a larger number of tunnels. The sudden increase of $\varv_3$ of $n$ and $n_{\mathrm{H}_2}$ (but surprisingly not $n_\mathrm{HI}$) fields at $z<0.5$ for intermediate and large values of density could be associated with a recent merging of filaments driven by AGN feedback; however, this claim demands a more careful study w.r.t smoothing, which here is kept constant through cosmic time.

\textbullet~ \textit{Metallicity:} Metal-poor domains ($Z\leq0.01Z_\odot$) maintain their morphology across time, with minor effect by AGN feedback, which makes them globally smoother (smaller $\varv_1$) and slightly flatter ($\varv_2$ closer to zero) especially for $z<2$ compared to the $Z$-field in NoAGN model, with a minor change to the topology at $z\lesssim1.5$. Conversely, the shape and topology of regions with higher metallicity, $Z=0.1Z_\odot$, and $Z=Z_\odot$, is strongly affected by AGN, especially for redshift respectively $z\lesssim2.5$ and $z<1.8$, mirroring the time evolution of the MFs of the {\sc H$_2$} field with some delay. These domains progressively occupy larger volume fractions, have more irregular shapes as indicated by the larger $\varv_1$, and form a network dominated by filaments, respectively, at $z\lesssim2.5$ and $z\lesssim0.5$.

% ---------------------------------------------
\subsection{Phase diagrams and morphology: Joint analysis}
\label{sec:discussion:phasediagrams}

{\sc Simba} simulations allow for a field-based classification of baryonic environments based on both thermodynamical and chemical properties while splitting the role of AGN and stellar feedbacks. Bivariate distributions, hereafter also dubbed phase diagrams, can be used to infer average scaling relations between temperature, density, pressure, and metallicity of gaseous components, including the scatter accounting for degeneracies with hidden variables or some stochasticity. Estimating a quantitative relationship between the bivariate distributions (one-point spatial statistics) and the morphology of the excursion sets encapsulated in MFs (higher-order spatial statistics) is a topic of interest: do the phase diagrams capture the same information as the MFs? Viz., are regions featuring a thermodynamical and chemical state described by some scaling relation characterised by a well-defined typical morphology?

The smoothing adopted here for the computation of MFs only allows for a tentative joint analysis based on the more representative phase diagrams, shown in Appendix~\ref{app:phasediagrams} at redshift $z=0,1,2,3,5$ and only for the Fiducial and NoAGN models (Figures~\ref{fig:PhaseDiagramsFiducial}-\ref{fig:PhaseDiagramsNoAGN}), intended to stress the effect of AGN feedback mechanisms on top of stellar feedback and hydrodynamics. By pin-pointing topologically motivated threshold values of the fields within the phase diagrams, one can attempt to find a relation between the thermo-chemical state of the gas and the percolation level and topology of its geometrical structure. The qualitative summary of results is the following (for more details, see aforementioned appendix):

\textbullet~ When AGN feedbacks are active, at $z\lesssim1$ warm regions with $10^5<T/\mathrm{K}<10^7$ resulting after smoothing from undistinguished warm-hot intergalactic medium (WHIM) and warm circumgalactic medium \citep[WCGM;][]{CGM2017} are almost dominated by tunnels, namely, they leave a network of cold filaments with $T\gtrsim10^6$K. The $T$-field is further populated by colder cavities, roughly corresponding to collapsed haloes, which dominate the total volume at $1\lesssim z\lesssim 2$ with temperature decreasing with time down to $10^3$K at $z=0$. When only stellar feedback is active, the volume is equally partitioned in warm and cold regions, the former dominated by bounded haloes and the latter by cavities, leaving a network of cold filaments characterised by a temperature two orders-of magnitude smaller than in presence of AGN feedbacks.

\textbullet~ The smoothed $P$-field at $z=0$ is characterised by a single phase out of the actual three when all the feedback mechanisms are active, with an effective equation of state $P\propto n^{1.2}$ fitting the polytropic laws for isothermal and ultra-relativistic gases. Half of the volume is occupied by domains with pressure $1\lesssim P/k_\mathrm{B}\mathrm{cm}^{-3}\mathrm{K} \lesssim 10^3$ and very noisy topology with vanishing Euler characteristic on average, namely, percolating the full volume. With AGN feedbacks turned off, while the one-point distribution of $P$ values is essentially unchanged qualitatively, the topology of the smoothed $P$-field is reacher, admitting a network dominated by low-pressure cavities with $P\lesssim 10k_\mathrm{B}\mathrm{cm}^{-3}\mathrm{K}$. At higher redshift, especially for $1<z<2$, the smoothed $P$-field seems less dominated by cavities when AGN feedbacks are operational. A more suited smoothing procedure is mandatory to deduce a solid conclusion about the morphology of the original $P$-field.

\textbullet~ The total, HI, and H2 smoothed density fields are the most affected by the smoothing process, which destroys the bimodal or trimodal distributions leaving only the high-density regions. This lifting is less severe for the $n_\mathrm{HI}$-field, at all redshifts and for the $n_{\mathrm{H}_2}$-field at $z>2$, which maintains its almost linear bis relation with the total gas $n$-field. The sparser regions surviving the smoothing occupy the largest fraction of the volume and are characterised by cavities. However, smoothing prevents the impact of feedback from morphology from being distinguished, which is otherwise quite clear instead in the original bivariate distributions.

\textbullet~ The joint analysis of the $Z$-field is perhaps the most intriguing. At all redshifts the one-point distribution of the smoothed and original field values are not very different and it does not distinguish between the Fiducial and NoAGN model; this is valid also for the $Z-n$ bivariate distributions. Instead, the MFs of excursion-sets suitably chosen to span the range of metallicities occupy very different regions of the smoothed $Z-n$ phase diagram. This is especially true at $z=0$, where the joint analysis suggests that the denser medium is an intricate network dominated by metal-rich isolated regions with $Z\gtrsim Z_\odot$ and filaments with only slightly smaller metallicity when also AGN feedback is present, or with filaments with $Z\simeq 0.1Z_\odot$ when only stellar feedback is active.

Altogether, MFs extend over the $T-n$, $P-n$, $n_\mathrm{HI}-n$, $n_{\mathrm{H}_2}-n$, and $Z-n$ phase diagrams tracing the mean values and scaling relations with small dispersion. The MFs analysis on smaller scales would allow us to capture the sparse domains removed by current smoothing (especially regions with small density and pressure, and very high temperature and metallicity) and relate the dispersion of phase diagrams to the fluctuations in MFs.

% ========================================
\section{Conclusions} 
\label{sec:conclusions}

This study serves as an initial exploratory examination into the influence of AGN and stellar feedback mechanisms on the global morphology of the cosmic web.
The primary objective is to comprehend how these small-scale processes shape the overall geometry and topology of thermo-chemical gaseous fields at circumgalactic and intergalactic scales across cosmic time. Investigating the morphology of gas fields, which constitute multiple facets of the cosmic web, is of significant scientific relevance. It allows us to discern the extent to which subgrid physics influences the geometry of the larger scales, irrespective of achievements at smaller scales,  with the ultimate goal of constraining differently feedback physics.\footnote{The subgrid physics is indeed, in part, tailored to fit the internal and clustering properties of galaxies, aiming at preventing any over-cooling catastrophe.} To achieve this goal, we analysed the hydrodynamical simulation suite {\sc Simba}, capturing the effect of feedback mechanisms including X-ray heating, AGN jets and winds, stellar, and hydrodynamical feedbacks on six fields, namely: the gas temperature, pressure, total density, {\sc Hi} density, {\sc H$_2$} density, and gas metallicity fields. A visual inspection of the diversity of 2D sections, as shown in Figure~\ref{fig:2dmaps_nosmooth_z0}, offers an insight into the topic.

To quantify the morphology of the $T$, $P$, $n$, $n_{\mathrm{HI}}$, $n_{\mathrm{H}_2}$, and $Z$ fields, we employed the MFs that are able to capture a more complex shape-based picture than that described by low-order statistics. Despite the absence of analytical models for the expectation value of MFs of highly non-Gaussian fields, such as the {\sc Simba} output, the numerical estimation of MFs still offer a powerful means to discriminate between the roles of AGN and stellar feedback.

Specifically, we performed an analysis of the MFs of the excursion sets of the `smoothed' fields at different redshifts, spanning the range of $0<z<5$ and progressively accounting for all the {\sc Simba} feedback mechanisms.\footnote{Note: smoothing is a necessary step for the computation of MFs, which limits the original spatial resolution and biases gas fields. The smoothed distribution and the corresponding excursion sets of the total density are impacted, especially at $z<3$, as traced by the lost bimodality of the one-dimensional probability distribution function. See appendices~\ref{app:bias} and \ref{app:phasediagrams}.} The results were given in Figures~\ref{fig:MFs_T_nonorm}-\ref{fig:MFs_metal_nonorm}. These Minkowski curves constitute the starting point for all subsequent analyses. The uncertainty of the estimated Minkowski curves (not shown) deserves a comment. An intrinsic source of error is the sample variance, which is expected to not dominate within observed domains as long as the structures traced by the excursion set do not span a significant fraction of the box size. Moreover, the specific estimator of the MFs and the algorithm employed for their computation, including the discretisation and smoothing schemes, introduce an additional error. However, since we are interested in a global description of the morphology (and not in the minute, local details), we have neglected this second contribution and considered the Minkowski curves to be sufficiently precise and accurate to attain our objectives.

From the Minkowski curves we deduced a ranking of feedback mechanisms with respect to their impact on the global morphology of the excursion sets. Taking into account all redshifts, the following qualitative picture emerges (see Table~\ref{tab:comparativemorpho}): Stellar feedback shapes all but the $T$ and $Z$ fields, whose morphologies are instead mostly determined by AGN jets that are otherwise marginal for the other fields. Furthermore, AGN winds comprise the second feedback mechanism regulating the morphology of the $P$, $n$, and $n_{\mathrm{H}_2}$ fields, while X-ray heating is for $T$ and $n_\mathrm{HI}$ fields, and stellar feedback is for the $Z$-field. The ranking could change if based on MFs for single values (thresholds) of the thermodynamical and chemical fields typical of specific systems, at some fixed redshift or in particular redshift intervals. However, such analysis goes beyond the scope of this study.

The analysis of the time-evolution of the MFs of excursion sets focusing on specific values of the thermodynamical and chemical fields allowed us to evaluate the strength of the feedback mechanisms as a function of cosmic time and to monitor whether they drive or alleviate morphological transitions (indicated by smooth trends or abrupt change of the MFs values; see the key  Figure~\ref{fig:MFtime}). The distinct features seen in that figure highlight the high information content encoded in the geometry and topology of the various physical fields beyond the total density commonly considered in past studies. Each feature likely encodes a specific process (as discussed in this work) that could eventually be used to infer the corresponding physical mechanism. This analysis highlights the potential of MFs as a probe for the cosmic evolution of the chemo-thermal cosmic web. For instance, the (smoothed) $T$-field at $10^5$K ($10^6$K) undergoes a topological transition around $z=1.2$ ($z=2$) only if AGN feedback is included; the $P$-field at $10^{4.8} k_\mathrm{B}\mathrm{cm}^{-3}$K and the $n$-field at $0.1\mathrm{cm}^{-3}$ are instead protected from a temporary topological transition occurring around $z=1.5$ and stopping at $z\simeq1$ if AGN feedback is active; the topology of the {\sc Hi}, {\sc H$_2$}, and metallicity fields for $n_\mathrm{HI}=10^{-2}\mathrm{cm}^{-3}$, $n_{\mathrm{H}_2}>10^{-2.5}\mathrm{cm}^{-3}$, and $Z>0.1Z_\odot$ are also strongly altered by AGN feedback around $z=1.5-2$. However, the topological and morphological transition of these excursions sets continue until $z=0$.

The analysis of the bivariate distributions or phase diagrams $T-n$, $P-n$, $n_\mathrm{HI}-n$, $n_{\mathrm{H}_2}-n$, and $Z-n$ jointly with MFs, shown in Figures~\ref{fig:PhaseDiagramsFiducial}-\ref{fig:PhaseDiagramsNoAGN}, illustrates the capability of the latter to capture striking features of the diagrams (for the smoothed fields), for instance using maxima, minima, and vanishing points of the Minkowski curves that account for specific morphologies. Alternatively, and more efficiently, one can choose MFs that correspond to field values (thresholds) spanning the probability distribution functions of the smoothed field. This, in turn, suggests that the geometry of the fields indeed reflects the thermo-chemical phases of the IGM and, therefore, the micro physics of star formation and feedback as well. Indeed, the location of field values in phase diagrams characterised by average scaling relations (e.g., dense, high-pressure, isothermal regions) corresponding to thresholds associated with specific values of MFs (e.g., values of $\varv_0$ and $\varv_3$) suggests relationships between the typical morphology of spatial regions associated with that scaling (in the example, dense, high-pressure, isothermal regions cover a tiny fraction of the volume and are isolated, as indicated by the small value of $\varv_0$ and the large, positive value of $\varv_3$). Besides, the covariance matrix of a few values extracted from the MFs needed for data modelling is, in principle, much simpler to estimate than that of the full distribution functions.

The current analysis is confined to {\sc Simba} suite of models, so it would be interesting to compare to other simulations such as {\sc IllustrisTNG}~\citep{pillepichetal18} or {\sc Flamingo}~\citep{Flamingo2023}. For instance, it is known that {\sc Simba}'s jet feedback causes wider baryon dispersal~\citep{Yang2022,Gebhardt2023} and more IGM heating~\citep{Christiansen2020} than other comparable simulations. Since jet feedback appears to drive the strongest features in the MFs, we would expect that comparisons to other large-volume cosmological simulations with different physical models would highlight areas of greatest discrimination between feedback prescriptions.

It would also be of interest to investigate the geometry of the baryonic cosmic web in simulations probing smaller scales such as {\sc New-Horizon} \citep{NewHorizon21}, {\sc Tempest} \citep{2019ApJ...882..156H}, {\sc Hestia} \citep{Hestia2020,2022MNRAS.512.3717D}, {\sc Romulus} \citep{Romulus2017,RomulusC2019,2023MNRAS.525.5677S}, or Project {\sc \small Gible} \citep{2023arXiv230711143R}, and address the issue of time delays between small-scale trigger and large-scale response in more details. One should parse the range of possible subgrid physics while considering different sets of simulations, ideally spanning the full range of realistic values. Jointly with small-scale sub-galactic analysis, this could provide us with more leverage to disentangle the various processes captured by present-day feedback models. 
Eventually, variations of the key Figure~\ref{fig:MFtime} should be used to fit observed data: the next morphological study will assess the MFs of observable fields instead of elementary thermodynamical or chemical fields, quantifying the corresponding biases. For instance, we can consider the {\sc Hi} optical depth $\tau\propto n_\mathrm{HI}$ probed by Lyman-$\alpha$ forest in high-resolution quasar spectra \citep[e.g.][]{McQuinn2016,Japelj2019A&A}, the X-ray luminosity $L_X\propto T^3$ probed observing galaxy clusters \citep[e.g.][]{Zou+2016,Bahar+2022,Poon+2023}, the Compton parameter $Y\propto \int\! d\ell\ P$ proportional to the line-of-sight integral of the pressure for ideal electron gas, probed by SZ-surveys \citep{Bleem2015,2022ApJ...926..179K}. We could also consider sub-mm, infrared, or optical luminosities that probe the molecular gas as done within the MUSE Analysis of Gas around Galaxies \citep[MAGG; ][]{MAGG}, which map the environment of high-redshift absorption line systems, or by probing specific metal contents such as [O/H], [Mg/H] or [Fe/H] \citep{KewleyNichollsSutherland2019}.

The present shape-based multi-field study could also be complemented by the analysis of the auto and cross-spectra or by the environment-dependent wavelet power spectrum \citep{env-WPS} of all chemo-thermal fields, to highlight which scale is most impacted by feedback and when; by studying multi-persistent homology \citep{MPP} and the sets of critical events available in the initial phases \citep{cadiou2020}, to assess the genuinely feedback-driven and gravity-driven effects and disentangle the impact of the geometry of the environment from internal processes. We also used complementary tools based on persistent homology \citep{Sousbie2011} and eventually focusing on each set of critical points by stacking the information in the frame set by eigenvectors of the curvature of the fields \citep{Kraljicetal2017}. This offers a solution that would jointly provide information on the local shape of the various gas fields.

% ========================================
\section*{Acknowledgements}
The authors thank D. Aubert, T. Buchert, G. De Lucia, C. Laigle, P. Monaco, and E. Nezri for their valuable comments and insights. This work has made use of the Horizon cluster on which the simulation was post-processed, hosted by the Institut d'Astrophysique de Paris; we thank S.~Rouberol for running it smoothly.
CS is grateful to INAF/Osservatorio Astronomico di Trieste for their hospitality. KK and CP thank the KITP for hosting the workshop
\href{https://www.cosmicweb23.org}{`CosmicWeb23: connecting Galaxies to Cosmology at High and Low Redshift'}.
This work is partially supported by the grant
\href{https://www.secular-evolution.org}{Segal} ANR-19-CE31-0017 of the French Agence Nationale de la Recherche and by the National Science Foundation under Grant No. NSF PHY-1748958.
%For the purpose of open access, the author has applied a Creative Commons Attribution (CC BY) license to any Author Accepted Manuscript version arising from this submission.

% ========================================

\bibliographystyle{aa}
\bibliography{MFs_SIMBA.bib}

% ========================================

\begin{appendix} 

% ----------------------------------------
\section{MFs of Gaussian, weakly Gaussian, and log-normal random fields}\label{app:analyticMFs}

A relatively small number of random fields admit analytical expressions for the expectation values of the MFs. The standard references are Gaussian random fields (GRF) $f$; in three dimensions, the MFs per unit volume of their excursion-set $C_\nu=\{\bm{x}\in D|f(\bm{x})\geqslant\nu\sigma_0\}$ have analytical expectation value given by \citep{Tomita1986}
\begin{subequations}\label{MFs-GRF}
\begin{align}
    \varv^\mathrm{GRF}_0(\nu) &= \frac{1}{2}\mathrm{erfc}(\nu/\sqrt{2}),\\
    \varv^\mathrm{GRF}_1(\nu) &= \frac{2}{3}\frac{\lambda}{2\pi}\mathrm{e}^{-\nu^2/2},\\
    \varv^\mathrm{GRF}_2(\nu) &= \frac{2}{3}\frac{\lambda^2}{(2\pi)^{3/2}}\nu\mathrm{e}^{-\nu^2/2},\\
    \varv^\mathrm{GRF}_3(\nu) &= \frac{\lambda^3}{(2\pi)^2}(\nu^2-1)\mathrm{e}^{-\nu^2/2}.
\end{align}
\end{subequations}
Here $\mathrm{erfc}(x)$ is the complementary error function and the amplitude parameter $\lambda^2 = \sigma_1^2/3\sigma_0^2$ is the ratio between the variance of the gradient of the field, $\sigma_1^2=\langle (\nabla f)^2\rangle$, and variance of the field itself, $\sigma_0^2=\langle f^2\rangle_\mathrm{c}
$, which scales like the square of the inverse zero-crossing distance \citep{Pogosyan2009}.

The mean MFs densities of weakly non-Gaussian random fields incorporate exponentially damped polynomial corrections into the Gaussian expressions, with coefficient proportional to linear combinations of generalised skewness parameters
\begin{align}\label{eq:MFs-wNGRF-skewness}
\hskip -0.2cm
S_0 &\!=\! \frac{\langle f^3 \rangle_c}{\sigma_0^4},\,
    S_1 \!=\! -\frac{3}{4}\frac{\langle f^2 \nabla^2f \rangle_c}{\sigma_0^2\sigma_1^2},\,
    S_2 \!=\! -\frac{9}{4}\frac{\langle (\nabla f\!\cdot\!\nabla f) \nabla^2f \rangle_c}{\sigma_1^4}.
\end{align}
Up to second-order in perturbation theory, they read \citep{Matsubara2003,GayPichonPogosyan2012,Matsubara+2022}
\begin{subequations}\label{eq:MFs-NGRF}
\begin{align}
   \hskip -0.2cm \varv_0(\nu) &= \varv_0^\mathrm{GRF}(\nu) + \frac{\sigma_0}{(2\pi)^{1/2}} \frac{S_0}{2} H_2 \mathrm{e}^{-\nu^2/2},\\
    \varv_1(\nu) &= \varv_1^\mathrm{GRF}(\nu)\! +\! \frac{\sigma_0\lambda}{9\pi}\left(\frac{1}{2}S_0H_3 + S_1H_1\right) \mathrm{e}^{-\nu^2/2},\\
    \varv_2(\nu) &\!=\! \varv_2^\mathrm{GRF}(\nu) \!+\! \frac{2\sigma_0\lambda^2}{3(2\pi)^{3/2}}\!\left(S_0H_4 \!+\! \frac{2}{3} S_1H_5 \!+\! \frac{1}{3}S_2\right) \!\mathrm{e}^{-\nu^2/2},\\
    \varv_3(\nu) &\!=\! \varv_3^\mathrm{GRF}(\nu)\! +\! \frac{\sigma_0\lambda^3}{(2\pi)^2}\!\left(\frac{1}{6}S_0H_5\! +\! S_1H_3\! +\! S_2H_1\right) \!\mathrm{e}^{-\nu^2/2},
\end{align}
\end{subequations}
where $\varv_\mu^\mathrm{GRF}(\theta)$ are given by Equations~\eqref{MFs-GRF} and $\{H_0,\dots,H_5\} = \{1,\theta,\theta^2-1,\theta(\theta^2-3),\theta^4-6\theta^2+3,\theta  (\theta^4-10\theta^2+15) \}$ are Hermite polynomials.
When the field $f$ is convolved with a Gaussian kernel, as routinely done in numerical computation of fields sampled on a regular lattice, the skewness parameters become the function of the smoothing length $R$ and of the linear power spectral density of the field $f$; see Equations~(13-16) in \citet{Nakagami+2004}.

\begin{figure}
\centering\includegraphics[width=8cm]{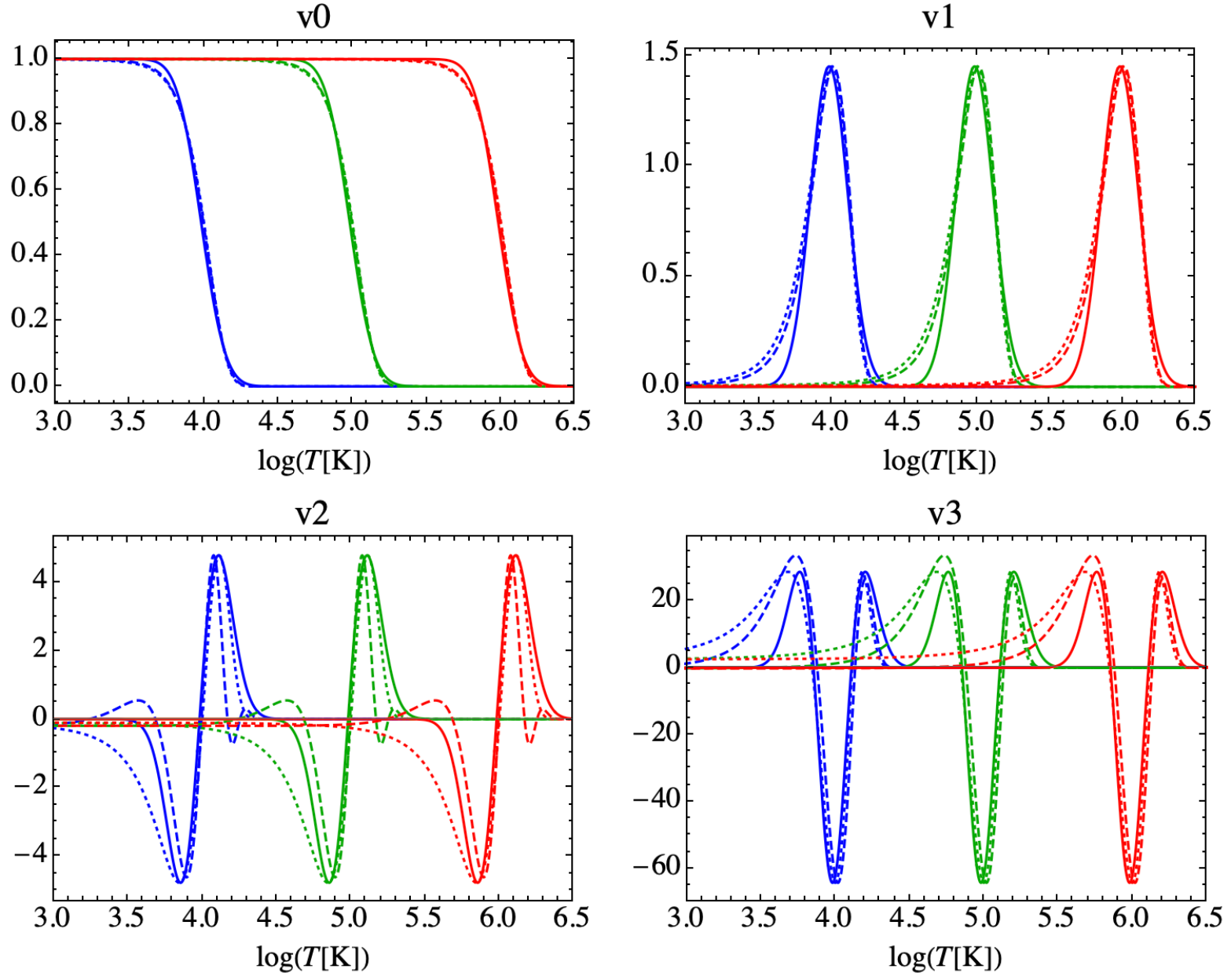}
\caption{Examples of Minkowski curves of Gaussian, weakly non-Gaussian, and log-normal temperature random field (dotted, dashed, and solid lines, respectively) for three values of the mean temperature ($T=10^4$~K, blue lines; $T=10^5$~K, green; $T=10^6$~K, red). The amplitude of the MFs of Gaussian and weakly non-Gaussian random fields are normalised to that of the LN random field; their apparent skewness is due to the logarithmic scale. For illustrative purpose, the parameters are fixed to $\sigma_0=0.3$, $(S_0,S_1,S_2)=(0.2,1,0.1)$, $\lambda=25$.
}
\label{fig:MFs-analytics}
\end{figure}

For a log-normal (LN) field $f=F(g)\equiv\exp(g-\sigma_g^2)-1$ 
related to a GRF $g$ with vanishing mean and variance $\sigma_g^2=\ln(1+\sigma^2)$, the excursion-sets $\{\bm{x}\in D|g(\bm{x})\geqslant\nu\sigma_g\}$ and $\{\bm{x}\in D|f(\bm{x})\geqslant F(\nu\sigma_g)\}$ are equivalent, namely, they share the same morphology. Accordingly, the average MFs per unit volume can be deduced from Equations~\eqref{MFs-GRF} by substituting the threshold and amplitude parameters with the following expressions \citep{MatsubaraYokoyama1996,Hikage+2003}:
\begin{subequations}\label{MFs-lognormal}
\begin{align}
    \nu_\mathrm{LN}(\nu) &= \frac{\ln\left[(1+\nu\sigma)\sqrt{1+\sigma^2}\right]}{\sqrt{\ln(1+\sigma^2)}}, \\
    \lambda_\mathrm{LN}(\lambda) &= \lambda \left[\frac{\sigma}{1+\sigma^2\ln(1+\sigma^2)}\right]^{1/2}.
\end{align}
\end{subequations}

Figure~\ref{fig:MFs-analytics} shows some examples of a temperature random field with parameters fixed to fictitious values for illustrative purposes (see caption). For other non-Gaussian random fields monotonic functions of a GRF, such as chi-square, Student, Fisher, Rayleigh, or Maxwellian \citep[see e.g.][]{Adler1981}.

% ----------------------------------------
\section{2D sections of smoothed gas fields}
\label{app:MF2dsections}

\begin{figure*}
\centering
\includegraphics[width=\textwidth]{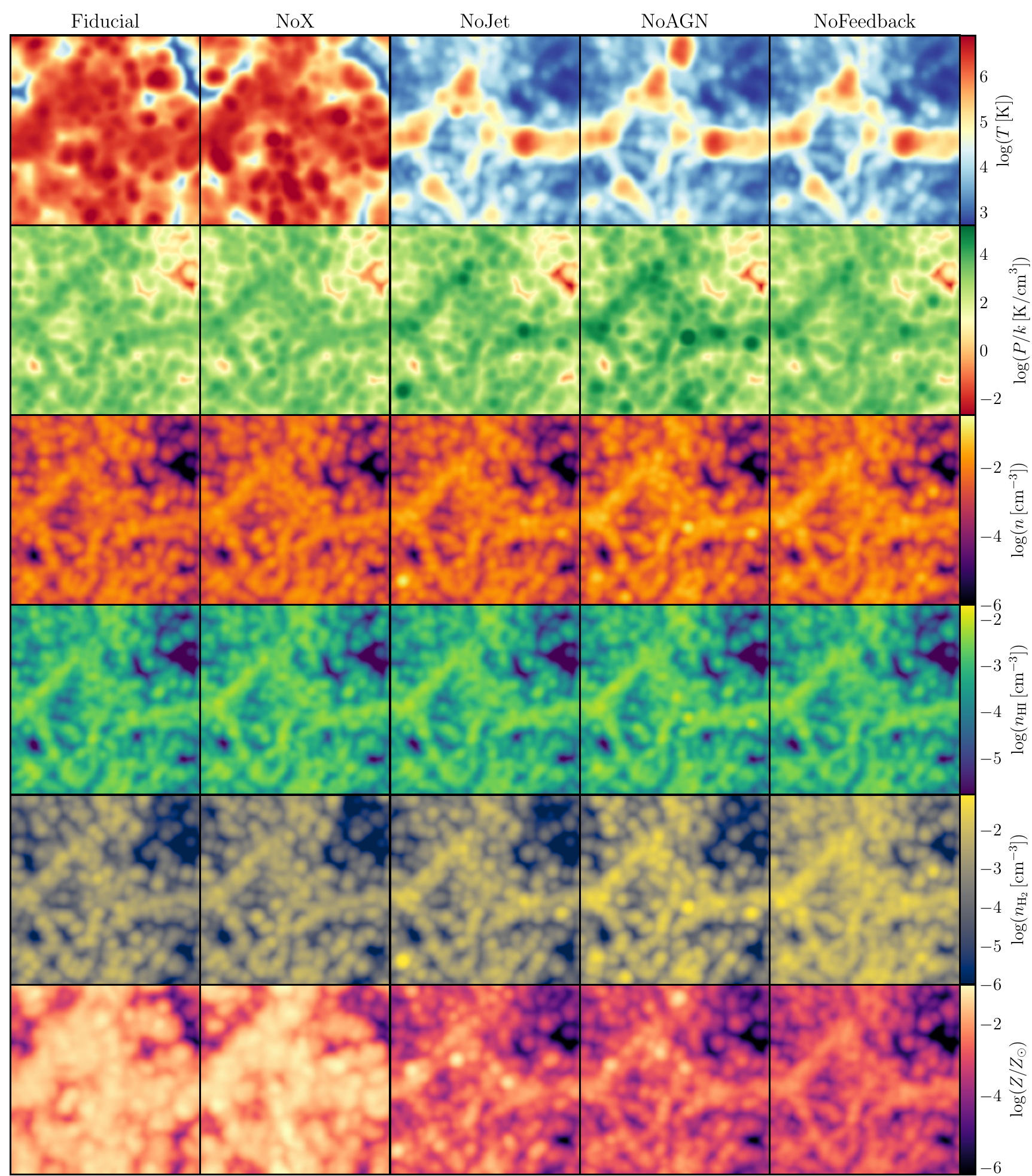}
\caption{Visualisation of the full box-size sections (8 Mpc/$h$ thick and 50 Mpc/$h$ wide in both directions) of the 3D gas temperature, pressure, density (total, {\sc Hi}, {\sc H$_2$}) and metallicity fields (from top to bottom) {at the smoothing scale used for the computation of MFs} and at $z=0$, for different {\sc Simba} models progressively excluding feedback mechanisms (from left to right; see Table~\ref{tab:models}). Similarly to Figure~\ref{fig:2dmaps_nosmooth_z0}, it illustrates how individual feedback processes operating on galactic scales leave different imprints on the gas fields at large scales.
}
\label{fig:2dmaps_smooth_z0}
\end{figure*}

\begin{figure*}
\centering
\includegraphics[width=\textwidth]{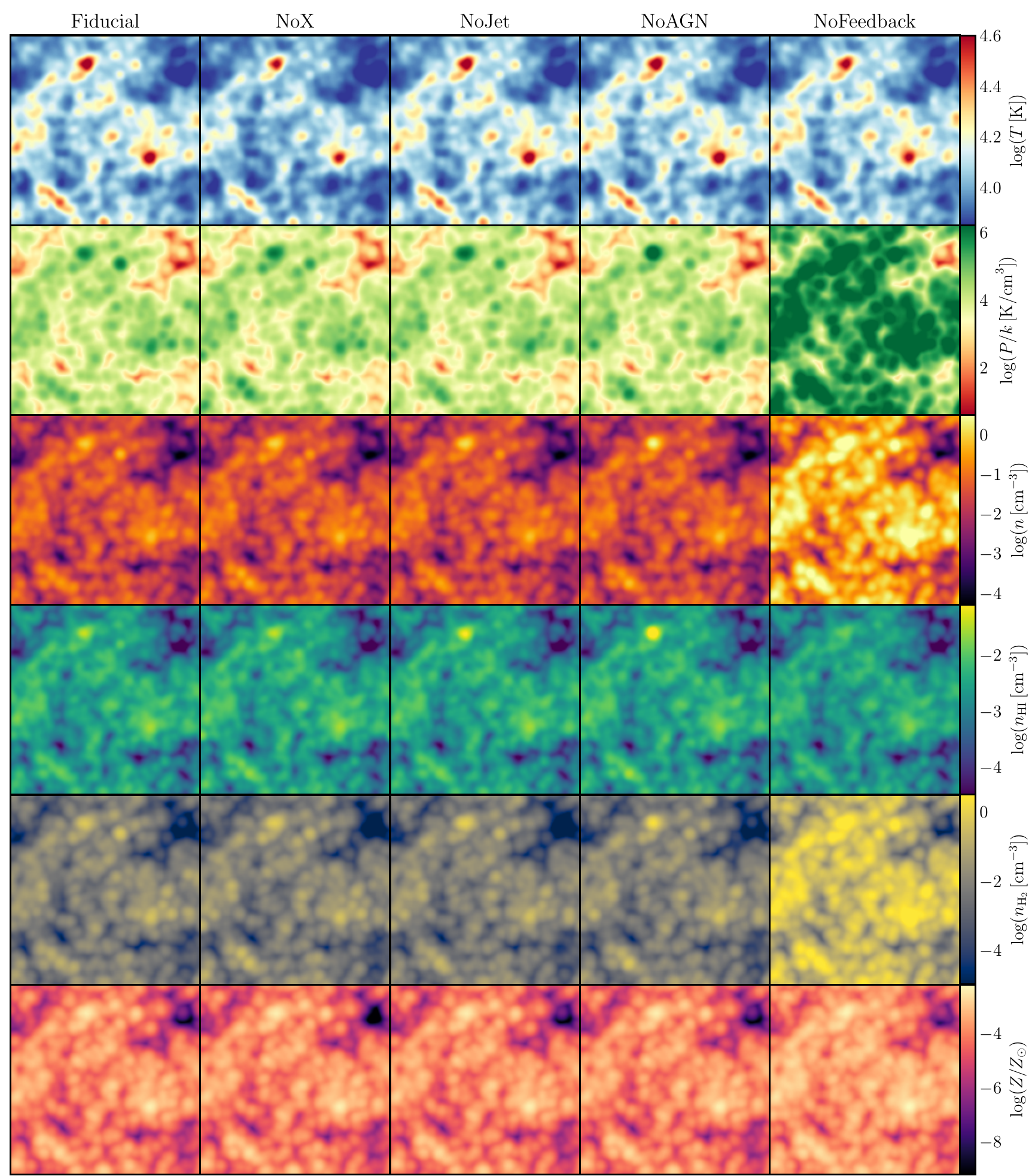}
\caption{Same as Figure~\ref{fig:2dmaps_smooth_z0}, but at $z=5$.}
\label{fig:2dmaps_smooth_z5}
\end{figure*}

Figures~\ref{fig:2dmaps_smooth_z0}-\ref{fig:2dmaps_smooth_z5} illustrate the large-scale imprint of the individual feedback processes operating on galactic scales on 2D sections (thickness: $8h^{-1}$Mpc) of the gas fields at $z=0$ and $z=5$, when the smoothing by grid sampling and convolution by Gaussian kernel (needed for MFs computations) are taken into account.

% ----------------------------------------
\section{Bias of smoothed fields}
\label{app:bias}

\begin{figure*}
\centering\includegraphics[width=17cm]{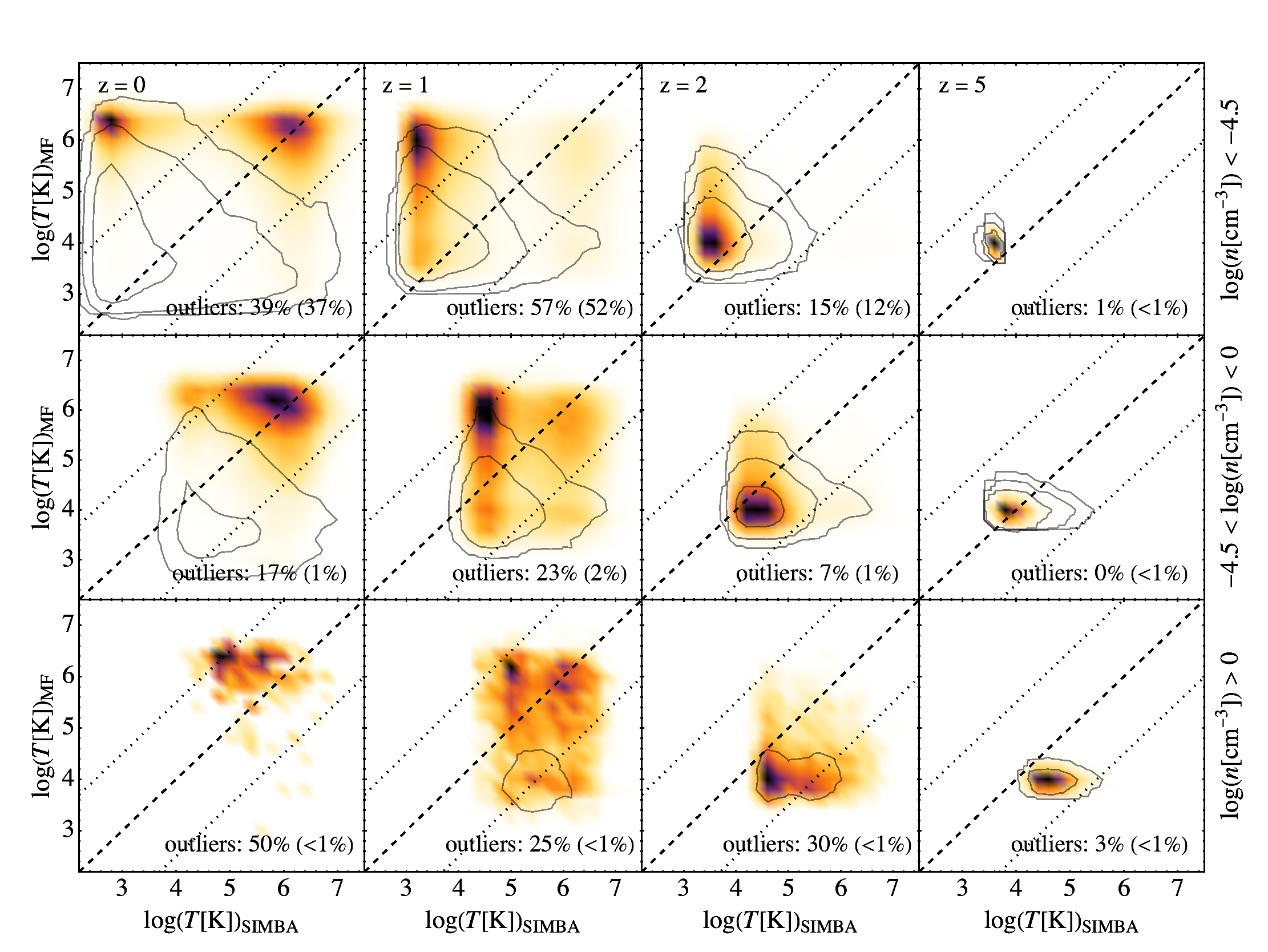}
\caption{Scattering plot mapping the point-wise (cell) values of original {\sc Simba} and smoothed temperature fields for the Fiducial and NoAGN models (colours and contours), as a function of redshift (left to right) and for three selections of the underlying original HI density field (top to bottom for low, mid, and high density; see legend). The fraction of outliers with $|\log T_\mathrm{MF}-\log T_\textsc{Simba}|>1.5$ is marked by dotted lines and indicated in each panel (in parenthesis, the fraction with respect to the full sample). Contours account for $100$, $1\,000$, and $10\,000$ counts, not shown if sparse.}
\label{fig:Tsimba-TMFs-nHI}
\end{figure*}

As stressed in the Guidelines (\S\ref{sec:results:guidelines}), the effective smoothing of $1.56 \, h^{-1}$Mpc resulting from the computation of the MFs biases the probed regions of the one-point of the fields. phase-space diagrams. This is illustrated for the $T$-field only in Figure~\ref{fig:Tsimba-TMFs-nHI}, which compares its values at $z=0,1,2,5$ for the Fiducial and NoAGN models before and after smoothing, shown respectively in colours and contours (the fraction of outliers with ${|\log T_\mathrm{MF}-\log T_\textsc{Simba}|>1.5}$ is 
quoted in each panel). At $z\leq1$ (first and second column panels) a Gaussian smoothing kernel with rms-variance $\sigma$ as large as 4 cells suppresses an important fraction of cold regions with temperature $T<10^4$~K, especially those with lowest total density, $n<10^{-4.5}\mathrm{cm}^{-3}$ (first row panels; contamination for about 40-50 percent of the total), which accordingly occupy sparse domains of size smaller than $\sim1.56h^{-1}$Mpc at this epoch. Warm regions with $T\gtrsim10^5$~K instead qualitatively maintain their original temperature also at $z=0$ where they occupy most of the volume, only marginally polluted by colder and hotter regions. At $z\leq2$ the smoothed and original $T$-field are quite similar. Therefore, although the MFs do not probe exactly the original $T$-field, they still yield important information about its global morphology for specific range of underlying density and especially at $z>1$. Smaller kernel width of size $\sigma=1$ cell would preserve the trace of cold domains, although diminishing their number by a factor of about 2.

Depending on the feedback mechanism, the original $P$-field in {\sc Simba} simulations is composed by three dominant populations at low redshift and two at high redshift. Smoothing mixes the intermediate and high pressure populations, suggesting that they occupy sparse domains with characteristic size smaller than $1.56 \, h^{-1}$Mpc. The difference is less important at $z>2$, where the smoothed $P$-field reasonably well represent the actual one.

The total, {\sc Hi}, and {\sc H$_2$} smoothed density fields are the most affected by smoothing, which destroy the bimodal or trimodal distributions leaving only the high-density regions. This lifting is less severe for the $n_\mathrm{HI}$-field, at all redshifts and for the $n_{\mathrm{H}_2}$-field at $z>2$ which maintains its almost linear bis relation with the total gas $n$-field. In particular, a similar suppression to the $T$-field occurs for the $n_\mathrm{HI}$-field (not shown). The Gaussian smoothing kernel with $\sigma=4$ cells totally erases sparse cells with density ${n_\mathrm{HI}<10^{-7}\mathrm{cm}^{-3}}$, which represent about 95 percent of the total sample volume; domains with such a low value of density behave as noise with a characteristic size smaller than $\sim1.56h^{-1}$Mpc at $z=0$, embedded in higher-density domains. Simultaneously, this Gaussian kernel shrinks the high-density tail from the original ${10^{-7}<n_\mathrm{HI}/\mathrm{cm}^{-3}<1}$ range to ${10^{-7}<n_\mathrm{HI}/\mathrm{cm}^{-3}<10^{-2}}$, qualitatively preserving its shape (skewness). Interestingly, contrary to the $T$-field a higher spatial resolution obtained with a Gaussian kernel with $\sigma=1$ cell is not sufficient to trace the original $n_\mathrm{HI}$-field; it would totally deform the shape of the original probability distribution function, artificially populating the $10^{-10}<n_\mathrm{HI}/\mathrm{cm}^{-3}<10^{-5}$ density range.

The one-point distribution of the $Z$-field is almost unchanged by smoothing, making it a privileged and challenging field for morphological analyses.

The computationally conservative choice of a Gaussian smoothing kernel with rms-variance $\sigma=4$ cells will therefore allow us to probe the global morphology of the WCGM and, to a smaller extent, of the network of (cold and dense) gaseous haloes.
More sophisticated mass assignment schemes and smoothing techniques such as adaptive kernels or wavelet denoising \citep{Puetter+2005,StarckMurtagh2006} could be more appropriate to maintain the morphology of the original field, at the cost of a likely more ambiguous interpretation of the probed scales \citep[e.g.][]{Martinez+2005}.

% ----------------------------------------
\section{Phase diagrams of original and smoothed fields}
\label{app:phasediagrams}

\begin{figure*}
\centering\includegraphics[width=17cm]{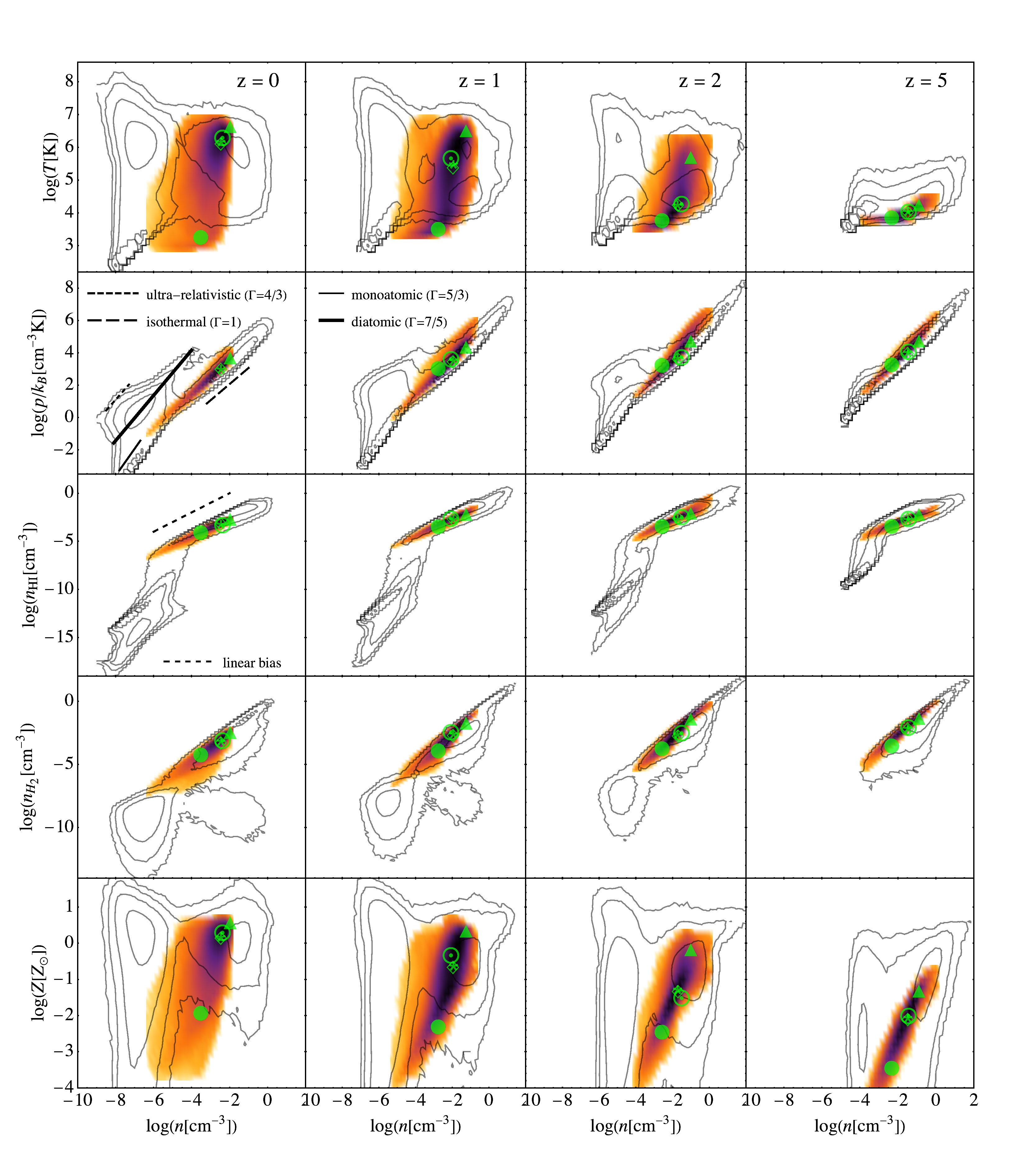}
\caption{Phase diagrams for the Fiducial model at redshift $z=0,1,2,5$ for the original and smoothed fields (contours and density plot, respectively; lines and colours account for log-counts over five decades between $10$ and $10^5$). As a reference, some polytropic laws are superposed to the $P-n$ diagram at $z=0$ (see legend) and the linear bias relation is indicated on the $n_\mathrm{HI}-n$ bivariate distribution at $z=0$.
Opaque symbols pinpoint threshold values where $\varv_0=0.5$ (star), $\varv_2=0$ (diamond), $\varv_3$ is maximum (triangle), minimum (dot circle), and has a second global maximum (filled circle), illustrating the relationship between thermodynamical and chemical states, possibly described by some scaling relation, and the average morphology of the corresponding excursion-set.}
\label{fig:PhaseDiagramsFiducial}
\end{figure*}

\begin{figure*}
\centering\includegraphics[width=17cm]{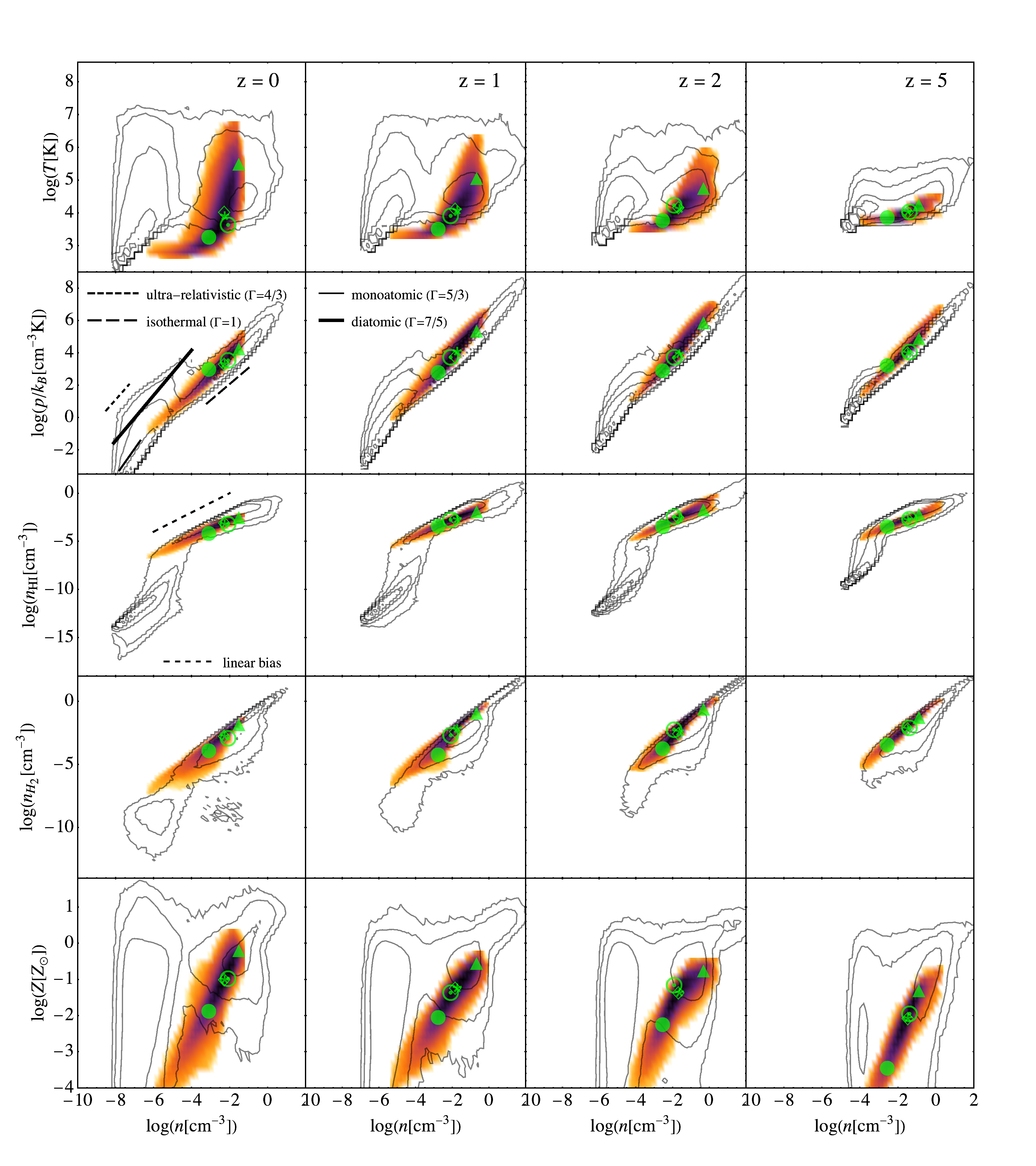}
\caption{Same as Figure~\ref{fig:PhaseDiagramsFiducial} but for the NoAGN model. As expected,  galactic nuclei impact jointly both the phase diagrams and their geometric and topological counterparts.
}
\label{fig:PhaseDiagramsNoAGN}
\end{figure*}

We describe here the analysis of the principal bivariate distributions or phase diagrams of the $T$, $P$, $n$, $n_\mathrm{HI}$, $n_{\mathrm{H}_2}$, and $Z$ fields for the Fiducial and NoAGN models at redshift $z=0,1,2,3,5$, before and after the smoothing required by the computation of MFs. This supports the tentative joint analysis of (smoothed) phase diagrams and morphology. Figures~\ref{fig:PhaseDiagramsFiducial}-\ref{fig:PhaseDiagramsNoAGN} report the logarithmic counts computed on the $128^3$ grid points both from the output of raw {\sc Simba} fields without additional smoothing (contour lines) and after smoothing (colours). The over-plotted symbols pinpoint field values $f_*$ defining excursion-sets with noteworthy MFs: excursion-sets filling half the volume, $\varv_0(f_*)=0.5$ (star); excursion-sets with vanishing integrated mean curvature, $\varv_2(f_*)=0$ (diamond); excursion-sets dominated by isolated regions, namely, with maximum $\varv_3(f_*)$ (triangle); excursion-sets dominated by tunnels, namely, with minimum $\varv_3(f_*)$ (dot circle); and excursion-sets with cavities, namely, where $\varv_3(f_*)$ has the second highest maximum (filled circle). We note that the $\varv_3(P)$ curve at $z=0$ is very noisy for low-pressure values and vanishing on average, showing only a global maximum and no minimum and no secondary maximum; the corresponding symbols are absent.

A detailed description of the phase diagrams follows:

\textbullet~ \textit{${T\!-\!n}$}: The temperature-density parameter space is commonly adopted to define the baryonic cosmic-web environments \citep{Fukugita+1998,Cen+1999,Dave+2001,Cen+2006,Powell2011}. Referring to the $z=0$ phase diagram of the Fiducial model (Figure~\ref{fig:PhaseDiagramsFiducial}, top-left panel), one can identify as hot medium the regions with $T>10^7$~K, typically under-dense or diffuse; warm-hot intergalactic medium (WHIM) and warm circumgalactic medium \citep[WCGM;][]{CGM2017} are regions with $10^5<T/\mathrm{K}<10^7$ and, respectively, $n\lesssim10^{-4}~\mathrm{cm}^{-3}$ and $10^{-4}\lesssim n/\mathrm{cm}^{-3}<0.13$; diffuse intergalactic medium and collapsed haloes are regions with $T<10^5$~K and same bounds on density as WHIM and WCGM; and star forming regions with $T<10^7$~K and $n>0.13~\mathrm{cm}^{-3}$. These bounds are consistent with \citet{Martizzi+2019,Martizzi+2020}, who performed a similar classification based on {\sc IllustrisTNG} simulations in the ${T\!-\!n_\mathrm{HI}}$ parameter space (on average $n_\mathrm{HI}\approx n$ for $n>10^{-5}\mathrm{cm}^{-3}$; see later). In agreement with the analysis of bias of smoothed field reported in the Appendix~\ref{app:phasediagrams} (see Figure~\ref{fig:Tsimba-TMFs-nHI}), smoothing erases the lower and higher-density regions ($n<10^{-6}\mathrm{cm}^{-3}$ and $n>10^{-2}\mathrm{cm}^{-3}$ at $z=0$) and the cold and hot regions ($T<10^{2.8}\mathrm{K}$ and $T>10^{6.8}\mathrm{K}$), so that MFs finally probe the morphology of excursion-sets that barely corresponds to the WCGM approximately at all redshift. The conclusion is qualitatively similar when X-ray heating is not included (not shown). 

If only stellar feedback is taken into account (see Figure~\ref{fig:PhaseDiagramsNoAGN}, first line, contour lines), the hot medium and the WHIM are less prominent, and the WCGM has an average lower temperature, especially at $z\leq1$; the phase-space diagrams of the Fiducial and NoAGN models are similar only at high redshift (see panel at $z=5$) when the AGN population is not sufficiently energetically powerful to affect the one-point statistics on large scales. The phase-space after smoothing for MFs (shaded areas in the figure) biases the raw bivariate distribution to a smaller extent in favour of a broader range of temperatures and densities than in the Fiducial model, here including colder and less dense regions especially at lower redshift, while it is qualitatively similar at $z\geq1$. Interestingly, without AGN feedback, the MFs defined by threshold values $(n_*, T_*)$ marked by symbols cover a wider portion of the phase-space, pinpointing also the collapsed haloes. This correspondence suggests that when investigating the feedback mechanisms operating on gas fields, a small number of MFs computed for appropriate excursion sets of $T$ and $n$-fields (namely, a few simple numbers, with associated small-size covariance matrix) provide complementary, valuable, and perhaps simpler information than one-point (bivariate) distributions (namely, continuous functions, with large non-trivial covariance matrices).

\textbullet~ \textit{$P\!-\!n_\mathrm{HI}$}: At all redshifts, and regardless of the feedback mechanism, the pressure-density diagram shown in Figure~\ref{fig:PhaseDiagramsFiducial} (second row, contours) indicates three phases: high-pressure domains in high-density regions ($P/k_\mathrm{B}>100\mathrm{cm}^{-3}\mathrm{K}$ and $n>10^{-4}~\mathrm{cm}^{-3}$ or $n_\mathrm{HI}>10^{-5}~\mathrm{cm}^{-3}$, at $z=0$), in which the gas is almost isothermal, namely, $p\propto n$; intermediate-pressure domains in low-density regions ($0.1 < P/k_\mathrm{B}\mathrm{cm}^{-3}\mathrm{K}<100$ and $n\simeq10^{-7}~\mathrm{cm}^{-3}$ or $n_\mathrm{HI}\simeq10^{-13}~\mathrm{cm}^{-3}$ at $z=0$), in which the gas has on average $P\propto n^\Gamma$, with a polytropic coefficient for diatomic molecules, $\Gamma=7/5$; and low-pressure domains in low-density regions ($P/k_\mathrm{B}<0.01\mathrm{cm}^{-3}\mathrm{K}$ and $n\simeq10^{-7}~\mathrm{cm}^{-3}$ at $z=0$), in which the gas behaves on average like monoatomic molecules with $\Gamma=5/3$. The smoothing for MFs computation (shaded in the figure) removes a large part of the second phase and completely erases the third one, allowing for a morphological description of the high-pressure domains only. However, this bias is less crucial for the smoothed $P$-field at higher redshift, almost absent at $z=5$. Very similar conclusions can be deduced for the NoAGN model.

\textbullet~ \textit{${n_\mathrm{HI}\!-\!n}$}: This diagram shows the relative bias of the neutral hydrogen with respect to the total gas density, which is almost monotonic; see the third rows in Figures~\ref{fig:PhaseDiagramsFiducial}-\ref{fig:PhaseDiagramsNoAGN}. Three phases coexist when all the AGN feedbacks are active; the one with lower density ($n_\mathrm{HI}\simeq10^{-15}~\mathrm{cm}^{-3}$ at $z=0$) is a genuine product of AGN activity, especially the X-ray heating and AGN jets, absent  only when stellar feedback is considered. Like in the pressure-density diagram, regardless of the feedback mechanism considered, the highest {\sc Hi} density domains survive to  smoothing for MFs and without any significant  bias; MFs can therefore accurately sample this phase, which on average is linearly biased to the underlying total gas density field (dashed line in the diagrams).

\textbullet~ \textit{${n_{\mathrm{H}_2}\!-\!n}$}: Analogously to the previous phase diagram, this one yields the relative bias of molecular hydrogen with respect to the total gas density; see the fourth rows in Figures~\ref{fig:PhaseDiagramsFiducial}-\ref{fig:PhaseDiagramsNoAGN}. As expected, proportionality holds only at high redshift and high density. AGN feedback mechanisms, especially X-ray heating and jets, increase the low-density phase, producing a third phase at $z<2$ with mean density $n_{\mathrm{H}_2}\approx10^{-9}~\mathrm{cm}^{-3}$ in the densest regions with $10^{-4}~\mathrm{cm}^{-3}<n<10^{-2}~\mathrm{cm}^{-3}$. Smoothing only drops the phase at a higher density, biasing the original population at $z=0$ but not at higher redshift. Consistently, MFs accurately probe this {\sc H$_2$} phase without important bias in the redshift range $1<z<5$.

\textbullet~ \textit{${Z\!-\!n}$}: The metallicity-density relation is shown in bottom panels of Figures~\ref{fig:PhaseDiagramsFiducial}-\ref{fig:PhaseDiagramsNoAGN}. Before smoothing, hydrodynamical interactions possibly augmented by stellar feedback produce two phases occupying regions with average density $n\simeq10^{-7}~\mathrm{cm}^{-3}$ and $10^{-2}~\mathrm{cm}^{-3}$. A third phase with high metallicity ($Z>Z_\odot$) is produced in the low-density regions ($10^{-8}~\mathrm{cm}^{-3}<n<10^{-6}~\mathrm{cm}^{-3}$) since $z=2$ by AGN feedback, specifically X-ray heating and jets. At $z=0$ this last population becomes one of the two dominant ones, with average metallicity $0.1<Z/Z_\odot<10$, at the expense of domains with lower values that survive only in the absence of AGN feedback (the relationship between the {\sc H$_2$} and the $Z$ phases pumped by AGN feedback at $z<1$  will be considered elsewhere). The distribution of the smoothed $Z$-field is almost 
unchanged, with phase mixing derived from smoothing of the $n$-field. Also in this case the MFs of well-chosen $Z$-fields pinpoint the high-density phase while fairly well reproducing the range of $Z$ values, at least until $z\simeq3$.

% ----------------------------------------

\end{appendix}

% ================================================

\end{document}